\documentclass[aps,pre,superscriptaddress,nofootinbib,twocolumn,10pt]{revtex4-2}

\usepackage{mathtools}
\usepackage[utf8]{inputenc} \usepackage[T1]{fontenc}    \usepackage{hyperref}       \usepackage{url}            \usepackage{booktabs}       \usepackage{amssymb,amsmath,amsfonts,dsfont}       \usepackage{nicefrac}       \usepackage{graphicx}       \usepackage{placeins}
\usepackage{float}
\makeatletter
\let\newfloat\newfloat@ltx
\makeatother
\usepackage{algorithm} 
\usepackage{algpseudocode} \usepackage{physics}
\usepackage[dvipsnames]{xcolor}
\usepackage{enumitem}
\usepackage{comment}

\newcommand{\bea}{\begin{eqnarray}}
\newcommand{\eea}{\end{eqnarray}}

\newcommand{\ff}{\tilde{\varphi}}

\newcommand{\vk}{{\vec{k}}}

\definecolor{dukeblue}{RGB}{1,33, 105}

\begin{document}

\title{The critical slowing down in training diffusion models}

\author{Luca Maria Del Bono}\thanks{Corresponding author: \href{mailto:lucamaria.delbono@uniroma1.it}{lucamaria.delbono@uniroma1.it.}}
\affiliation{Dipartimento di Fisica, Sapienza Università di Roma, Piazzale Aldo Moro 5, Rome 00185, Italy}
\affiliation{CNR-Nanotec, Rome unit, Piazzale Aldo Moro 5, Rome 00185, Italy}

\author{Giulio Biroli}
\affiliation{Laboratoire de Physique Statistique, École normale supérieure, PSL Research University, 24 rue Lhomond, 75005 Paris, France}

\author{Patrick Charbonneau}
\affiliation{Department of Physics, Duke University, Durham, North Carolina 27708, USA}
\affiliation{Department of Chemistry, Duke University, Durham, North Carolina 27708, USA}

\author{Marylou Gabrié}
\affiliation{Laboratoire de Physique Statistique, École normale supérieure, PSL Research University, 24 rue Lhomond, 75005 Paris, France}

\begin{abstract}
Computational sampling has been central to the sciences since the mid-20th century. While machine-learning-based approaches have recently enabled major advances, their behavior remains poorly understood, with limited theoretical control over when and why they succeed. Here we provide such insight for diffusion models---a class of generative schemes highly effective in practice---by analyzing their application to the $O(n)$ model of statistical field theory in the Gaussian limit $n \to \infty$. In this analytically tractable setting, we show that training a score model with a one-layer network architecture matching the exact solution exhibits a form of critical slowing down in parameter learning. This slowing down also impacts the generation process, indicating that the well-known difficulties of sampling near criticality persist even for learned generative models.  To overcome this bottleneck, we consider the power of architectural depth. We find that using a two-layer architecture drastically reduces the critical slowing down, with the training time scaling logarithmically rather than quadratically with system size. Using a Fourier implementation of the architecture, we further show that this acceleration in training time can be achieved without drastically increasing operational complexity. Taken together, these results demonstrate that diffusion models can overcome the critical slowing down through appropriate architectural design, and establish a controlled framework for understanding and improving learned sampling methods in statistical physics and beyond.

\end{abstract}

\maketitle

\section{Introduction}

The advent of electronic computers in the mid-20th century revolutionized numerical sampling~\cite{battimelli2020computer}. Markov chain Monte Carlo (MCMC) methods, in particular, have since become a cornerstone of modern scientific inquiry. 
They enable not only rigorous tests of theoretical predictions, but also the exploration of analytically intractable systems, with applications spanning fields as diverse as drug discovery, materials design, and financial risk analysis. 
Sampling can also serve as a powerful optimization tool---simulated annealing \cite{Kirkpatrick1983} being a foundational example---by recasting cost functions as energies 
and then evolving dynamics on the corresponding landscape. 

Despite the broad impact of sampling and optimization, their computational cost remains prohibitive for many systems of interest. In extreme cases, such hurdles have even motivated the development of custom-built hardware~\cite{BaityJesi2014}. 
Among the many sampling challenges encountered, few are as obfuscating as phase transitions---the very source of physical interest for many systems (like ferromagnetism and 
superconductivity)
and an unwelcome feature for most others (like the SAT-UNSAT transition in satisfiability problems \cite{monasson1997statistical,mezard2002analytic} and the emergence of spurious memories and catastrophic forgetting in associative memories \cite{amit1985spin, amit1985storing}). The standard Ising model of ferromagnetism, 
for instance, exhibits a critical slowing down upon approaching its disorder-to-order (paramagnetic-to-ferromagnetic) transition temperature.
The time required for simple MCMC schemes to decorrelate configurations then grows rapidly and diverges in the thermodynamic limit of infinitely large models. Although in simple cases clever schemes---such as Swendsen--Wang cluster updates---can mitigate this slowing down by capturing the growth of the underlying correlation length (see, e.g., ~\cite{newman1999,landaubinder2015}), the resulting gains are fragile and can be undone by introducing infinitesimal frustration in spin couplings~\cite{AlfaroMiranda2026}, {with system-specific cluster updates only partially mitigating the effect \cite{kandel1992cluster, Houdayer2001,zheng2022}% and suffering from lack of generality
}.

Recent years have opened a promising new route for overcoming sampling difficulties.   
Building on remarkable advances in machine-training methodologies \cite{Carleo2019, Senior2020, Dawid2022},  
\emph{generative models} have been developed to sample 
hard-to-study distributions. After proper training, these schemes can produce configurations (approximately) drawn from the target distribution. 
These approaches have shown promising results in a variety of problems ranging from optimization (i.e. ground state search) of complex systems~\cite{del2025demonstrating} to sampling of multimodal targets~\cite{noe2019boltzmann, invernizzi2022skipping, noble2024learned, schonle2025efficient, grenioux2025improving}. 
Different architectures have correspondingly been proposed, including autoregressive networks~\cite{wu2019solving,mcnaughtonBoostingMonteCarlo2020,del2025nearest, wang2025enhancing}, normalizing flows \cite{noe2019boltzmann, albergo2019flow,  kanwar2020equivariant, de2021scaling, gabrieAdaptiveMonteCarlo2022, gerdes2022learning, singha2023conditional}, and stochastic interpolants~\cite{chen2025scale,potaptchikTiltMatchingScalable2025}. Over the last decade, diffusion models---a subset of generative models rooted in statistical physics \cite{SohlDickstein2015,songScoreBasedGenerativeModeling2020,hoDenoisingDiffusionProbabilistic2020}---have set the bar for state-of-the-art image and video generation \cite{rombach2022high,DhariwalNichol2021, Saharia2022Imagen, sordo2025review, ma2025controllable}. More recently, they have also found uses in physical contexts \cite{BiroliMezard2023, noble2024learned, bae2025diffusion, sanokowski2025scalable}.

Although many practical refinements have been proposed \cite{matthewsContinualRepeatedAnnealed2022a,tanScalableEquilibriumSampling2025, schonle2025efficient}, physical understanding of these approaches remains underdeveloped. 
Efforts have thus far mainly focused on capturing the behavior of perfectly trained models by studying the generation dynamics for simple statistical physics systems \cite{BiroliMezard2023} and the limitations of their generation capability for systems with a random first order transition or in inference models with a hard phase \cite{ghio2024sampling}. Work on the training process has mainly been carried out, again, for simple statistical physics models for the specific class of autoregressive models \cite{del2025performance}, for energy based models in the Gaussian case \cite{aarts2024scalar, catania2025theoretical}, and for Gaussian mixture models \cite{soletskyiTheoreticalPerspectiveMode2025,foglianiAnnealingVariationalInference2026}.
However, key sampling challenges---such as the treatment of structural heterogeneity and extended spatial correlations---are left with no clear path forward. The critical slowing down, in particular, remains largely beyond the reach of even the most powerful methods. In cases where progress has been made, such as through the use of special encoding of spatial hierarchies at criticality~\cite{marchand2023}, generalizations are not obviously at hand.

In this work, we obtain physical insights into the capabilities of diffusion models for surmounting the critical slowing down by studying an archetype  from statistical field theory, the $n$-vector model in the large $n$ limit, $O(n\rightarrow\infty)$. The simplicity of this model brings fairly involved dynamical schemes within analytical reach, thus providing a controlled description of the training dynamics. 
Interestingly, 
we find that this model has a \emph{score function}---the central quantity of diffusion models---that is linear, and can therefore be fully represented by a simple, one-layer linear neural network of the same size as the system, $L^d$ in $d$ dimensions. 
Other key findings include (see Fig.~\ref{fig:summary}):
\begin{itemize}
    \item For a single linear layer architecture that perfectly captures the exact score, the training (or learning) dynamics exhibits a critical slowing down. 
    For gradient descent optimization, the slowing down of the training scales similarly as the per-spin relaxation time of simple MCMC schemes (and standard Langevin dynamics), $L^2$. 
    \item An overparameterized architecture---specifically, replacing the single linear layer with two such layers---dramatically reduces the scaling of the training time for gradient descent optimization from $L^2$ to $\log L$. 
    \item Although the critical regime calls for a score kernel that is inherently non-local, an implementation of the above architecture in Fourier space limits the overall computational cost of the forward pass to $N \log N \propto dL^d \log L$ ($N = (L/a)^d$, where $a$ is the lattice spacing used to discretize the system), instead of $N^2 \propto L^{2d}$ for a (na\"ive) real space implementation.
\end{itemize}
As a result, in this model the critical slowing down affecting Langevin dynamics, that has a total computational cost scaling as $NL^2\propto L^{d+2}$, can be overcome in all dimensions, a clear advance for statistical physics.

Our work also sheds light on less well-understood aspects of machine learning. Although linear networks have been extensively studied in both shallow and deep regimes---including convergence~\cite{arora2018convergence, eftekhari2020training}, training dynamics{~\cite{saxe2014exact, arora2018optimization, saxe2019mathematical, tarmoun2021implicit, labarriere2024optimization, wang2026analytical}}, and implicit biases~\cite{gunasekar2018implicit, gidel2019implicit, varre2023spectral}, even in Gaussian settings{~\cite{pierret2025diffusion, merger2025generalization}}---the critical slowing down of their training (and corresponding generation process) has remained largely unexplored. Leveraging datasets from statistical physics here provides a controlled setting in which to investigate this phenomenon. %Additionally, we demonstrate the effectiveness of local kernels in statistical physics applications. The role of locality in diffusion models has only recently begun to be studied \cite{lukoianov2026locality}, particularly in connection with generalization properties \cite{kamb2025an} and even more recently in statistical-physics settings, in relation with the  $k$-SAT and $k$-XORSAT problems~\cite{bhatt2026generating}. However, these studies have largely been empirical. Our framework, by contrast, provides a principled prescription for how locality should scale with system size and dimension in order to control approximation errors, a clear advance for machine learning.

The rest of the paper is organized as follows. Section~\ref{sec:background}  introduces the problem setting, in particular the $O(n)$ model from statistical field theory (Sec.~\ref{sec:On_model}) and the functioning of diffusion models (Sec.~\ref{sec:diffusion}). Section~\ref{sec:theory} presents the results for the one-layer network architecture. In particular, Sec.~\ref{sec:exact_score} derives the exact score of the model, Sec.~\ref{sec:training} describes the training dynamics under gradient descent, 
and Sec.~\ref{sec:generation} considers the generation dynamics. 
Section \ref{sec:twolayers} describes the training dynamics under gradient descent of the two-layer case, and Sec.~\ref{sec:fourier} motivates and assesses the implementation of the score in Fourier space. Section \ref{sec:conclusions} concludes by recapitulating the findings and offering perspectives on future developments.

\section{Background}\label{sec:background}
In this section, we review first the physical behavior of the $n$-vector model, $O(n)$, in the Gaussian limit $n\to\infty$ and then that of score-based generative models. Note that because both the physical $O(n)$ model and its diffusion probabilistic counterpart are customarily referred to as \emph{models}, some terminological ambiguity may appear unavoidable at times. We have nevertheless been careful throughout the text to distinguish between the two contexts whenever necessary.

\subsection{$O(n)$ model and critical slowing down}\label{sec:On_model}
The $O(n)$ model was first proposed by Eugene Stanley in the late 1960s~\cite{stanley1968dependence} and has since become a staple of statistical field theory (see, e.g., Refs.~\cite{itzykson1991statistical,ZinnJustin2002,mussardo2020statistical}). Its action reads
\begin{equation}\label{eq:fullaction}
\begin{split}
\mathfrak{S}(\vec{\varphi}) &= \frac{1}{2}\times{}\\
&\int_V d^d \vec{x}\,\left[
\sum_{\substack{\nu=1,\\a=1}}^{d,n}\bigl(\partial_\nu \varphi_a\bigr)^2
+ m^2 \sum_{a=1}^n \varphi_a^2
+ \frac{g}{2n}\left(\sum_{a=1}^n \varphi_a^2\right)^2
\right],
\end{split}
\end{equation}
where $\vec{\varphi} = \vec{\varphi}(\vec{x})$ is a $n$-components vector field over a $d$-dimensional volume $V$, and $m^2$ and $g>0$ control the model behavior with the \emph{mass} parameter $m$ playing a role analogous to temperature. In the limit $n\rightarrow\infty$, the action  becomes Gaussian (see App.~\ref{app:lambda_eq_derivation})
\begin{equation}\label{eq:lambda_action}
\mathfrak{S}_\Lambda(\vec{\varphi}) = \frac{1}{2}\int_V d^d \vec{x} \left[ 
 \sum_{\substack{\nu=1,\\a=1}}^{d,n} (\partial_\nu \varphi_a)^2 
+ (m^2 + \Lambda g) \sum_{a=1}^{n} \varphi_a^2 
\right],
\end{equation}
with parameter $\Lambda$ self-consistently determined by
\begin{equation}
\label{eq:lambdadefinition}
\Lambda = 
\frac{1}{(2\pi)^d} \int d^d \vec{k} \,
\frac{1}{\vec{k} \cdot \vec{k} + m^2 + \Lambda g},
\end{equation}
as in Ref.~\cite[Eq.~(17.5)]{Fradkin2021} up to some constants in the definition of the action.
In this limit, the effective mass, $m^2_\mathrm{eff} = m^2+\Lambda g$, fully controls the model behavior. In particular, a phase transition takes place at the critical point $m_\mathrm{eff}=0$.
Hereafter, we focus on the high-temperature regime down to that point, i.e., $m_\mathrm{eff} \geq0$.\footnote{The ferromagnetic phase expected for $m_\mathrm{eff}^2<0$ is here unreachable, because the Gaussian $\mathcal{O}(n \to\infty)$ model 
is then ill-defined.}

\begin{figure*}[th]
    \centering
    \includegraphics[width=0.85\linewidth]{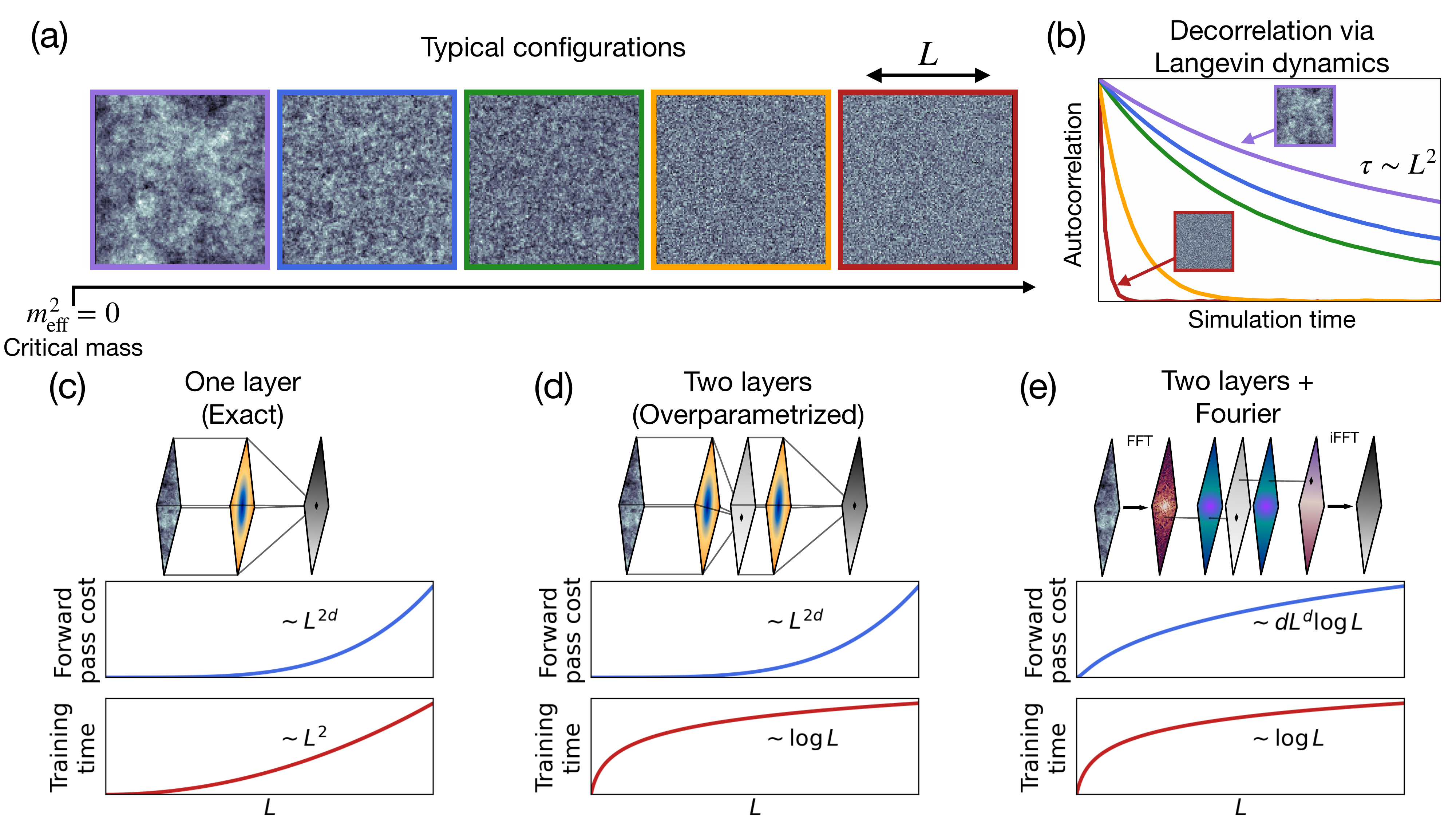}
    \caption{\textbf{Top row: Schematic of the physical setup and critical slowing down.} 
    (\textbf{a}) Equilibrium $d=2$ configurations of one of the field components for systems with an action as in Eq.~\eqref{eq:lambda_action}, shown for different effective masses (frame colors). As $m_{\mathrm{eff}}$ decreases, the correlation length increases, thus giving rise to spatially extended field fluctuations. (\textbf{b}) Time decay of the field autocorrelation function, Eq.~\eqref{eq:autocorrelation}, for the same systems as in (\textbf{a}). As $m_{\mathrm{eff}}$ decreases, the autocorrelation decays increasingly slowly and hence its \textbf{autocorrelation time} $\tau$ increases. At criticality, $\tau\sim L^2$, where $L$ is the linear size of the system. \textbf{Bottom row: Schematic of the main results.} (\textbf{c}) A one-layer network architecture that matches the structure of the exact score  exhibits the same $L^2$ slowing down at criticality as simple MCMC sampling and Langevin dynamics. (\textbf{d}) By contrast, introducing overparameterization via a second linear layer reduces the time scaling to $\log L$. (\textbf{e}) Fourier space convolution reduces the number of operations for the forward pass of the score to $dL^d\log L$. This operation entails (fast) Fourier transforming the configuration, computing the convolution, and then inverse (fast) Fourier transforming.}
    \label{fig:summary}
\end{figure*}

In this regime, we are interested in generating configurations of the fields $\varphi(\vec{x})$ with the Boltzmann--Gibbs probability
\begin{equation}\label{eq:GB}
    P(\vec{\varphi}) = \frac{1}{\mathcal{Z}}e^{-\mathfrak{S}_\Lambda(\vec{\varphi})},
\end{equation}
where $\mathcal{Z}=\int \mathcal{D}\vec{\varphi}e^{-\mathfrak{S}_\Lambda(\vec\varphi)}$ is the partition function. Upon sampling this distribution with a standard local dynamics (such as the Metropolis--Hastings MCMC sampling scheme---the Swiss-knife of numerical simulations---or Langevin dynamics), a critical slowing down is observed upon approaching $m_\mathrm{eff}=0$. This marked sluggishness of the sampling dynamics 
can be gleaned from the characteristic decay time of the 
self-correlation function 
\begin{equation}\label{eq:autocorrelation}
    C(\vec{x},\vec{x};\mathcal{T})=\frac{1}{V}\int_V d^dx\;\Big\langle \varphi(\vec{x},\mathcal{T})\,\varphi(\vec{x},0)\Big\rangle,
\end{equation}
where $\langle\dots\rangle$ denotes averaging over the distribution in Eq.~\eqref{eq:GB} and $\mathcal{T}$ the simulation runtime as measured by, e.g., the number of simulation steps per unit volume. 
As $m_\mathrm{eff}\to 0$, this correlation time increases (see Fig.~\ref{fig:summary}); for $m_\mathrm{eff}= 0$, it grows with systems size and hence diverges in the thermodynamic (infinite $V$) limit. In particular, for an hypercubic system of side $L$, the characteristic decay time of the correlation function in Eq.~\eqref{eq:autocorrelation} diverges as $L^z$, where the \textit{dynamical critical exponent} $z=2$ for standard local dynamics.\footnote{This setup corresponds to the \textit{Model A} of Ref.~\cite{hohenberg1977theory}, for which $z=2+\eta$ with the \textit{anomalous dimension} $\eta$ scaling as $\eta\sim 1/n$.} 
The critical slowing down---a physical phenomenon related to the emergence of long-range spatial correlations---therefore hampers simulation methods based on variants of physical dynamics. 
In this work, we investigate whether diffusion models can overcome this limitation and provide an alternative framework for sampling physical systems.

Note that although the condition for criticality is well-defined, it cannot be said \textit{a priori} that there exist a finite $m^2$ such that, upon solving Eq.~\eqref{eq:lambda_action}, $m_\text{eff}^2$ vanishes. For example, $d = 1$  yields 
\begin{equation}
    m_\text{eff}^2 = m^2+\frac{g}{2}\frac{1}{\sqrt{m_\text{eff}^2}},
\end{equation}
which for $m_\text{eff}^2 \to 0$ implies a (negatively) diverging $m^2$,  i.e., $m^2 \sim -1/\sqrt{m_\text{eff}^2}$.\footnote{For statistical field theoretical models, $m^2$ can be negative. Although the parameter is then no longer related to a physical temperature, the model remains well-defined.} Analogously, after introducing an ultraviolet (UV) cutoff,\footnote{This cutoff amounts to excluding Fourier modes larger than a cutoff value $k_{\max}=2\pi/a$, thus setting a minimum distinguishable length scale following the discretization of space in (hyper)cubes of side $a$.} $d = 2$  yields a divergence of the form $m^2 \sim \log{m_\text{eff}^2}$. In other words, no finite-$m$ (i.e., finite-temperature) phase transition takes place in $d \leq 2$, as expected from the Hohenberg--Mermin--Wagner theorem 
\cite{mermin1966absence, halperin2019hohenberg}. (For $d\geq 3$, by contrast, a finite-$m$ solution always exists.) 
For convenience, we henceforth {work directly with the Gaussian field theory and use $m_\mathrm{eff}$ as the control parameter in} Eq.~\eqref{eq:lambda_action}, thus allowing us to probe criticality by imposing {$m_\mathrm{eff}\to0$} even for $d\leq2$. 

\subsection{Diffusion models}\label{sec:diffusion}

Diffusion models seek to generate data according to a target distribution $P^*(\vec{\varphi})$~\cite{SohlDickstein2015, songScoreBasedGenerativeModeling2020, hoDenoisingDiffusionProbabilistic2020, BiroliMezard2023}. They consist of (i) a forward (noising) and (ii) a backward (denoising) process.  First, data $\vec{{\varphi}}(\vec{x},0)$ {extracted according to $P^*(\vec{\varphi})$} is noised through an Ornstein--Uhlenbeck process,
\begin{align}
&\vec{{\varphi}}(\vec{x},0) \sim P^*(\vec{\varphi}) \\
\label{eq:ODE}
&\frac{\partial \vec{{\varphi}}(\vec{x},t)}{\partial t}  = -\vec{\varphi}(\vec{x},t)+\vec{\zeta}(\vec{x},t),
\end{align}
where $\vec{{\varphi}}(\vec{x},t)$ is a time-dependent vector field and $\vec{\zeta}$ is a white Gaussian noise {with covariance $\left\langle \zeta(\vec{x},t)\zeta(\vec{x}',t') \right\rangle
=
2\,\delta^{(d)}(\vec{x}-\vec{x}')\,\delta(t-t')$, where $\delta^{(d)}$ is the $d$-dimensional Dirac delta function}. 
Denoting $P_t$ the instantaneous distribution of the noised fields at time $t$, we note that at large times $t_{\max} \gg 1$, $P_{t_{\max}}$ evolved from $P_0 = P^*$ is approximately normally distributed. 
Second, the result is denoised running the reverse-process ordinary differential equation (ODE) from $t_{\max}$ back to $t=0$:\footnote{A stochastic differential equation (SDE) approach is also possible. The backward diffusion (denoising) equation is then
\begin{equation}\label{eq:ODEdenoising}
    -\partial_t \vec{\varphi}(\vec{x}, t) = \vec{\varphi}(\vec{x}, t) + 2\vec{\mathcal{F}}(\vec{\varphi}(\vec{x}, t)) + \vec{\zeta}(\vec{x}, t).
\end{equation}}
\begin{equation}\label{eq:ode}
    -\partial_t \vec{\varphi}(\vec{x}, t) = \vec{\varphi}(\vec{x}, t) + \vec{\mathcal{F}}_t(\vec{\varphi}(\vec{x}, t))
\end{equation}
where $\vec{\mathcal{F}}_t(\vec{\varphi})$ is the \textit{score} function,
\begin{equation}\label{eq:score}
\vec{\mathcal{F}}_t(\vec{\varphi}) = \frac{\delta \log P_t(\vec{\varphi})}{\delta \vec{\varphi}(\vec{x})} .
\end{equation}
If this backward (denoising) process is solved exactly, then one is guaranteed to obtain data sampled from the target probability distribution back at time $t=0$~\cite{Anderson1982}. In other words, data can be generated approximately according to $P^* = P_0$ by first generating data according to $P_{t_{\max}}$---easily achieved if one approximates $P_{t_{\max}}$ as a Gaussian---and then solving Eq.~\eqref{eq:ode} backward in time. In order to do so, however, one needs the score  defined in Eq.~\eqref{eq:score}, which in most cases cannot be obtained  analytically. Therefore, for  machine training applications $\vec{\mathcal{F}}_t(\vec{\varphi})$ is typically parametrized via a deep neural network $\vec{F}_t$ that is trained by minimizing the score matching loss, 
\begin{equation}
\label{eq:matchingloss}
        \mathcal{L}_t(\vec{F}_t) = \frac{1}{2}\int \mathcal{D}\vec{\varphi} \, P_t(\vec{\varphi}) \left\| \vec{F}_t(\vec{\varphi}) - \vec{\mathcal{F}}_t(\vec{\varphi}) \right\|^2,
\end{equation}
using straight in lieu of calligraphic fonts to distinguish trained from exact functions. 
{Because Eq.~\eqref{eq:matchingloss} involves the exact score $\vec{\mathcal{F}}_t$, the approach might look circular. However, it can be shown that, apart from constants that do not influence optimization, the loss in Eq.~\eqref{eq:matchingloss} is equivalent to} 
\begin{equation}\label{eq:bg_diffusion_dsm_loss}
    {\widehat{\mathcal{L}}_t(\vec{F}_{t})
    =
    \frac{1}{2} \mathbb{E}_{\vec{\varphi}_0 \sim P_0}
    \int \mathcal{D}\vec{\varphi}\,
    P_t(\vec{\varphi}|\vec{\varphi}_0)
    \left\|
        \vec{F}_{t}(\vec{\varphi})
        +
        \frac{\vec{\varphi}-e^{-t}\vec{\varphi}_0}{\Delta_t}
    \right\|^2 ,}
\end{equation}
{where the expectation is taken over the initial field, $\vec{\varphi}_0$, extracted according to $P_0$, and $P_t(\vec{\varphi}|\vec{\varphi}_0)$ is the probability that $\vec{\varphi}_0$ is noised as the field $\vec{\varphi}$ at time $t$, which is Gaussian and therefore known. The (denoising score matching) loss \eqref{eq:bg_diffusion_dsm_loss}, therefore, does not require knowing $\vec{\mathcal{F}}_t$. Because here $\vec{\mathcal{F} }_t$ can be computed exactly---as shown in Sec.~\ref{sec:exact_score}---the form in Eq.~\eqref{eq:matchingloss} is used to describe training.} %of the score $\vec{F}_t$. }

\subsection{Errors in diffusion models}
The procedure described above is affected by multiple sources of error that arise at different stages.
\begin{itemize}
    \item {\bf Training dynamics.} The error resulting from training the score $\vec{F}_t(\vec{\varphi})$ for a finite time is discussed in Sec.~\ref{sec:theory}. For a one-layer architecture, this contribution dominates because it is affected by a critical slowing down. This effect also propagates to the generated fields. This error is the main focus of this work.
    \item {\bf Sample complexity.} The error resulting from the finiteness of the training dataset is analyzed in App.~\ref{app:discretizations}. Maintaining a fixed relative accuracy in the score (and hence in generated fields) does not require increasing the dataset size with system size. This contribution is always subdominant.
    \item {\bf Discretization error.} The error resulting from discretizing Eq.~\eqref{eq:ode} is also analyzed in App.~\ref{app:discretizations}. The scaling of this error depends on the choice of numerical integration scheme. A simple Euler scheme suffices to keep the relative error on the generated fields constant. The absolute error then scales linearly in the linear size of the system and is therefore controlled by growing the number of integration steps with $L$. Although for a one-layer architecture this scaling is subdominant, it is not so for multi-layer architectures. Higher-order integrators can nevertheless markedly reduce this scaling.
\end{itemize}

The error that results from these approximations  can be variously measured. Metrics commonly used in computer science and applied mathematics include the Wasserstein distance, the Kullback--Leibler (KL) divergence, and the total variation distance, each computed between the distribution of the generated fields and the target distribution. 
Because these metrics act on the full probability distribution, they are broadly applicable. However, they can be difficult---or even impossible---to evaluate in practice, and their connection to physical observables is often not straightforward. In physics, by contrast, error estimates are typically tied to specific observables of interest, such as the action, the energy, the magnetization, or the power spectrum.
An advantage of the Gaussian $\mathcal{O}(n\rightarrow\infty)$ model is that it offers an immediate link between a physically meaningful quantity---the covariance matrix of  the generated fields---and the KL divergence. When evaluating the error in the distribution of generated fields, we therefore mainly focus on this covariance. 
The error on the distribution of generated fields is also related to the error on the score, which is usually more analytically tractable. We will therefore sometimes consider both the total error on the score and the error on individual Fourier modes in the reciprocal space of field configurations. Nevertheless, standard errors may not properly weigh the physically relevant features, so particular attention is paid to this effect. Indeed, while controlling the latter allows to describe the generation dynamics under the usage of an approximated score (Sec. \ref{sec:generation}), bounding the former does not suffice to guarantee the reproduction of all the meaningful physical characteristics of the model (App. \ref{app:local_score}).

\section{One-layer neural network}\label{sec:theory}
The simplicity of the action for the $O(n\rightarrow\infty)$ model, Eq.~\eqref{eq:lambda_action}, allows for both the score and the generation process to be extracted, thereby providing important insights into the behavior of the associated diffusion model. In this section, we first show that the score is linear. As a result, it can be exactly represented by a linear one-layer network. In this setting, we can then study the training dynamics, the different sources of error on generation, and how these two depend on the distance from the critical point. 

In particular, we show that upon training the (approximate) score using gradient descent, the training exhibits a slowing down akin to that of the local dynamics in the limit $m_\mathrm{eff} \to 0$. We additionally show that the error induced by this slowing down propagates to the distribution of generated fields. 

\subsection{Computation of the exact score}\label{sec:exact_score}

As can be gleaned from Eq.~\eqref{eq:lambda_action}, the various field components, $\varphi_a$, of the $O(n\to \infty)$ model are effectively decoupled, so considering a single one suffices. Dropping the subscript $a$ to lighten the notation, the resulting probability distribution at time $t$ of the forward diffusion process is (see App.~\ref{app:score_computation})
\begin{multline}\label{eq:forward_process_prob}
P_t (\tilde{\varphi}) \propto
\exp\left\{ -\frac{(2\pi)^d}{2}\times\right. \\ 
\left. \int d^d \vec{k}\,
\left[
\frac{k^2 + m_{\text{eff}}^2}
{\Delta_t\left(k^2 + m_{\text{eff}}^2\right) + e^{-2t}}
\right]\,
\tilde{\varphi}(\vec{k})\,\tilde{\varphi}(-\vec{k})
\right\}.
\end{multline}
where $\tilde{\varphi}$ is the Fourier transform of the field $\varphi$ and 
$\Delta_t = 1-e^{-2t}$.
The \emph{exact} score in Eq.~\eqref{eq:score} is then
\begin{equation}\label{eq:exact_score}
\mathcal{F}{ _t}(\varphi) =  -\int d^d \vec{y} \; \mathcal{S}{ _t}(x,y) \varphi(\vec{y}),
\end{equation}
thus implicitly defining the \emph{exact} \textit{score kernel}
\begin{widetext}
\begin{equation}\label{eq:sigma}
\begin{gathered}
\mathcal{S}_t(\vec{x},\vec{y})
=
\int \frac{d^d \vec{k}}{(2\pi)^d}\,
\frac{k^2 + m_{\mathrm{eff}}^2}{\Delta_t (k^2 + m_{\mathrm{eff}}^2) + e^{-2t}}\,
e^{i \vec{k}\cdot(\vec{y}-\vec{x})} = 
\frac{1}{\Delta_t}\,\delta^{(d)}(\vec{y}-\vec{x})
-
\frac{e^{-2t}}{(2\pi)^{d/2}\Delta_t^2}\,
\left(
\frac{M_t}{|\vec{y}-\vec{x}|}
\right)^{\frac{d}{2}-1}
K_{\frac{d}{2}-1}\!\left(M_t |\vec{y}-\vec{x}|\right),
\end{gathered}
\end{equation}
\end{widetext}
where $M_t^2=m_{\mathrm{eff}}^2+\frac{e^{-2t}}{\Delta_t}$ and $K_\nu$ is the modified Bessel function of the second kind of order $\nu$. $\mathcal{S}_t(\vec{x},\vec{y})$ has a strictly local part (given by the delta function) and a tail part. At short separations, $|\vec{x}-\vec{y}|\ll M_t^{-1}$, the tail part exhibits the standard Yukawa behavior, reducing to a power-law singularity $\mathcal{S}_t(\vec{x},\vec{y})\sim |\vec{x}-\vec{y}|^{-(d-2)}$ for $d>2$ (and to a logarithmic one in $d=2$). By contrast, at large separations, $|\vec{x}-\vec{y}|\gg M_t^{-1}$, the tail part is exponentially suppressed as $\mathcal{S}_t(\vec{x},\vec{y})\sim e^{-M_t |\vec{x}-\vec{y}|}/|\vec{x}-\vec{y}|^{(d-1)/2}$. Note that because $M_t$ diverges for $t \to 0$ and decreases exponentially (to $m_\mathrm{eff}$) as $t$ increases, score truncation is more significant at smaller forward diffusion times and less so as $t$ increases, especially for $m_\mathrm{eff}^2 = 0$.

As expected for a Gaussian physical model, the score is linear in the fields $\varphi_t$. Moreover, as expected for a translationally invariant model, the score kernel only depends on the spatial separation, $\vec{x}-\vec{y}$. The score is therefore equivalent to a linear convolutional layer with kernel size equal to the system’s linear dimension with a single channel and circular padding so as to keep the output size equal to the system size.\footnote{Recall that in convolutional networks each output channel is a feature map obtained by its own set of weights; a single channel corresponds to the output having only one scalar-valued field. Padding extends the input beyond its boundaries before convolution; circular padding amounts to wrapping the field periodically.} In addition, the reciprocal space score kernel is diagonal in that it depends on a single momentum,
\begin{equation}\label{eq:momentum_kernel}
    \tilde{\mathcal{S}}{ _t}(\vec{k}) =  
\frac{k^2 + m^2_{\text{eff}}}{\Delta{_t}(k^2 + m^2_{\text{eff}}) + e^{-2t}}.
\end{equation}

Note that we here first took the limit $n \to \infty$ of the action in Eq.~\eqref{eq:fullaction}, thus obtaining a Gaussian model, and then performed the forward (noising) process to obtain Eq.~\eqref{eq:forward_process_prob}. Doing the opposite---first noising the action in Eq.~\eqref{eq:fullaction} and then taking the limit $n \to \infty$---yields the same result, as shown in App.~\ref{sec:check_exchange}.

\subsection{Critical slowing down in training dynamics}\label{sec:training}
As mentioned in Sec.~\ref{sec:diffusion}, in a typical setting  the exact score is not known and has to be learned from data. This is the procedure we study here.  We learn, via gradient descent, an \emph{approximate score} parameterized in the same form as the exact score---namely, a single linear convolutional layer with kernel size equal to the system’s linear dimension,
\begin{equation}
\label{eq:approxscore}
F_t(\varphi)= -\int d^d \vec{y} \; S_t(\vec{x},\vec{y}) \varphi(\vec{y}).
\end{equation}
The resulting trained score can then be evaluated by direct comparison with the exact expression in Eq.~\eqref{eq:sigma}. 
Let us first suppose that we have access to the true distribution of the data, $P_t(\varphi)$, and use these data to train the kernel $S_t(x,y)$ of the approximate score, Eq.~\eqref{eq:approxscore}. For this task, we first consider a gradient flow optimization of $S_t$ based on the score matching population loss of Eq.~\eqref{eq:matchingloss} for a single field.
Written in terms of the kernels, that loss is
\begin{equation}
\mathcal{L}_t = \frac{1}{2} \int \mathcal{D}\varphi \, P_t(\varphi) 
\left\| \int d^d\vec{y} \left[ S_t(\vec{x}, \vec{y}) - \mathcal{S}_t(\vec{x}, \vec{y}) \right] \varphi(\vec{y}) \right\|^2,
\end{equation}
and, by definition, under the gradient flow the training kernel evolves as
\begin{equation}
    \frac{dS_t(\vec{x}\,', \vec{y}\,')}{d\bar{t}} = - \eta \frac{\delta \mathcal{L}_t}{\delta S_t(\vec{x}\,', \vec{y}\,')},
\end{equation}
where $\eta$ is the learning rate and $\bar{t}$ is the \textit{training} time (not to be confused with the \textit{noising} time $t$). We therefore obtain the (operator) equation
\begin{equation} \label{eq:operator_equation}
\frac{dS_t}{d \bar{t}} = -{\eta}(S_t - \mathcal{S}_t) C_t,
\end{equation}
expressed using the two-point correlation function
\begin{equation}
C_t(\vec{x}, \vec{y}\,)=\langle \varphi(\vec{x}) \varphi(\vec{y}\,) \rangle_t ,
\end{equation}
where $\langle \cdot \rangle_t$ denotes averaging over $P_t$.
Starting from a completely untrained kernel, $S_{t}(\bar{t}=0) = 0$, the solution to Eq.~\eqref{eq:operator_equation} is
\begin{equation} \label{eq:solution_score_full}
S_t(\bar{t}) = \mathcal{S}_{t} \left(1 - e^{-{\eta}C_t\bar{t}} \right).
\end{equation}
The timescales $\tau_t$ that control the training of the kernel are therefore set by the inverse of the eigenvalues $\tilde{C}_t$ of $C_t$, which, because $C_t$ is translationally invariant, can be obtained from its Fourier transform,\footnote{Indeed, they are simply related by a factor $(2 \pi)^d$ arising from the Fourier transform convention chosen.} and the field correlation in Fourier space, 
\begin{equation}
    \langle \tilde{\varphi}(\vec{k})\tilde{\varphi}(\vec{q})\rangle_t = \frac{\delta^{(d)}(\vec{k}+\vec{q})}{(2\pi)^d}\tilde{C}_t(\vec{k}),
\end{equation}
for which one finds
\begin{equation}
\label{eq:solution_score}
\tilde{C}_t(\vec{k}) = \frac{\Delta{_t}(\vec{k}^2 + m_{\text{eff}}^2) + e^{-2t}}{\vec{k}^2 + m_{\text{eff}}^2},
\end{equation}
and the corresponding timescales
\begin{equation}\label{eq:timescale_momentum}
\tau_t(\vec{k}) = \frac{\vec{k}^2 + m_{\text{eff}}^2}{\Delta{_t}(\vec{k}^2 + m_{\text{eff}}^2) + e^{-2t}}.
\end{equation}
Note that in the limit of long noising times, all timescales tend to unity,
\begin{equation}
    \lim_{t\rightarrow\infty} \tau_t(\vec{k})=1,
\end{equation}
and hence the trained kernel converges to the exact one on all spatial scales concurrently. In the short-noising-time limit, by contrast, training timescales are more broadly distributed,
\begin{align}\label{eq:char_time}
\lim_{t \to 0} \tau_t(\vec{k}) = \vec{k}^2 + m_{\text{eff}}^2.
\end{align}
{These $t \to 0$ timescales exactly match, mode-by-mode, the autocorrelation timescales of the Langevin dynamics.}
{The smallest training timescale is therefore the one associated with the smallest momentum $k_{\mathrm{min}} = |\vec{k}_{\mathrm{min}}|$:}
\begin{equation}
    {\tau_{\min} = k_{\mathrm{min}}^2 + m_{\text{eff}}^{2}.}
\end{equation}

\begin{figure}[t]
    \centering
    \includegraphics[width=\columnwidth]{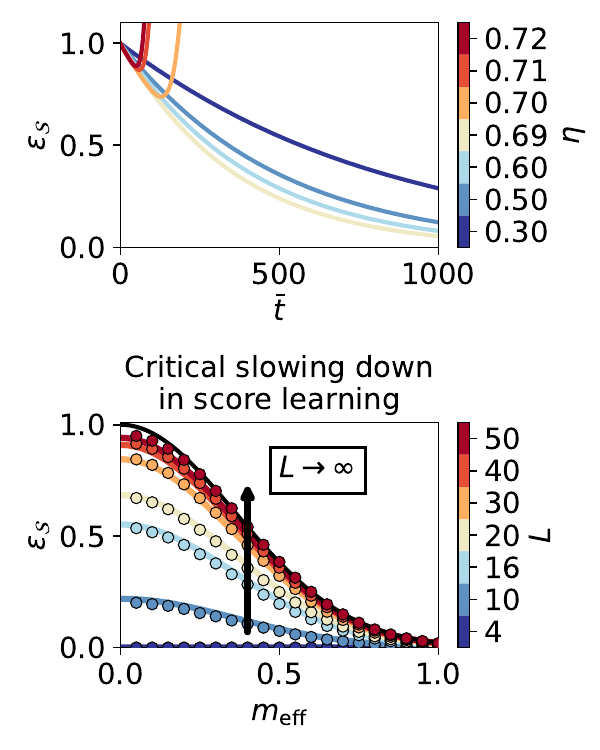}    
    \caption{\textbf{Top:} Relative error in the trained score kernel, ${\varepsilon_{\mathcal{S}}}$, Eq.~\eqref{eq:relative_error_score}, {as a function of training time $\bar{t}$ (measured in epochs)} under gradient descent evolution for $k_{\min} = 2\pi/L$ with $L=20$ and {effective mass} $m_\text{eff} = 1/2$. The evolution eventually becomes uncontrolled if $\eta>\eta_\mathrm{max}=2(k_{\min}^2+m_\mathrm{eff}^2) = 0.697$. \textbf{Bottom}: Same quantity (colored data points) as a function of $m_\text{eff}$ for different system sizes $L$ at fixed training time $\bar{t} = 500$. The learning rate is taken to be near the largest possible one,  $\eta_L = 2 f\left( m_{\mathrm{eff}}^2 + \left( {2\pi}/{L} \right)^2 \right)$ with $f = 0.95$. 
    Theoretical predictions (colored {full} lines) take the form ${\varepsilon_{\mathcal{S}}} = e^{-\bar{t}/\tau_L^\text{eff}}$ with $ \tau_L^\text{eff} = \frac{m_\text{eff}^2+({\pi}/{a})^{2}}{2f(m_\text{eff}^2 + ({2\pi}/{L})^{2})}$, including for $L \to \infty$     (black {full}  line). The agreement between theory and numerics is near quantitative. \textbf{Setting:} $d=1$, $a = 0.5$, noising time $t = 10^{-4}$, and training dataset size $M = 10^5$.}
    \label{fig:csd_one}
\end{figure}

Upon discretizing the gradient flow so as to obtain a practical gradient descent scheme, algorithmic convergence requires using a sufficiently small time step or, equivalently,  a sufficiently small learning rate $\eta$. Classical results from {one-dimensional} convex optimization prescribe {an optimal learning rate $\eta_\mathrm{opt} \simeq \tau$ and maximum learning rate for convergence $\eta \lesssim 2 \tau$} (see, e.g., \cite{Saad2003} and \cite[Eq.~(18)]{mehta2019high}) {for a single mode with timescale $\tau$}. {In our multi-dimensional linear case, the maximum possible learning rate {$\eta_\mathrm{max}$} is set by the smallest timescale across all modes, $\eta < 2 \tau_{\min} = \eta_\mathrm{max}$}.\footnote{{The optimal learning rate $\eta_\mathrm{opt}$ in the minimax-optimal sense, on the other hand, is given by \cite{nesterov2013introductory}
\begin{equation}
    \eta_\mathrm{opt} = \frac{2}{\tau_{\min}^{-1} + \tau_{\max}^{-1}}.
\end{equation}
Because $\eta_\mathrm{max}/2\leq \eta_\mathrm{opt} \leq \eta_\mathrm{max}$, the optimal learning rate has the same scaling as the maximum learning rate. Moreover, the two almost coincide for large $L$ and small $m_\mathrm{eff}^2$ because then $\tau_{\min} \ll \tau_{\max}$.}} Larger learning rates lead to the divergence of the relative error of the trained score,
\begin{equation}\label{eq:relative_error_score}
    {\varepsilon_{\mathcal{S}}} = \frac{\sum_{x,y}|S(x,y)-\mathcal{S}(x,y)|}{\sum_{x,y}|\mathcal{S}(x,y)|},
\end{equation}
as the training time grows (see Fig.~\ref{fig:csd_one}). For $\eta = \eta_{\mathrm{max}}$, at small noising times, $t \simeq 0$,  %with the choice of 
the effective training time $\tau^\text{eff}(\vec{k})$ taken to learn mode $\vec{k}$ is therefore given by
\begin{equation}\label{eq:eff_timescale}
    \tau^\text{eff}(\vec{k})= {\frac{\tau(\vec{k})}{\eta_{\mathrm{max}}}= \frac{\tau(\vec{k})}{2\tau_{\min}}= \frac{m_\text{eff}^2 + \vec{k}^2}{2 (m_\text{eff}^2+k_{\mathrm{min}}^2)}}.
\end{equation}

{The smallest training timescale, which is associated with the smallest momentum $k_{\min} = 0$, leads to a divergence of the effective timescale for every $k = |\vec{k}| > 0$ as $1/m_{\mathrm{eff}}^2$. In other words, when the range $(k_\mathrm{min},k_\mathrm{max})$ results in a ratio of timescales $\tau^{\mathrm{eff}}(k_\mathrm{max})/\tau^{\mathrm{eff}}(k_\mathrm{min})$ that diverges, gradient descent is no longer possible on all $k$ scales at once. This divergence of the training timescale is a first form of critical slowing down in training. However, %the $1/m_{\mathrm{eff}}^2$ scaling is very fragile. Indeed, since that 
because the zero mode, which corresponds to the mean of the configurations, can be learned separately, we can set it to zero without loss of generality.\footnote{{Analogously, we could consider a the natural size-dependent infrared cutoff in the statistical field theory sense.}} For a system of finite linear size $L$ the smallest training timescale is then associated with mode $|\vec{k}|=2 \pi/L$, which gives }
\begin{equation}
\tau_{\min} = (2\pi/L)^2 + m_{\text{eff}}^{2}.
\end{equation}
As a result, at $m_\mathrm{{eff}}^2 = 0$ the effective time is finite but scales as $\tau^\text{eff} \sim L^2$ for any $k = \mathcal{O}(1)$. \footnote{Notice that, because the subsequent $k$ modes scale as $1/L$, removing the mode $k = 2\pi/L$ does not cure the diverging behavior of the effective training time, at variance with the $k = 0$ mode. The $L^2$ scaling is robust against this type of operations.}

Recalling from Sec.~\ref{sec:background} that the dynamical critical exponent for Langevin dynamics for this model is $z=2$, we find that the critical scaling of the training timescale is therefore the same as for the critical slowing down of local dynamics. In other words, in the critical regime the standard approach to training the score kernel is no more efficient than standard sampling schemes.

The diverging effective training time implies that for a constant training time---such as fixing the amount of computational resources devoted to training the score kernel---the training procedure worsens as the system size increases (see Fig.~\ref{fig:csd_one}).\footnote{Here and in all the numerical experiments, although the system is defined on a discretized lattice of lattice spacing $a$, we use the continuum dispersion relation and associate each Fourier mode with the inverse variance $k^2+m_{\mathrm{eff}}^2$, rather than with its lattice counterpart $\widehat{k}^{\,2}+m_{\mathrm{eff}}^2$, where
$
\widehat{k}^{\,2} = 
\frac{4}{a^2}\sum_{\mu=1}^{d}
\sin^2\left(\frac{a k_\mu}{2}\right).
$
We make this choice to facilitate a direct comparison with the corresponding continuum theory. No qualitative change is expected when using the discrete symbols, as the low-momentum modes that control the critical behavior are approximately the same.} Interestingly, this behavior is consistent with what has been reported in both practical applications and theoretical studies of comparable systems \cite{deldebbio2021, del2025performance}, thus highlighting the general hardness of training statistical physics models upon approaching their critical temperature.~\footnote{The non-monotonic behavior reported in Ref.~\cite[Fig.~4]{deldebbio2021} in the ferromagnetic phase is here inaccessible because $m_\mathrm{eff}^2\geq 0$ for the Gaussian $\mathcal{O}(n\rightarrow\infty)$ model (see Sec.~\ref{sec:background}).}.

We have checked that similar difficulties arise for other optimization algorithms. For instance, (optimally tuned) momentum improves the scaling of the effective training time to $L$, as expected from classical optimization theory \cite{Polyak1964}, thus weakening albeit not eliminating the critical slowing down. By contrast, more advanced algorithms like Adagrad \cite{Duchi2011Adagrad} or Adam \cite{KingmaBa2015Adam} altogether fail to converge to the exact kernel score---in this simple convex case---when optimization is carried out in real space ({see Supplementary Material~\cite{SM}}), as previously reported \cite{Reddi2018AdamConvergence}. Therefore, although the precise scaling depends on the choice of algorithm, an effective critical slowing down persists even when considering optimization schemes more advanced than gradient descent.

{Note that the above discussion is agnostic to the choice of physical dynamics. The training time only depends on (i) the machine-learning choices of optimization procedure and score architecture, and (ii), crucially, the equilibrium measure of the physical system of interest. For example, if instead of using classical Langevin dynamics one used conserved-order-parameter dynamics (i.e., \textit{Model B} of Ref.~\cite{hohenberg1977theory}), then the dynamical critical exponent would be $z = 4$. But because standard Langevin dynamics  (\textit{Model A}) and Model B share the same equilibrium measure, the critical scaling of the score training would be the same for both cases.}

\subsection{Generation (denoising) with infinite data}\label{sec:generation}
We now consider whether the distribution of generated fields is also affected by a critical slowing down. To do so, we consider the denoising (backward diffusion) of the fields in Fourier space for three different cases: (1) the \emph{exact score} $\mathcal{F}_t$, Eq.~\eqref{eq:score}, which is the exact reverse process of the noising (forward diffusion) equation, Eq.~\eqref{eq:ODE}; (2) the \emph{approximate score from a fixed training time} $\bar{t}$, which corresponds to considering, for each diffusion time $t$, a score trained up to $\bar{t}$; (3) the \emph{approximate score from a fixed relative error} $\bar\varepsilon$, which corresponds to considering, for each diffusion time $t$, a score $F_t = (1-\bar{\varepsilon})\mathcal{F}_t$ trained so as to reach $\bar{\varepsilon}$ relative error to the true score.

\subsubsection{Exact score}\label{sec:generation_exact_score}

For the exact score in Eq.~\eqref{eq:exact_score} with the Fourier space kernel in Eq.~\eqref{eq:sigma}, we find that the denoising  process Eq.~\eqref{eq:ode} becomes 
\begin{equation}\label{eq:diffusion_ode_exact}
    \partial_t \tilde{\varphi}^*(\vec{k}, t) = -\tilde{\varphi}^*(\vec{k}, t) \left [1-\frac{\vec{k}\cdot \vec{k}+m_\text{eff}^2}{\Delta_t(\vec{k}\cdot \vec{k}+m_\text{eff}^2)+e^{-2t}}\right ],
\end{equation}
where in this section we use $\tilde\varphi^*$ to denote the field coming from the exact backward diffusion equation. Equation~\eqref{eq:diffusion_ode_exact} can be integrated exactly, yielding
\begin{equation}\label{eq:solution_backward}
    \tilde{\varphi}^*(\vec{k}, t;t_\mathrm{max}) = \frac{
 \sqrt{e^{-2t} + \Delta_{t} \, (\vec{k} \cdot \vec{k}+m_\text{eff}^2)}
}{
\sqrt{e^{-2t_{\max}}+ \Delta_{t_{\max}} \, (\vec{k} \cdot \vec{k}+m_\text{eff}^2)}
}\tilde{\varphi}^*(\vec{k}, t_{\max}),
\end{equation}
where $\tilde{\varphi}^*(\vec{k}, t_{\max})$ is the starting field generated at the maximum diffusion time $t_{\max}$ according to $P_{t_\mathrm{max}}$. 
Taking the limit ${t_{\max} \to \infty}$ and starting with a white-noise field $\tilde{\varphi}^*(\vec{k}, \infty)$ gives
\begin{align}
\label{eq:exact-score-denoising-field-infinite-time}
\tilde{\varphi}^*(\vec{k}, t;\infty)  &= 
\lim_{t_\mathrm{max}\rightarrow\infty}\tilde{\varphi}^*(\vec{k}, t;t_\mathrm{max}) \nonumber \\
&= \frac{
 \sqrt{e^{-2t}+ \Delta_{t} \, (\vec{k} \cdot \vec{k}+m_\text{eff}^2)}
}{
\sqrt{\vec{k} \cdot \vec{k}+m_\text{eff}^2}
}\tilde{\varphi}^*(\vec{k}, \infty),
\end{align}
which is exactly distributed according to $P_t$. Therefore, the process remains Gaussian at all times with the variance of the Fourier components matching Eq.~\eqref{eq:forward_process_prob}.
More importantly, starting from an isotropic normal distribution at $t_\mathrm{max} < \infty$ introduces but an exponentially small error for all $m_\mathrm{eff}$.

\subsubsection{Approximate score from a fixed training time $\bar{t}$}\label{sec:generation_fixed_training_score}

For the approximate score $S_t$ trained for a time $\bar{t}$ with learning rate $\eta$---as given by Eq.~\eqref{eq:solution_score_full}---the generation dynamics Eq.~\eqref{eq:diffusion_ode_exact} becomes
\begin{align}\label{eq:diffusion_ode_approx}
    \partial_t \tilde{\varphi}(\vec{k}, t) =& \nonumber\\
    -\tilde{\varphi}(\vec{k}, t)& \left [1-\frac{\vec{k}\cdot \vec{k}+m_\text{eff}^2}{\Delta_t(\vec{k}\cdot \vec{k}+m_\text{eff}^2)+e^{-2t}}\left(1-e^{-\frac{\eta\bar{t}}{\tau_t(\vec{k})}}\right)\right ],
\end{align}
where $\tau_t(\vec{k})$ is given by Eq.~\eqref{eq:timescale_momentum}.
For $\tilde \varphi^*(\vec{k}, t_{\max})$ the starting field generated at the maximum diffusion time $t_{\max}$, this dynamics can be explicitly integrated to give
\begin{equation}\label{eq:solution_fixed_training_time}
\begin{multlined}
\tilde{\varphi}(\vec{k}, t; t_{\max})
=\exp\!\Biggl\{
\tfrac12\,e^{-\eta\bar{t}}
\Biggl[
\operatorname{Ei}\!\Bigl(
e^{-2t}\,\eta\bar{t}\,
\frac{k^2+m_\mathrm{eff}^2-1}{k^2+m_\mathrm{eff}^2}
\Bigr)
\\[2pt]
-\operatorname{Ei}\!\Bigl(
e^{-2t_{\max}}\,\eta\bar{t}\,
\frac{k^2+m_\mathrm{eff}^2-1}{k^2+m_\mathrm{eff}^2}
\Bigr)
\\[2pt]
+e^{\eta\bar{t}}
\Biggl(
-\operatorname{Ei}\!\Bigl(
\eta\bar{t}\left[
e^{-2t}\frac{k^2+m_\mathrm{eff}^2-1}{k^2+m_\mathrm{eff}^2}-1
\right]
\Bigr)
\\[2pt]
+\operatorname{Ei}\!\Bigl(
\eta\bar{t}\left[
e^{-2t_{\max}}\frac{k^2+m_\mathrm{eff}^2-1}{k^2+m_\mathrm{eff}^2}-1
\right]
\Bigr)
\Biggr)
\Biggr]
\Biggr\}
\,\tilde{\varphi}^*(\vec{k}, t; t_{\max}) \,,
\end{multlined}
\end{equation}
where $\tilde{\varphi}^*$ is given by Eq.~\eqref{eq:solution_backward} and $\mathrm{Ei}(x)$ is the exponential integral,
\begin{equation}
    \operatorname{Ei}(x) = -\int_{-x}^{\infty} \frac{e^{-t}}{t} \, dt
\quad \text{for } x > 0.
\end{equation}
As expected, this expression reduces to Eq.~\eqref{eq:solution_backward} in the limit $\bar{t} \to \infty$.

\subsubsection{Approximate score from a fixed error $\bar{\varepsilon}$ }\label{sec:generation_fixed_error_score}
If instead of fixing the training time $\bar{t}$ one fixes the error $\bar{\varepsilon}$ made in training the score, $S_t = \mathcal{S}_t (1-\bar{\varepsilon})$, Eq.~\eqref{eq:diffusion_ode_exact} becomes
\begin{equation}\label{eq:solution_fixed_error}
    \partial_t \tilde{\varphi}(\vec{k}, t) = -\tilde{\varphi}(\vec{k}, t) \left [1-\frac{\vec{k}\cdot \vec{k}+m_\text{eff}^2}{\Delta_t(\vec{k}\cdot \vec{k}+m_\text{eff}^2)+e^{-2t}}\left(1-\bar{\varepsilon}\right)\right ],
\end{equation}
which for the same initial condition gives
\begin{multline}\label{eq:fixed_error}
\tilde{\varphi}(\vec{k}, t;t_\mathrm{max}) =\\
\shoveleft{\frac{e^{-\bar{\varepsilon}t}}{e^{-\bar{\varepsilon}t_{\max}}}
\left(
\frac{
 e^{-2t} + \Delta_{t}\,(\vec{k}\cdot\vec{k}+m_\text{eff}^2)
}{
 e^{-2t_{\max}} + \Delta_{t_{\max}}\,(\vec{k}\cdot\vec{k}+m_\text{eff}^2)
}
\right)^{\frac{1-\bar{\varepsilon}}{2}}
\,\tilde{\varphi}(\vec{k}, t_{\max}).}
\end{multline}
Here again, this expression reduces to Eq.~\eqref{eq:solution_backward} for $\bar{\varepsilon} \to 0$.

\begin{figure}
    \centering
    \includegraphics[width=\columnwidth]{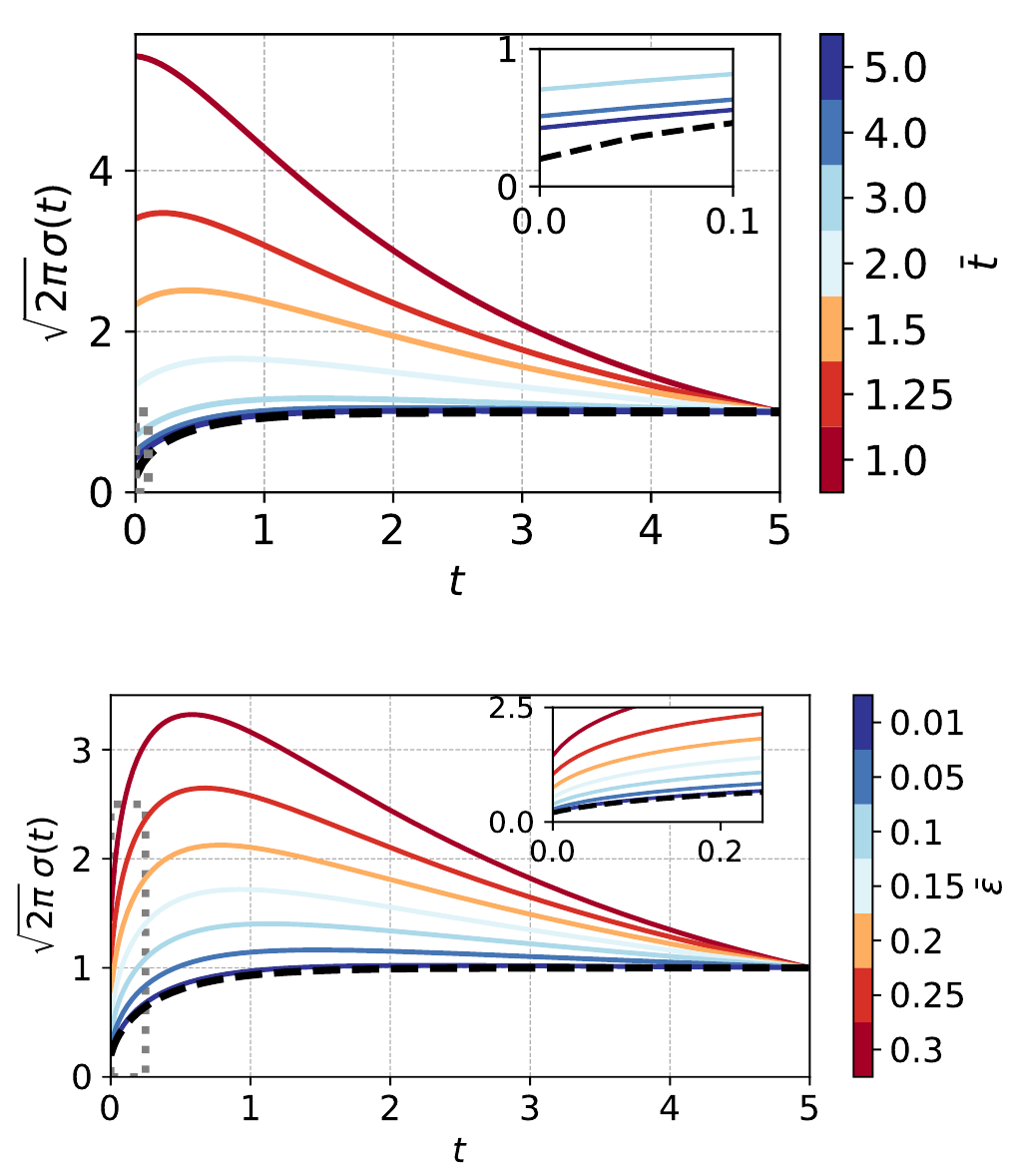}
    \caption{Standard deviation $\sigma(t)$ of the Fourier components of the fields from approximate (colored lines) and exact (Eq.~\eqref{eq:solution_backward}; dashed black lines) generation (denoising) dynamics for: (\textbf{top}) a fixed training time $\bar{t}$,  Eq.~\eqref{eq:solution_fixed_training_time}, and (\textbf{bottom}) a fixed relative score error $\bar{\varepsilon}$, Eq.~\eqref{eq:solution_fixed_error}. (insets) Enlarged low-$t$ region.     \textbf{Setting:} $d = 1$, $k = 5$, $m_\mathrm{eff} = 0.2$, $t_{\max} = 5$.}\label{fig:std_trained_time_and_error}
\end{figure}

Figure~\ref{fig:std_trained_time_and_error} illustrates the behavior of Eq.~\eqref{eq:solution_fixed_training_time} (for finite $\bar{t}$) and Eq.~\eqref{eq:fixed_error} (for finite $\bar{\varepsilon}$). In both cases, deviations from ideality monotonically increase the variance after complete denoising, at $t=0$. For small $\bar{t}$ or large $\bar{\varepsilon}$, however, the denoising evolution of the variance is non-monotonic.  The trained score kernel becomes larger or smaller than 1 as $t$ varies---as determined by the exact score kernel growing as $t$ decreases (when $k^2 +m_\mathrm{eff}^2 >1$)---while the relative error grows less quickly or stays constant. In other words, the overshoot reflects the trained score not correctly counter-balancing the exponential growth from the reverse-Ornstein--Uhlenbeck term $\varphi$.

\subsection{Generation (denoising) error}\label{sec:error_generation}

The results of Sec.~\ref{sec:generation} make possible an overall evaluation of the generation error. Because the fields remain Gaussian throughout the reverse diffusion process, they are also Gaussian at generation time, $t = 0$; and because the mean is zero (for a linear score without biasing), the distribution of the fields generated by the backward diffusion (denoising) process is completely characterized by its covariance. Consequently, the difference between the generated and the exact covariances provides a natural measure of generation error. We here quantify this error by considering the Frobenius error on the reconstructed covariance matrix $C$ for fixed training time $\bar{t}$ starting from Eq.~\eqref{eq:solution_fixed_training_time}:
\begin{equation}\label{eq:error_covariace}
    \varepsilon_C(\bar{t}, L) \;=\; \frac{\left\|\,C_{\mathrm{gen}}(\bar{t}, L) - C_*(L)\,\right\|_F}{\left\|\,C_*(L)\,\right\|_F},
\end{equation}
where $C_*$ and $C_{\mathrm{gen}}(\bar{t})$ are the covariance matrices estimated on data generated using the data distribution $P^*$ and on data generated by the diffusion model trained for $\bar{t}$ epochs, respectively, and $||\cdot||_F$ denotes the Frobenius norm.

One can expand the evolution of the fields in Eq.~\eqref{eq:solution_fixed_training_time} for $m_\mathrm{eff} = 0$, $t = 0$ (generation time), $t_\mathrm{max} \gg 1$, and $\eta\bar{t} \approx 0$. To leading order, the typical relative error in the estimate of the standard deviations in reciprocal space scales as $e^{-\eta\bar{t}t_\mathrm{max}} = e^{-{\bar{t}t_\mathrm{max}}/{L^2}}$, when $\eta$ is taken to scale as the maximal learning rate, $\eta = 1/L^2$ (see Sec.~\ref{sec:training}). Put differently, the critical slowing down on training is reflected in the critical slowing down of generating data with the correct covariance matrix. 

\begin{figure*}[t!]
    \centering
    \includegraphics[width=1\linewidth]{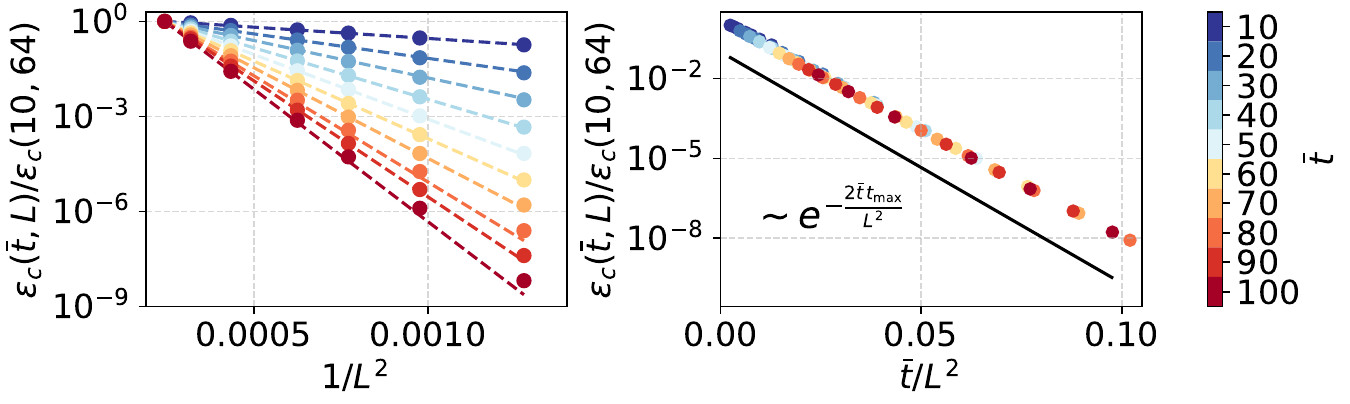}
    \caption{Critical slowing down in configuration generation. \textbf{Left:} System size dependence of the Frobenius error of the matrix, $\varepsilon_C$ from Eq.~\eqref{eq:error_covariace}, scaled by that error for $\bar{t} = 10$ and $L = 64$ (points) for different training times. Results are fitted to the form $Ae^{-C/L^2}$ (dotted line). \textbf{Right}: Same  results rescaled as $\bar{t}/L^2$. The ensuing collapse follows the predicted scaling $\sim e^{-\frac{2\bar{t}\, t_{\mathrm{max}}}{L^2}}$ (black curve). Small discrepancies are due to the various approximations. \textbf{Setting:} $d=2$, $m_\mathrm{eff} = 0.001$, and $a = 1$.    }
    \label{fig:curves_frob_analytical}
\end{figure*}

This prediction can be variously evaluated. We first test it by generating fields in reciprocal space from Eq.~\eqref{eq:solution_fixed_training_time}, reverse-Fourier transforming them, and then comparing the empirical covariance with the exact one. Figure~\ref{fig:curves_frob_analytical} shows the Frobenius error, Eq.~\eqref{eq:error_covariace}, as a function of system size, together with exponential fits of the form $Ae^{-C/L^2}$. In all cases, $C \simeq 2t_{\max} \bar{t}$. A similar trend applies to other error estimates, as exemplified for the Wasserstein-1 error of the probability distribution of the fields~\cite{SM}. We can also compare errors in the generated distribution with known bounds. In particular, Ref.~\cite[Thm.~1]{benton2024error} gives that the Wasserstein-2 error for the $k$-th mode obeys, for a fixed error $\bar{\varepsilon}$,
\begin{equation}\label{eq:wasserstein_bound}
W_{2}(k)
\le
|\bar{\varepsilon}|
\sqrt{\frac{t_{\max}}{2}\log A_k}\,
\exp\!\left(\int_0^{t_{\max}} L_k(t)\, dt\right) .
\end{equation}
where $L_k(t)$ is the Lipschitz constant
\begin{equation}
L_k(t)
=
\left|
-1 + (1-\bar{\varepsilon})\,
\frac{k^2 + m_{\mathrm{eff}}^2}{e^{-2t} + \Delta_t(k^2 + m_{\mathrm{eff}}^2)}
\right| .
\end{equation}
In our case, we obtain
\begin{equation}\label{eq:exact_wasserstein}
W_{2}(k)
=
e^{t_{\max}} A_k^{-1/2}
\left| A_k^{\bar{\varepsilon}/2} - 1 \right|,
\end{equation}
where $A_k = 1 + \left(e^{2t_{\max}} - 1\right) (k^2 + m_{\mathrm{eff}}^2)$. As expected, this error is always smaller or equal to that in Eq.~\eqref{eq:wasserstein_bound}. However, because the ODE case has been less studied than the SDE one, associated bounds are expected to be relatively weak, as is here the case.

\subsection{Summary}
For a one-layer network, in the limit $m_\mathrm{eff} \to 0$, a critical slowing down affects the gradient descent training of the score. This effect directly leads to an imperfect denoising (generation) dynamics, which in turn leads to the same scaling of the error in the final generated distribution of fields. Although these results have here been derived for an infinite number of training configurations and by analytically solving the generation dynamics, neither the use of a finite dataset nor the discretization of the generation dynamics change scaling of the resulting error (see App.~\ref{app:discretizations}). In the one-layer case, the error in the score approximation therefore dominates other effects.

Note that the above analysis could be extended to the case in which the exact score is not linear, but is learned using a linear form. In this scenario, the linear score will not learn the exact one, but will instead learn the best linear approximation, which can be written in terms of the covariance of the data. The same reasoning on the training timescales can therefore be applied to this general case as well by simply taking into account the covariance matrix of the true data distribution.

\section{Two-layer neural network}\label{sec:twolayers}

\begin{figure*}
    \centering
    \includegraphics[width=\linewidth]{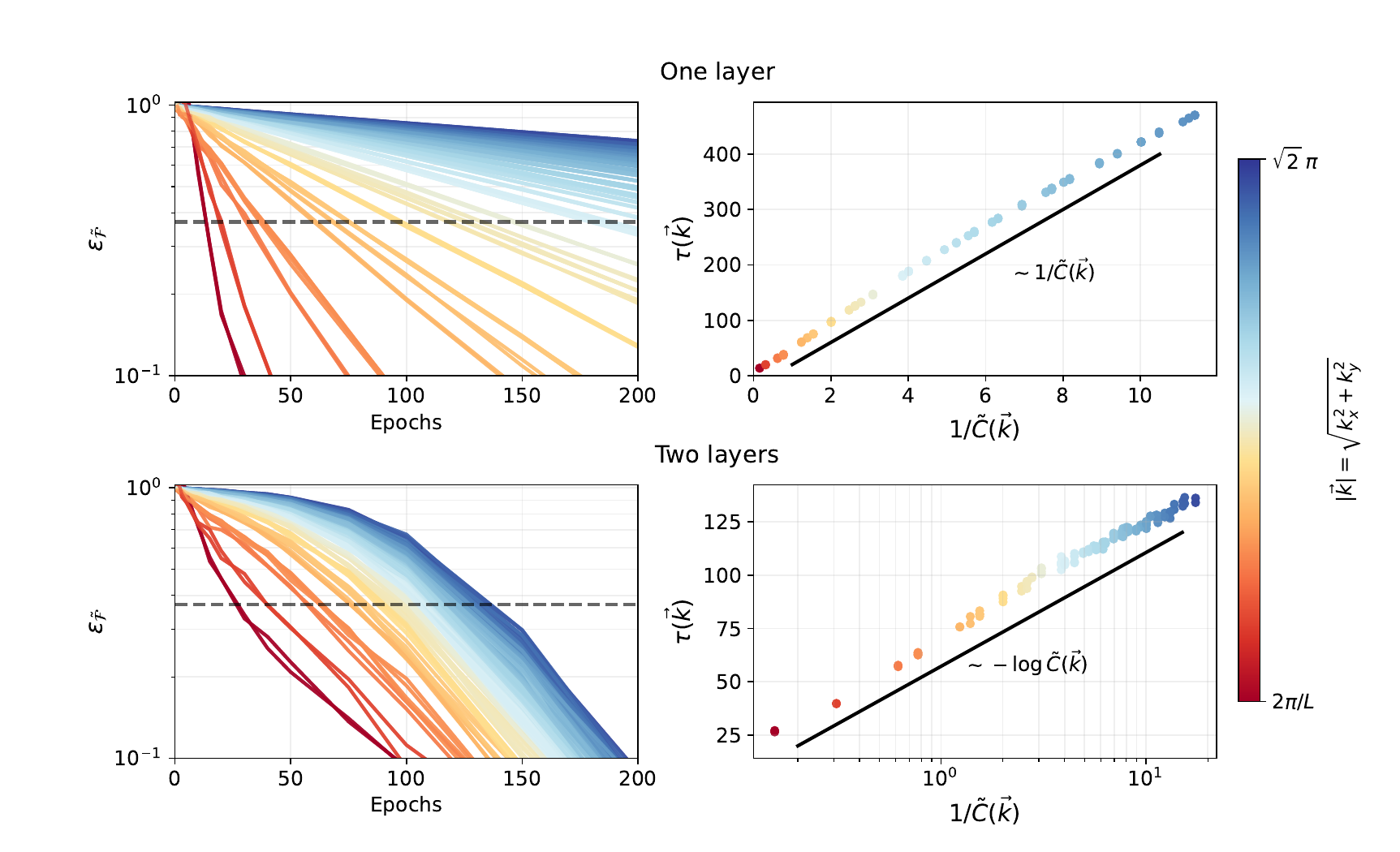}
    \caption{Error analysis for the one-layer network architecture (\textbf{top row}) and the two independent layer architecture (\textbf{bottom row}). \textbf{Left column:} relative error in the real part of the score, $\varepsilon_{\widetilde{\mathcal F}}(\vec{k})$ (Eq. \eqref{eq:fourier_score_error}), for different Fourier modes as a function of training time. {The dashed black lines corresponds to an error $e^{-1} \simeq 0.37$.} \textbf{Right column:} Characteristic decay time $\tau(\vec{k})$ of the covariance eigenvalue for different modes, $\widetilde C(\vec{k}) = 1/k^{2}$. In the one-layer case, the decay time scales linearly with the covariance eigenvalue; in the two-layer case, that scaling is logarithmic {(as evidenced by the reference full black lines)}. \textbf{Setting:}  $d=2$, $m_\mathrm{eff} = 0$, $a = 1$, $t = 0.01$, $L = 16$, and (top) $\eta = 0.01$ and (bottom) $\eta = 0.005$. The  optimization performed using gradient descent over 1024 training configurations, and the average is taken over 500 simulations. Part of the code used for the superstructure of the diffusion model (e.g. the integration algorithm) was adapted from Ref.~\cite{wang2024diffusion}, using a standard machine training implementations via \texttt{PyTorch} convolutions.}
    \label{fig:1vs2layers_sim}
\end{figure*}

A key motivation for overparameterized networks is their potential to accelerate learning. Here, we investigate whether this acceleration can alleviate the critical slowing down identified in Sec.~\ref{sec:theory}. To do so, we model the score with a minimal overparameterized architecture: a network composed of two linear convolutional layers (i.e., without activation functions), each with a kernel spanning the entire system. Because both layers are linear and capture the full range of the exact score in Eq.~\eqref{eq:exact_score}, this model has the same functional expressivity as the one-layer network considered in Sec.~\ref{sec:theory}. We find, however, that gradient-descent training on this overparameterized architecture leads to markedly improved convergence.

\subsection{Gradient-descent dynamics}
Consider an overparameterized linear score $F^{\mathrm{ov}}_{t}$ modeled via two translationally invariant kernels, $S^{(1)}_{t}$ and $S^{(2)}_{t}$,
\begin{equation}
F^{\mathrm{ov}}_{t}(\varphi)
= -\int d^d\vec{z}\,d^d\vec{y}\;
S^{(2)}_{t}(\vec{x},\vec{z})\,S^{(1)}_{t}(\vec{z},\vec{y})\,\varphi(\vec{y}).
\end{equation}
The corresponding gradient flow equations are then
\begin{align}
\frac{dS^{(1)}_{t}}{d\bar{t}}
&= -\eta\,\big(S^{(2)}_{t}\big)^{\dagger}\Big(S^{(2)}_{t}S^{(1)}_{t}\,C_{t}-\mathbb{I}\Big)\\
\frac{dS^{(2)}_{t}}{d\bar{t}}
&= -\eta\,\Big(S^{(2)}_{t}S^{(1)}_{t}\,C_{t}-\mathbb{I}\Big)\,\big(S^{(1)}_{t}\big)^{\dagger},
\end{align}
where the identity matrix $\mathbb{I}$ follows from Eqs. \eqref{eq:momentum_kernel} and \eqref{eq:solution_score}, which give  \(\mathcal{S}_{t}=C_{t}^{-1}\). (See also \cite[Eq.~(8)]{BiroliMezard2023} and App.~\ref{app:score_computation} for details.) In Fourier space, these expressions diagonalize, and the evolution of each mode is described by a pair of coupled differential equations (assuming the Fourier modes are real),
\begin{align}
\frac{d\tilde{S}^{(1)}_{t}(\vec{k})}{d\bar{t}}
&= -\eta\,\tilde{S}^{(2)}_{t}(\vec{k})
\left(\tilde{S}^{(2)}_{t}(\vec{k})\,\tilde{S}^{(1)}_{t}(\vec{k})\,\tilde C_{t}(\vec{k})-1\right),\\
\frac{d\tilde{S}^{(2)}_{t}(\vec{k})}{d\bar{t}}
&= -\eta\,
\left(\tilde{S}^{(2)}_{t}(\vec{k})\,\tilde{S}^{(1)}_{t}(\vec{k})\,\tilde C_{t}(\vec{k})-1\right)
\tilde{S}^{(1)}_{t}(\vec{k}),
\end{align}
which have a fixed point at \(\tilde{S}^{(2)}_{t}(\vec{k})\,\tilde{S}^{(1)}_{t}(\vec{k})\,\tilde C_{t}(\vec{k})=1\), i.e., \(S^{(2)}_{t}S^{(1)}_{t}\to \mathcal{S}_{t}\). Note that given the overparameterized nature of this setup, no unique expression determines the optimal $\tilde{S}^{(1)}_{t}$ and $\tilde{S}^{(2)}_{t}$ separately; multiple choices equivalently minimize the loss function.

Interestingly, this dynamics is closely related to that studied in Ref.~\cite{saxe2014exact}, but differs in one key respect. In that work, training inputs are whitened, so that the input covariance matrix is the identity, while the input--output correlation is a general matrix. By contrast, in our setting the input--output correlation is fixed to the identity by construction, whereas the input covariance matrix is nontrivial. This distinction reflects the underlying structure of the two models: whitening the inputs is a generic preprocessing step that can almost always be carried out; the identity input--output correlation arises specifically from the Gaussian structure of diffusion models.

\subsection{Simplified case}\label{sec:simple_case}

In order to tease out the behavior of the two-layer architecture,  we first consider a simplified version of the above coupled equations. As in Ref.~\cite{saxe2014exact}, we assume that the two layers are the same at initialization and are therefore the same at all subsequent times, as can easily be shown. 

Dropping the \(\vec{k}\) and $t$ dependencies, {\( s= \tilde{S_t}^{(1)}(\vec{k})=\tilde{S_t}^{(2)}(\vec{k})\)} and setting \(c= \tilde C_t(\vec{k})>0\) reduces the dynamics of each mode to\footnote{{Notice that, if instead of initializing the two layers to be equal we repeated a single layer twice, we would have an additional factor 2. This factor can be however reabsorbed in the learning rate without changing the following discussion.}}
\begin{equation}\label{eq:gf_single_twice}
{\dot{{s}}=\eta\, s\,(1-c\, s^{2}).}
\end{equation}
Setting {\(u(t)=s(t)^2\)} gives the logistic equation
\begin{equation}\label{eq:logistic}
\dot u=2\eta \,u\,(1-cu),
\end{equation}
with solution for \(u_0= s(0)^2\)
\begin{equation}
u({\bar{t}})=\frac{u_0\,e^{2\eta \bar{t}}}{1-cu_0+cu_0\,e^{2\eta \bar{t}}},
\end{equation}
and therefore
\begin{equation}
{ s({\bar{t}})}=\mathrm{sgn}({ s(0)})\,
\sqrt{\frac{u_0\,e^{2\eta \bar{t}}}{1-cu_0+cu_0\,e^{2\eta \bar{t}}}}.
\end{equation}
Note that the eigenvalue \(c=\tilde C_t(\vec{k})\) does not appear inside the exponential factor, unlike in the one-layer evolution, Eq.~\eqref{eq:solution_score_full}, where the rate is proportional to \(c\). Consequently, for fixed initialization, reaching a prescribed accuracy takes a time that depends logarithmically rather than linearly on \(c\). The spread of effective training times across Fourier modes is therefore controlled by the ratio of logarithms of the extreme values of \(\tilde C_t(\vec{k})\), rather than by those of the condition number \(C_t\) itself. {Additionally, convergence of gradient descent with $\eta = \mathcal{O}(1)$ requires a small initialization, $c u_0 = \mathcal{O}(1)$ (otherwise one needs $\eta \lesssim 1/cu_0$, which leads to the same bad behavior shown in the one layer case). This can be achieved for all modes by having $s(t = 0) = \mathcal{O}(1/L)$ (see also the next section for a discussion on initialization), which leads to an overall training time scaling as $\log L$, as announced.}

\begin{comment}
{An hand-wavy argument of why the above procedure works goes as follows. Intuitively, \eqref{eq:logistic} can be seen as an exponential equation similar to \eqref{eq:operator_equation}, but with an effective, weight-dependent learning rate $\eta_\mathrm{eff}(u) = u\eta$. This $\eta_\mathrm{eff}(u)$ is implicitly tuned during training and therefore acts as an adaptive learning rate in Fourier space. In particular, near convergence $u \simeq 1/c$, therefore $\eta_\mathrm{eff}  \simeq \eta/c$. The convergence arguments in Sec. \ref{sec:training} suggest non-rigorously that convergence $\eta_\mathrm{eff}  \simeq \eta/c \lesssim 1/c$. Therefore, a learning rate $\eta =\mathcal{O}(1)$ would still lead to convergence, and $\eta$ does not have to be taken vanishingly small as in the one layer case. In this sense, the change in the architecture, and therefore the change in the loss, acts as a dynamical regularization akin to preconditioning (see Sec. \ref{eq:summary} for connections with the literature on linear networks).}
\end{comment}

\subsection{General case}
We now consider the more general case in which the two layers are different at initialization. The treatment is then more involved, but the solution can still be written in closed form. (At variance from Ref.~\cite{saxe2014exact}, we here give the expression without transforming to hyperbolic coordinates.)

For a fixed mode \(\vec{k}\), let
\begin{equation*}
c = \tilde C(\vec{k})>0,\qquad 
s_1(\bar{t})= \tilde{S}^{(1)}_t,\qquad
s_2(\bar{t})= \tilde{S}^{(2)}_t,
\end{equation*}
and define the layer mismatch parameters 
\begin{equation*}
d_0:= s_1^2(0)-s_2^2(0),\qquad \delta:=|d_0|,\qquad \alpha:=\frac{\delta}{2}.
\end{equation*}
By defining the product
\begin{equation}
u(\bar{t}):=s_1(\bar{t})s_2(\bar{t})
\end{equation}
and
\begin{equation}
\kappa:=c\alpha=\frac{c\delta}{2},
\qquad
A:=\sqrt{1+\kappa^2}
=\sqrt{1+\frac{c^2 d_0^2}{4}},
\end{equation}
we obtain that the solution is described by the following set of equations:
\begin{equation}
\label{eq:twolayerset}
\begin{aligned}
& q_0=\frac{u_0}{\sqrt{u_0^2+\alpha^2}+\alpha},\\[2mm]
& R_0=\frac{q_0+\kappa-A}{q_0+\kappa+A},\\[2mm]
& R(\bar{t})=R_0\,e^{-2A\eta \bar{t}},\\[2mm]
& q(\bar{t})=\frac{(A-\kappa)+(A+\kappa)R(\bar{t})}{1-R(\bar{t})},\\[2mm]
& u(\bar{t})=\frac{\delta\,q(\bar{t})}{1-q(\bar{t})^2}.
\end{aligned}
\end{equation}
For notational convenience, we here wrote the solution as a set of equations, but note that one \textit{need not} solve them recursively, because each only depends on the previous one.

From Eqs.~\eqref{eq:twolayerset}, we find that the exponential part of the convergence is controlled by \(A\eta\), so that the characteristic decay time is
$\tau(\vec{k})\sim 1/\eta A$.
Therefore, as long as the product $c^2\,d_0^2$ is at most $\mathcal{O}(1)$, the dependence of the timescale on {$L$ does not become polynomial}. Because the maximum $c$ scales as $L^2$, the above equation requires $d_0 \lesssim 1/L^2${---a condition also required for keeping a learning rate $\mathcal{O}(1)$). This condition can be achieved by taking (as a sufficient condition) $s_1$  and $s_2$ of order $1/L$. %This scaling is the same presented in the previous section. 
(By contrast, standard machine learning initialization schemes for weights, such as the Kaiming initialization~\cite{he2015}, keep $s_1$ and $s_2$ of order one. This choice would grow increasingly problematic for learning as the systems size increases.)}
{%Because thdo not implement the correct order of magnitude. 
%Therefore, for large system sizes an initialization based on the above scaling argument should be used. 
Following this prescription, one finds that the typical time to learn all the modes scales logarithmically in $L$, as announced. However, in practice, fairly large systems need to be considered for the effect to be noted (see Fig.~\ref{fig:1vs2layers_sim}). %long as the system size is not too large (and therefore the largest $c$ not too large), even for classical machine learning implementations the product $c d_0$ is still small for most modes, (as we see in the numerical experiments of the next section).
}

\begin{comment}
    Interestingly, this condition precisely corresponds to the standard working regime for neural networks. Standard initializations for machine training, such as Kaiming initialization~\cite{he2015}, indeed keep the weights of $\mathcal{O}(1/\sqrt{N}) = \mathcal{O}(1/L^\frac{d}{2})$ in order to keep the variance of the activations during the forward pass $\mathcal{O}(1)$. Therefore $d_0 = \mathcal{O}(1/L^d)$ and
\begin{equation}
    {C_t(\vec{k}_\mathrm{min})\,d_0 \sim L^2d_0 = \mathcal{O}(1/L^{d-2})}
\end{equation}
\begin{equation}
        {C_t(\vec{k}_\mathrm{max})\,d_0 \sim d_0 = \mathcal{O}(1/L^d)},
\end{equation}
so the condition is satisfied for every $d \geq 2$.
\end{comment}

\subsection{Numerical validation}
To contrast the one-layer and two-layer network architectures in practice, we consider their behavior for a $d=2$ system. Figure~\ref{fig:1vs2layers_sim} depicts the reciprocal space score kernel error 
\begin{equation}\label{eq:fourier_score_error}
\varepsilon_{\widetilde{\mathcal F}}(\vec{k})
=
\left(
\frac{\overline{\left|\widetilde{F}_t^{\mathrm{ov}}[\varphi_t (\vec{k})]-\widetilde{\mathcal F}_t[\varphi_t (\vec{k})]\right|^2}}
{\overline{\left|\widetilde{\mathcal F}_t[\varphi_t (\vec{k})]\right|^2}}
\right)^{1/2},
\end{equation}
where the overbar denotes the empirical average over a set of configurations $\varphi_t \sim P_t$, for each of the Fourier modes and
\begin{equation}
    \widetilde{\mathcal F}_t[\varphi (\vec{k})] = -\tilde{\mathcal{S}_t}(\vec{k})\tilde{\varphi}(\vec{k}),
\end{equation}
with an analogous expression for $\widetilde{F}_t^{\mathrm{ov}}[\varphi_t (\vec{k})]$. We here consider the total score rather than the score kernel, as it is the quantity directly measurable in real neural networks. (Also, given the overparameterized nature of the network, the score of individual layers is not uniquely defined.)
As expected, the characteristic decay time of the error with training epoch grows as $1/\tilde{C}(\vec{k})$  for a one-layer network and as $-\log\tilde{C}(\vec{k})$ for the two-layer one. As a result, the spread of the training dynamics of the different modes is markedly smaller in the two-layer case. As expected, the same holds for the simplified case~\cite{SM}. 

\subsection{Beyond two layers}

{Can further overparametrization be even more beneficial? To explore this question, we  consider the simplified case from Sec.~\ref{sec:simple_case}, with all layers the same at initialization and subsequent times---following Ref.~\cite{saxe2014exact}. For $N_\ell$ trainable layers, we define the quantity $u=s^{N_\ell}$. Equation~\eqref{eq:logistic} then becomes}
\begin{equation}
    {\dot{u}=\eta N_\ell u^{2-\frac{2}{N_\ell}}(1-cu)}.
\end{equation}
{Repeating the above analysis, we get that for local convergence to be attained a learning rate $\mathcal{O}(L^0)$ suffices. However, this scaling requires that the layers be suitably initialized with
\begin{equation}
    {s_0 \sim L^{-\frac{2}{N_\ell}}.}
\end{equation}
{Additionally, the training times $\tau_\mathrm{eff}$ then scale as}\footnote{{This results corresponds to the typical timescale of growth near small initialization and the exponential convergence around the linearized fixed point.}}
\begin{equation}
    {\tau_\mathrm{eff} \sim L^{2-\frac{4}{N_\ell}}.}
\end{equation}
{The initialization condition for convergence therefore becomes weaker as the number of layers increases and tends to the standard machine learning initializations, $\mathcal{O}(1)$, as $N_\ell\to\infty$. %(see the discussion at the end of the previous section). 
For any $N_\ell>2$, however, yields a \emph{worse} scaling than the two-layer case---even ignoring the additional computational cost of handling $N_\ell$ layers. In particular, for $N_\ell\geq3$, the training time again becomes a power of $L$, losing the logarithmic improvement of the two-layer case. The scaling further worsens as $N_\ell$ increases. In particular, it tends to $L^2$, the same scaling as in the one-layer case, for $N_\ell\to\infty$. %The two-layer case  equivalently the training dynamics described by Eq.~\eqref{eq:gf_single_twice}, represents an optimal choice in terms of training time.}

The slower dynamics of networks with more than two layers was already observed in Ref.~\cite{saxe2014exact}. In that case, however, the single-layer model converges faster than its two-layer counterpart, whereas in our case the $N_\ell=2$  architecture is theoretically optimal---at least for the balanced case. For the imbalanced case, the situation is less clear, due to the progressive intractability of the equations for the increasing number of layers, no definite answers have yet been obtained\cite{arora2018optimization, min2023convergence}. In any case, we here restrict our analysis to $N_\ell=2$, %since we have already shown that increasing the depth from one to two layers yields a substantial improvement, 
which suffices to reduce the training-time slowing down from a power law to a logarithm in $L$. Little more than a prefactor improvement could then be expected. %Thus, even if additional layers were to further accelerate training (a possibility for which we currently have no supporting evidence) the resulting improvement could affect only the logarithmic dependence and would therefore be of limited practical significance.

\subsection{Summary}\label{eq:summary}
The addition of a linear layer to the minimal network considered in Sec.~\ref{sec:background} qualitatively alters the training dynamics. Most notably, the scaling of the effective training time is reduced from $L^2$ to $\log L$, thereby overcoming the critical slowing down. %In other words, all multi-layer architectures should exhibit a similar behavior, distinct from the one-layer case.

This picture is consistent with the use of overparameterization in linear models as a form of effective preconditioning~\cite{arora2018optimization, labarriere2024optimization}. {By  %, in the training dynamics, 
implicitly transforming the training directions, which makes the optimization problem easier to solve}, the training is accelerated in certain regimes~\cite{tarmoun2021implicit}. It differs with observations in Ref.~\cite{rahaman2019spectral}, where a stronger separation of timescales is required to learn different modes, including in nonlinear networks, and with Ref.~\cite{saxe2014exact}, where the timescale grows linearly (rather than logarithmically) with the eigenvalues. In both cases, the specific input--output structure is responsible for the difference. {A related difference appears with respect to \cite{wang2026analytical}. In that case, however, the origin of the different scaling follows from the choice of parametrization: in our work the network directly parametrizes the score; Ref.~\cite{wang2026analytical} considers a denoiser parametrization.}

Although the use of a two-layer architecture eliminates the critical slowing down in the training dynamics, each score estimate requires more operations than is needed for the classical Langevin dynamics. Convolutional neural networks whose kernel spans the whole system (as the one that are characterized by the analytical study of the previous sections) require $\mathcal{O}(N^2)$ operations for each forward pass (recall $N=(L/a)^d$); Langevin dynamics, by contrast, only requires $\mathcal{O}(NL^2)$ at criticality (and $\mathcal{O}(N)$ away from it). Unfortunately, the divergence of the correlation length at criticality does not allow to use a local score to limit the number of parameters, see App.~\ref{app:local_score}. Therefore, from the point of view of computational complexity, which is the main focus of this work, it remains unclear whether a two-layer architecture could ever be beneficial. We address---and solve---this problem in the next section. 

\section{Fourier space implementation of the convolution}\label{sec:fourier}

%Because the convolution kernels considered in Secs.~\ref{sec:theory} and~\ref{sec:twolayers} span the whole system, the neural networks considered require a number of parameters that grows with system size, $N = (L/a)^d$. As a result, each score estimate requires $\mathcal{O}(N^2)$ operations. While this scaling is an improvement over standard Langevin dynamics in $d=1$, for $d \geq 2$ it is significantly worse. Despite overcoming the critical slowing down in training, the resulting diffusion models maintain an unfavorable overall computational cost. 

In this section, to surmount the remaining issue of computational complexity, we introduce a Fourier implementation of the convolution. More specifically, we leverage the well-known property that convolutions are multiplications in Fourier space. In this way, although the neural network is modeled by the same amount ($N$) of parameters as classical convolutional layers, each of these parameters appears in but one multiplication. The total cost of the convolution in Fourier space is then $\mathcal{O}(N)$, subdominant with respect to the cost of the Fourier transform, $\mathcal{O}(N\log N)$, which therefore dominates the cost of evaluating the score. The total cost is therfore much smaller than the cost of a convolutional neural network call, which instead is $\mathcal{O}(N^2)$. 

Because the diffusion process here takes place in real space, we proceed as follows: for each call to the score, we first transform the input in Fourier space via a fast Fourier transform (FFT); we then apply the layer(s) via simple multiplication in Fourier space; finally, we convert the output in real space via the inverse FFT (iFFT) (see Fig. \ref{fig:summary}e). The computational cost of (i)FFT is $\mathcal{O}(N\log N) = \mathcal{O}(dL^d\log L)$, thus coinciding with the complexity of the whole call of the score. %, since the multiplication part's cost is just $\mathcal{O}(N)$.

This scaling is verified in Fig.~\ref{fig:timescale_cnn_fourier}, where the forward-pass time of a two-layer convolutional network implemented in \texttt{PyTorch} is compared with that of our Fourier-space implementation. As expected, the Fourier-based network is faster than its real-space convolutional counterpart and scales more efficiently with $N = (L/a)^d$.

\begin{figure}[t]
    \centering
    \includegraphics[width=1\linewidth]{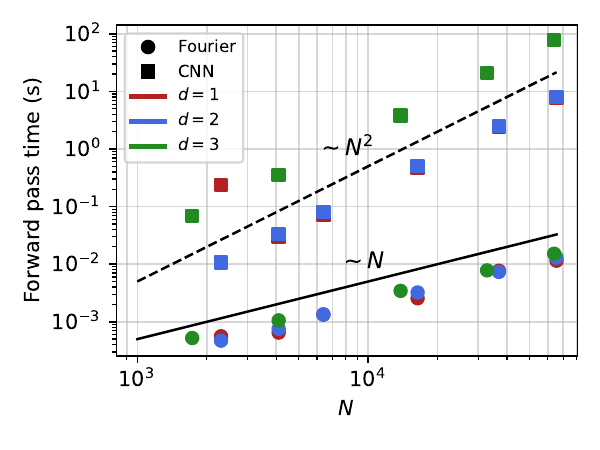}
    \caption{Forward-pass time of a two-layer convolutional network implemented in \texttt{PyTorch} (squares) with that of our Fourier-space implementation (circles). The comparison is performed on batches of 1024 samples for increasing system sizes $L$ ($ a = 1$) and spatial dimensions $d$=1 (red), 2 (blue) and 3 (green). Expected scalings $\sim N$ (full line) and $\sim N^2$ (dashed line) are included. The $\log L$ correction to the Fourier-space  scaling is not observed over this range of system sizes. Wall times refer to runs on a NVIDIA Tesla V100-SXM2-32GB GPU. }
    \label{fig:timescale_cnn_fourier}
\end{figure}

We also validate the architecture by generating configurations and testing the accuracy on physical observables. Our procedure is different from the one used for standard neural-network architectures in which score modeling typically incorporate the diffusion time through a time embedding, allowing a single network to represent the score at all diffusion times. By contrast, and in order to directly implement the theory developed in the previous sections, we employ a separate network at each fixed diffusion time $t$, with each network trained independently. The diffusion time is therefore not provided as an input of the network. Nevertheless, because our linear model has a small number of parameters and can be trained rapidly, the additional computational overhead associated with training independent networks at different times remains treatable. The diffusion time interval goes from $t_{\max}=5$ down to $t_{\min}=10^{-4}$ and is discretized into 3000 time points. %, and an independent network is trained at each fixed diffusion time $t$. 
Each model has weights initialized as Gaussian random variables of zero mean and standard deviation $1/L$ and is trained by gradient descent on a dataset of size $M=4096$, using a learning rate $\eta=0.025$.\footnote{This learning rate is smaller than the maximum learning rate acceptable for convergence. Moreover, it was chosen keeping into account the residual plateau of the score error due to the interplay between $\eta$ and finite $M$ (the training dataset size) used in this section. Indeed, quick numerical tests seem to indicate that the residual score error scales as $\sqrt{\eta/M}$ (not depending on $L$ and therefore not influencing the critical slowing down discussion).} The backward diffusion equation is integrated numerically using a fourth-order Runge--Kutta scheme with 1500 integration steps.\footnote{The number of integration steps is half the number of trained models because each Runge--Kutta step also requires evaluating the score at the midpoint of the corresponding time interval.}

%and examining their physical properties. 

We first consider systems of sizes $L=16,\,32,\,64$ and $128$, with progressively smaller effective masses, $m_{\mathrm{eff}}=3/L$. We use 175, 200, 225, and 250 training epochs for $L=16,\,32,\, 64,$ and $128$, respectively, increasing the number of epochs linearly for each doubling of the system size, in conformity with the $\log L$ scaling of Sec.~\ref{sec:twolayers}. %For each system size, we train the two-layer Fourier network introduced above. 
%We use the trained network to generate configurations for the different sizes. 

The first physical observable we focus on is the averaged power spectrum: 
\begin{equation}\label{eq:powerspectrum}
 P(k)=\langle |\tilde{\varphi}(\mathbf{k})|^2\rangle,
\end{equation}
which is shown at low $k$ is shown in Fig.~\ref{fig:masses_fits} 
for both the target and generated ensembles. Fits to 
\begin{equation}
\label{eq:fitPk}
1/P(k)\simeq \mathfrak{a}k^2+\mathfrak{b}, 
%1/P(k)\simeq \mathfrak{a} k^2+\mathfrak{b}
\end{equation}
provide an independent estimate of the effective mass. %, $m_{\mathrm{fit}}=\sqrt{\mathfrak{b}/\mathfrak{a}}$. 
Results are numerically indistinguishable from target values, thus showing that the generated configurations correctly reproduce the system behavior upon approaching criticality.
 %? agreement with the . %The effective mass (which is related to the correlation length of the system) in the following way. We consider the inverse power spectrum in the low-momentum regime.  Therefore, we fit the lowest nonzero momentum shells of the generated configurations to this linear form and extract the mass $m_\mathrm{fit}$ from these relations.

%We find that the fitted masses agree with the target values within their statistical uncertainties, 

\begin{figure*}
    \centering
    \includegraphics[width=1\linewidth]{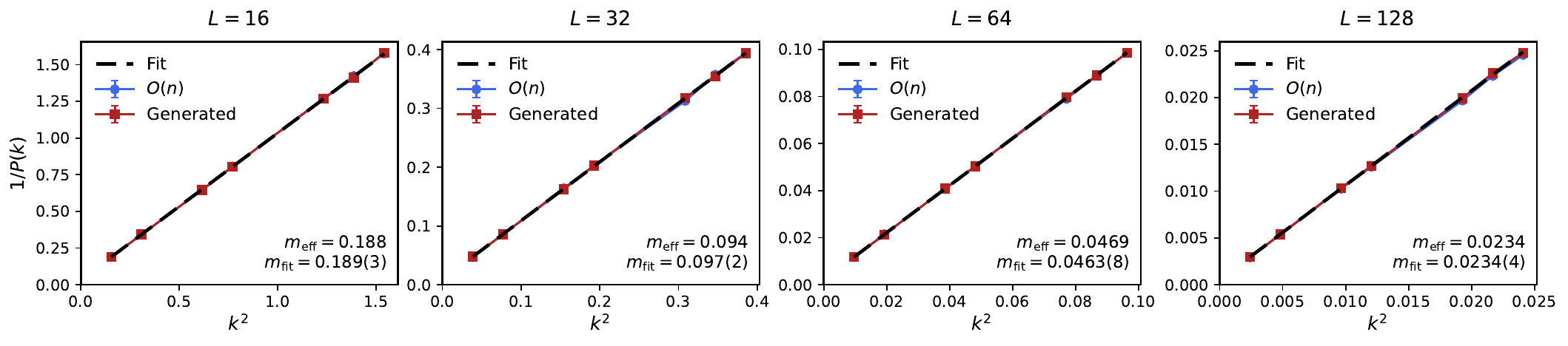}
        \caption{
        Inverse power spectrum $1/P(k)$ for the real (blue) and generated (red)
        configurations in the low-$k$ regimes, together with fits of the
        latter points to the functional form
        $\mathfrak{a}k^2+\mathfrak{b}$.
        Each curve has been obtained using $20000$ configurations.
        The fitted masses,
        $m_{\mathrm{fit}}=\sqrt{\mathfrak{b}/\mathfrak{a}}$,
        are in very good agreement with the true values of the effective mass
        $m_{\mathrm{eff}}=3/L$.
        \textbf{Setting:} $d=2$, $a=1$.
    }
    \label{fig:masses_fits}
\end{figure*}

We next consider several observables measured on similarly generated configurations at $m_{\mathrm{eff}}=0$ in Fig.~\ref{fig:complete}. %We train the network using the same procedure described above for four system sizes, $L=16,\, 32,\, 64$ and 128. 
Results from generated configurations are indistinguishable from theoretical expectations.

\begin{figure*}
    \centering
    \includegraphics[width=1\linewidth]{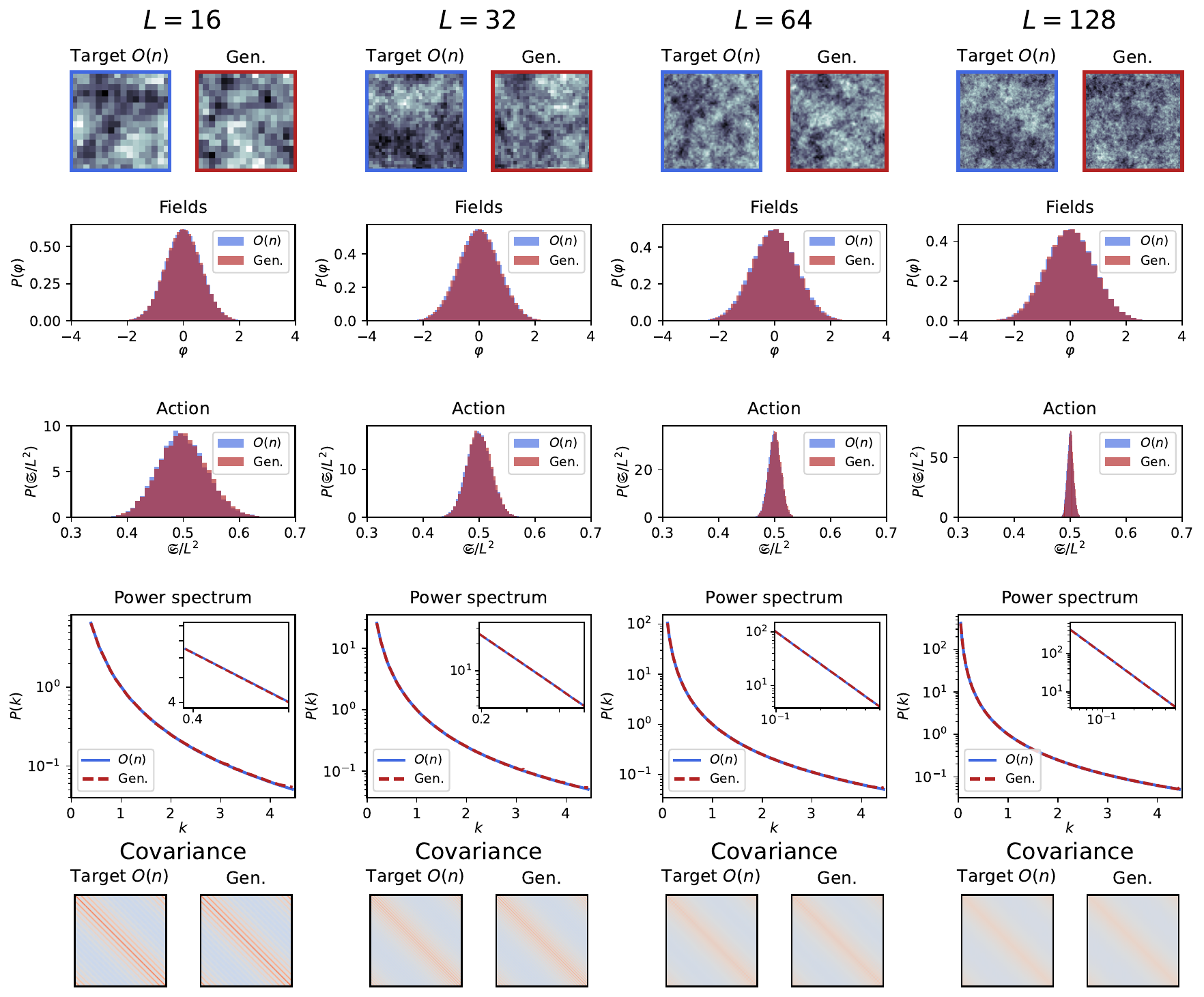}
    \caption{Generated configurations at $m_\mathrm{eff}=0$ of the $O(n)$ model in the limit $n\rightarrow\infty$ for $L = 16, 32, 64, 128$. \textbf{Top row}: Sample configurations extracted using Eq.~\eqref{eq:GB} (blue frames) and generated via the diffusion model (red frames). \textbf{Middle rows}: Comparison between real and generated configurations of the distribution of a single component of the fields, $P(\varphi)\propto e^{\frac{-\varphi^2}{2\sigma^2(L)}}$, where the variance $\sigma^2(L)$ only weakly depends on $L$, and of the action $\mathfrak{S}$, Eq.~\eqref{eq:lambda_action}, as well as the power spectrum, $P(k)=\langle  \tilde{\varphi}(\vec{k}) |^2 \rangle_{|\vec{k}| = k}$ (the inset corresponds to a zoom-in in the low-$k$ region in log-log scale). \textbf{Bottom row}: Covariance matrices of the real and generated distributions. All the observables have been computed using 20000 configurations. \textbf{Setting:} $d=2$, $m_\mathrm{eff} = 0$, and $a =1$. }
    \label{fig:complete}
\end{figure*}

\section{Conclusions}\label{sec:conclusions}

We have investigated diffusion models as generative tools for configurations of the $O(n)$ model in the limit $n \to \infty$, where it becomes Gaussian and analytically tractable. In this setting, the exact score admits a closed-form expression that can be represented by a particularly simple neural network architecture, thus placing the training dynamics of diffusion models on firm theoretical ground. Within this controlled framework, a clear picture emerges: as the system approaches criticality, training the score becomes increasingly difficult. The process exhibits a pronounced slowing down that mirrors the critical slowing down of a standard local dynamics. This effect also impacts generation, with errors in the learned score propagating to the sampling process and leading to the same characteristic scaling behavior. In this sense, the diffusion model faithfully inherits the dynamical bottlenecks of the physical system it is trained to reproduce.

This behavior, however, is \emph{not} universal across architectures. In particular, overparameterizing the score network through the introduction a second linear layer fundamentally alters the training dynamics, leading to a much more favorable scaling of the training time. The training dynamics then becomes well-conditioned. But while neural networks can mitigate the critical slowing down with respect to standard simulation methods, they introduce a computational overhead that could, in principle, neutralize those benefits or even increase the overall computational complexity. To address this limitation, we demonstrate that an architecture in which convolutions are implemented in Fourier space provides a good overall scaling of the computational complexity.

Taken together, these results provide a coherent picture of how architectural choices interact with the intrinsic dynamics of diffusion models. They show that while critical slowing down is a natural consequence of the underlying physical structure, it is not an unavoidable limitation: appropriate parameterizations can eliminate its impact on training and generation.

These results also open up a number of research directions. The formal analysis could be extended beyond the Gaussian $O(n\rightarrow\infty$) model studied here. For instance, one could consider finite-$n$ model by expanding perturbatively around the Gaussian limit or tackle other systems for which the exact score is known, such as the Curie--Weiss model, thus extending the analysis of Ref.~\cite{BiroliMezard2023} to the training process. {Studying these systems, in which, at variance with the case studied here, it is possible to explore the ferromagnetic phase, would allow to study the behavior of diffusion models under symmetry breaking. This would also make it possible to investigate the interplay between training and speciation times \cite{biroli2024dynamical}.} {The study of a broader array of of models could further provide physical intuition, which is presently lacking, on why overparametrization with two layers proves so useful in the setting studied here.}

Strategies for weakening the critical slowing down through a suitable choice of architecture could also be explored. 
Our theoretical analysis suggests that faster training could be achieved in Fourier space. 
While our Fourier architecture is simply a clever way of representing convolutions whose kernel spans the whole system, the usage of Fourier space could be extended beyond this simple setup. In more realistic multi-scale settings modes are coupled, but using a Fourier architecture could be helpful to take into account interactions across scale in a parsimonious and effective way as previously suggested~\cite{Tivnan2023FourierDiffusion, masuki2025generative, chang2025flow}. Working in Fourier space could leverage strategies that are not available in real space, including the physics of the renormalization group \cite{masuki2025generative, singha2025multilevel}. 
A more aggressive numerical approach could be to leverage existing insights to sample more complex models at criticality directly. In all cases, it is important to note that we have here taken the training dataset as given. In most practical sampling applications, one is given the form of the energy and not an actual dataset, and the main objective is obtaining the latter from the former. This problem, which is more generic and has been the focus of many recent works, is also a promising avenue of research.

These developments may ultimately enable a new paradigm for sampling, in which learned generative dynamics complement---or even replace---traditional MCMC approaches.

\section*{Data availability}
The data that support the findings of this article are openly available~\cite{RDR} [Data will be made openly available upon manuscript acceptance, but a DOI is not available at the time of submission.]

\acknowledgments We thank Davide Carbone, Sebastian Goldt, Misaki Ozawa, and Gilles Tarjus for useful discussions. LMDB is grateful to École normale supérieure for its hospitality during his stays, during which part of this work was conducted. This study was conducted using the DARIAH HPC-AI cluster at CNR-NANOTEC in Lecce, funded by the "MUR PON Ricerca e Innovazione 2014-2020" project, code PIR01\_00022 and H2IOSC Project - Humanities and cultural Heritage Italian Open Science Cloud funded by the European Union – NextGenerationEU – NRRP M4C2 - Project code IR0000029.

\renewcommand{\thesection}{\Alph{section}} \renewcommand{\thefigure}{\thesection\arabic{figure}} \setcounter{figure}{0}
\counterwithin{figure}{section}

\appendix

\section{Details on the analytical computations}
This appendix details the calculation sketched in Sec.~\ref{sec:theory}.

\subsection{Self-consistent equation for parameter $\Lambda$}\label{app:lambda_eq_derivation}

We start by deriving the explicit form of the mean-field expression for $\Lambda$, which can be written as
\begin{equation}\label{eq:eq_lambda}
\Lambda = \frac{1}{n}\sum_a\langle \varphi_a(\vec{x}) \cdot \varphi_a(\vec{x}) \rangle.
\end{equation}
To proceed with the computation, recall that the real space correlation $\langle \varphi_a(\vec{x}) \, \varphi_a(\vec{x}) \rangle$ can be rewritten in terms of the momentum space correlation $ \langle \tilde{\varphi}_a(\vec{q}) \, \tilde{\varphi}_a(\vec{k}) \rangle$ as
\begin{align}\label{eq:correlation_realspace}
\langle \varphi_a(\vec{x}) \, \varphi_a(\vec{x}) \rangle
&= \int d^d \vec{q} \, d^d \vec{k} \,
e^{-i (\vec{k} + \vec{q}) \cdot \vec{x}} \langle \tilde{\varphi}_a(\vec{q}) \, \tilde{\varphi}_a(\vec{k}) \rangle,
\end{align}
where inverse Fourier transform is
\begin{equation}
\varphi_a(\vec{x}) = \int d^d \vec{k} \, e^{-i \vec{k} \cdot \vec{x}} \, \tilde{\varphi}_a(\vec{k}).
\end{equation}
To compute $\langle \tilde{\varphi}_a(\vec{q}) \, \tilde{\varphi}_a(\vec{k}) \rangle$, we introduce a source term $J_a(\vec{x})$ to the action in Eq.~\eqref{eq:lambda_action}
\begin{align}
\mathfrak{S}_\Lambda =& \int d^d \vec{x} \left( 
\frac{1}{2} \sum_{a, \nu} (\partial_\nu \varphi_a)^2 + \frac{1}{2} (m^2 + \Lambda g) \sum_a \varphi_a^2 
\right)\nonumber
\\&
+ \int d^d \vec{x} \sum_a J_a(\vec{x}) \varphi_a(\vec{x}),
\end{align}
and transform to Fourier space
\begin{align}
&\mathfrak{S}_\Lambda =\\& \int d^d \vec{x} \, \frac{1}{2} \Bigg[
\sum_{a, \nu} \Bigg( 
    - \int d^d \vec{k} \, d^d \vec{q} \, 
    k_\nu q_\nu \,
    \tilde{\varphi}_a(\vec{k}) \tilde{\varphi}_a(\vec{q}) \,
    e^{-i \vec{x} \cdot (\vec{k} + \vec{q})}
\Bigg) \nonumber\\
&\quad + m_\text{eff} \sum_a \int d^d \vec{k} \, d^d \vec{q} \,
e^{-i \vec{x} \cdot (\vec{k} + \vec{q})}
\tilde{\varphi}_a(\vec{k}) \tilde{\varphi}_a(\vec{q})
\Bigg] \nonumber\\
&\quad + \int d^d \vec{x} \sum_a \int d^d \vec{k} \, d^d \vec{q} \,
e^{-i \vec{x} \cdot (\vec{k} + \vec{q})}
\tilde{J}_a(\vec{q}) \tilde{\varphi}_a(\vec{k}).\nonumber
\end{align}
Using the delta-function identity
\begin{equation}
\label{eq:deltaidentity}
\int d^d \vec{x} \, e^{-i \vec{x} \cdot (\vec{k} + \vec{q})}
= (2\pi)^d \, \delta^{(d)}(\vec{k} + \vec{q}),
\end{equation}
we can write
\begin{widetext}
\begin{align}\label{eq:action_fourier}
\mathfrak{S}_\Lambda &= \frac{(2\pi)^d}{2} \int d^d \vec{k} \, d^d \vec{q} \,
\left( (\vec{k} \cdot \vec{k} + m^2_\text{eff}) \left( \sum_a \tilde{\varphi}_a(\vec{k}) \,
\delta^{(d)}(\vec{k} + \vec{q}) \, \tilde{\varphi}_a(\vec{q}) \right) \right) + \int d^d \vec{q} \, d^d \vec{k} \, \sum_a \tilde{\varphi}_a(\vec{k}) \, \tilde{J}_a(-\vec{k}).
\end{align}
\end{widetext}
The momentum space correlation can then be rewritten in terms of the partition function, $\mathcal{Z} =  \int \mathcal{D}\tilde{\varphi} e^{-\mathfrak{S}_\Lambda}$, as
\begin{equation}
\frac{\delta^2 \log \mathcal{Z}}{\delta \tilde{J}_a(-\vec{k}) \, \delta \tilde{J}_a(-\vec{q})} \biggr\rvert_{\tilde{J} = 0}
=\langle \tilde{\varphi}_a(\vec{k}) \tilde{\varphi}_a(\vec{q}) \rangle.
\end{equation}
Performing the Gaussian integration over $\tilde{\varphi}$ gives
\begin{align}
&\mathcal{Z} = \int \mathcal{D} \tilde{\varphi} \, e^{-\mathfrak{S}_\Lambda}\times\\
&\left[
\mathcal{Z}(\tilde{J} = 0) \cdot \exp\left(
\frac{1}{2} \sum_a\int \frac{d^d \vec{k} \, d^d \vec{q}}{(2\pi)^d} \,
\frac{ \tilde{J}_a(\vec{k}) \, \delta^{(d)}(\vec{k} + \vec{q}) \, \tilde{J}_a(\vec{q}) }
     { \vec{k} \cdot \vec{k} + m^2_\text{eff} }
\right)
\right]\nonumber
\end{align}
and, therefore, the correlations in momentum space can be written as
\begin{align}
\langle \tilde{\varphi}_a(\vec{k}) \tilde{\varphi}_a(\vec{q}) \rangle &= \frac{\delta^2 \log \mathcal{Z}}{\delta \tilde{J}_a(-\vec{k}) \, \delta \tilde{J}_a(-\vec{q})}\nonumber
\\&= \frac{1}{(2\pi)^d} \, \frac{\delta^{(d)}(\vec{k} + \vec{q})}{\vec{k} \cdot \vec{k} + m^2_\text{eff}}.
\end{align}
From Eq.~\eqref{eq:correlation_realspace}, we then have that in real space
\begin{align}
\langle \varphi_a(\vec{x}) \, \varphi_a(\vec{x}) \rangle
&= \int d^d \vec{q} \, d^d \vec{k} \,
e^{-i (\vec{k} + \vec{q}) \cdot \vec{x}} \cdot
\frac{1}{(2\pi)^d} \,
\frac{ \delta^{(d)}(\vec{k} + \vec{q}) }{ \vec{k} \cdot \vec{k} + m^2_\text{eff} }\nonumber \\
&= \frac{1}{(2\pi)^d} \int d^d \vec{k} \,
\frac{1}{\vec{k} \cdot \vec{k} + m^2 + \Lambda g}.
\end{align}
Note that the final expression does not depend on the subscript $a$. The self-consistent Eq.~\eqref{eq:eq_lambda} for $\Lambda$ is then
\begin{equation}\label{eq:boxed_lambda}
\boxed{\Lambda = 
\frac{1}{(2\pi)^d} \int d^d \vec{k} \,
\frac{1}{\vec{k} \cdot \vec{k} + m^2 + \Lambda g}.
}
\end{equation}
This result matches \cite[Eq.~(17.5)]{Fradkin2021} up to constants in the definition of the action. 

The integral in Eq.~\eqref{eq:eq_lambda}, however, has an UV divergence for $d \geq 2$, which is typically addressed through a renormalization approach. Alternatively, in this case, we can consider a UV cutoff $\lambda$ that is consistent with a simulation with discretization in space with steps $a = 1/\lambda$:
\begin{align*}
\Lambda &= \frac{1}{(2\pi)^d} \int d^d \vec{k} \,
\frac{1}{\vec{k} \cdot \vec{k} + m^2 + \Lambda g} \\
&= \frac{1}{(2\pi)^d} \Omega(d) \int_0^\lambda dk\,
\frac{k^{d-1}}{k^2 + m^2_\text{eff}}
\end{align*}
with $\Omega(d)$ the surface of a $d$-dimensional unit sphere. Because $m_\text{eff}^2 = 0$ at criticality, we then have
\begin{align}
\Lambda &= \frac{1}{(2\pi)^d} \, \Omega(d) \int_0^\lambda dk \, k^{d-3} \nonumber\\&= 
\begin{cases}
\frac{1}{(2\pi)^d} \, \Omega(d) \, \frac{\lambda^{d-2}}{d-2} & \text{for } d > 2\\
\frac{1}{2\pi} \, \ln \lambda & \text{for } d =2
\end{cases}
\end{align}

\subsection{Score calculation}\label{app:score_computation}
Consider now a single component of the $n$-dimensional field, dropping the subscript $a$ for notational simplicity. The  score $\mathcal{F}$ is related to the probability distribution of the forward diffusion (noising) process,
\begin{equation*}
P_t(\varphi) \propto \int \mathcal{D}\psi(\vec{x}) \, \, e^{-\mathfrak{S}_\Lambda(\psi(\vec{x})) - \int d^d \vec{x} \, \frac{1}{2} \left( \frac{\varphi(\vec{x}) - \psi(\vec{x}) e^{-t}}{1 - e^{-2t}} \right)^2},
\end{equation*}
as
\begin{equation}
\mathcal{F}(\vec{\varphi}) = \frac{\delta \log P_t(\vec{\varphi})}{\delta \varphi(\vec{x})} .
\end{equation}
To write down the exact score for the model, we consider Eq.~\eqref{eq:forward_process_prob} and again move to Fourier space.

First, we consider the argument of the exponential. The action $\mathfrak{S}_\Lambda$ can be rewritten in momentum space as in Eq.~\eqref{eq:action_fourier}. For the argument of the integral,
\begin{align}
&-\frac{1}{2} \frac{\left( \varphi(\vec{x}) - \psi(\vec{x}) e^{-t} \right)^2}{\Delta{ _t}} 
\\
&= \frac{1}{2\Delta{ _t}} \left( \varphi(\vec{x})^2 + \psi(\vec{x})^2 e^{-2t} - 2\varphi(\vec{x})\psi(\vec{x}) e^{-t} \right),\nonumber
\end{align}
where $\Delta{ _t} = 1-e^{-2t}$, Fourier transforming gives
\begin{widetext}
\begin{equation}
-\frac{1}{2\Delta{ _t}} \int d^d \vec{k} \, d^d \vec{q} \left[ \tilde{\varphi}(\vec{k}) \tilde{\varphi}(\vec{q}) + \tilde{\psi}(\vec{k}) \tilde{\psi}(\vec{q}) e^{-2t} - 2 \tilde{\varphi}(\vec{k}) \tilde{\psi}(\vec{q}) e^{-t} \right] e^{-i(\vec{k} + \vec{q}) \cdot \vec{x}},
\end{equation}
and integrating over $\vec{x}$ gives
\begin{equation}\label{eq:exp_arg_fourier}
- \frac{(2\pi)^d}{2\Delta{ _t}} \int d^d \vec{k} \, d^d \vec{q} \; \delta^d(\vec{k} + \vec{q}) \left[ \tilde{\varphi}(\vec{k}) \tilde{\varphi}(\vec{q}) + \tilde{\psi}(\vec{k}) \tilde{\psi}(\vec{q}) e^{-2t} - 2 \tilde{\varphi}(\vec{k}) \tilde{\psi}(\vec{q}) e^{-t} \right].
\end{equation}
The term $\tilde{\varphi} \tilde{\varphi}$ which will therefore subsist after integrating over $\psi$,
\begin{equation}\label{eq:phi_term}
- \frac{(2\pi)^d}{2\Delta{ _t}} \int d^d \vec{k} \, d^d \vec{q} \; \tilde{\varphi}(\vec{k}) \tilde{\varphi}(\vec{q}) \delta^d(\vec{k} + \vec{q}).
\end{equation}
Equation \eqref{eq:forward_process_prob} can now be written as
\begin{equation}\label{eq:forward_process_prob_fourier}
P_t = e^{- \frac{(2\pi)^d}{2\Delta{ _t}} \int d^d \vec{k} \, d^d \vec{q} \; \tilde{\varphi}(\vec{k}) \tilde{\varphi}(\vec{q}) \delta^d(\vec{k} + \vec{q})}\frac{1}{\mathcal{Z}}\int \mathcal{D}\psi(\vec{x}) \, \, e^{-\mathfrak{S}_\Lambda(\psi(\vec{x})) - \frac{(2\pi)^d}{2\Delta{ _t}} \int d^d \vec{k} \, d^d \vec{q} \; \delta^d(\vec{k} + \vec{q}) \left[  \tilde{\psi}(\vec{k}) \tilde{\psi}(\vec{q}) e^{-2t} - 2 \tilde{\varphi}(\vec{k}) \tilde{\psi}(\vec{q}) e^{-t} \right]},
\end{equation}
and therefore the action $\mathfrak{S}_\psi$ for field $\psi$ to be integrated over is
\begin{equation}\label{eq:action_intermediate}
\mathfrak{S}_\psi = - \frac{(2\pi)^d}{2} \int d^d \vec{k} \, d^d \vec{q} \;
\left[
\tilde{\psi}(\vec{k}) \left( \vec{k} \cdot \vec{k} + m_{\text{eff}}^2 + \frac{e^{-2t}}{\Delta{ _t}} \right) \tilde{\psi}(\vec{q})
- \tilde{\varphi}(\vec{k}) \left( 2 \cdot \frac{1}{\Delta{ _t}} \cdot e^{-t} \right) \tilde{\psi}(\vec{q})
\right] \delta^d(\vec{k} + \vec{q}).
\end{equation}
The Gaussian integration yields:
\begin{equation}
\frac{1}{\mathcal{Z}} \int \mathcal{D}\psi(\vec{x}) \, 
\exp \bigg[ 
    -\mathfrak{S}_\Lambda(\psi(\vec{x})) 
    - \frac{(2\pi)^d}{2\Delta{ _t}} \int d^d \vec{k} \, d^d \vec{q} \; 
    \delta^d(\vec{k} + \vec{q}) 
    \left(  
        \tilde{\psi}(\vec{k}) \tilde{\psi}(\vec{q}) e^{-2t} 
        - 2 \tilde{\varphi}(\vec{k}) \tilde{\psi}(\vec{q}) e^{-t} 
    \right) 
\bigg] \propto
\end{equation}
\begin{equation}
\propto \exp \bigg\{ 
    \int d^d \vec{k} \; \frac{1}{2} (2\pi)^d \,
    \frac{\frac{e^{-2t}}{\Delta{ _t}^2}}{\vec{k} \cdot \vec{k} + m_{\text{eff}}^2 + \frac{e^{-2t}}{\Delta{ _t}}} \,
    \tilde{\varphi}(\vec{k}) \tilde{\varphi}(-\vec{k})  
\bigg\}
\end{equation}
\begin{equation}
= \exp \left \{ \frac{(2\pi)^d}{2\Delta{ _t}} \int d^d \vec{k} \;
\frac{e^{-2t}}{\Delta{ _t} \left( \vec{k} \cdot \vec{k} + m_{\text{eff}}^2 \right) + e^{-2t}} \;
\tilde{\varphi}(\vec{k}) \tilde{\varphi}(-\vec{k}) \right \}
\end{equation}
Summing this result with Eq.~\eqref{eq:phi_term} gives
\begin{equation}\label{eq:probability_phi}
P_t \propto \exp \left \{ \frac{(2\pi)^d}{2\Delta{ _t}} \int d^d \vec{k} \left[
\frac{e^{-2t}}{\Delta{ _t} \left( \vec{k} \cdot \vec{k} + m_{\text{eff}}^2 \right) + e^{-2t}} - 1
\right] \tilde{\varphi}(\vec{k}) \tilde{\varphi}(-\vec{k}) \right \}
\end{equation}
Note that in the limit $t \to 0$, the original Gaussian distribution of the data is recovered.

We can now finally compute the score:
\begin{align}
\mathcal{F}{ _t}& = - \frac{\delta}{\delta \varphi(x)} \left[
\frac{(2\pi)^d}{2\Delta{ _t}} \int d^d \vec{k} \,
\frac{\vec{k} \cdot \vec{k} + m_{\text{eff}}^2}{\vec{k} \cdot \vec{k} + m_{\text{eff}}^2 + \frac{e^{-2t}}{\Delta{ _t}}} \,
\tilde{\varphi}(\vec{k}) \tilde{\varphi}(-\vec{k})
\right]\\
&= - \frac{\delta}{\delta \varphi(x)} \frac{(2\pi)^d}{2\Delta{ _t}} \int d^d \vec{k} \,
\frac{\vec{k} \cdot \vec{k} + m_{\text{eff}}^2}{\vec{k} \cdot \vec{k} + m_{\text{eff}}^2 + \frac{e^{-2t}}{\Delta{ _t}}}
\left[
\int \frac{d^d \vec{y}}{(2\pi)^d} \, \frac{d^d \vec{z}}{(2\pi)^d} \;
e^{i \vec{k} \cdot (\vec{y} - \vec{z})}
\varphi(\vec{y}) \varphi(\vec{z})
\right]
\\&= -\frac{1}{\Delta{ _t}} \int \frac{d^d \vec{k}}{(2\pi)^{d}}  \,
\frac{k^2 + m_{\text{eff}}^2}{k^2 + m_{\text{eff}}^2 + \frac{e^{-2t}}{\Delta{ _t}}}
\int d^d \vec{y} \; e^{i \vec{k} \cdot (\vec{y} - {\vec{x}})} \varphi(\vec{y})
\\&= -\int d^d \vec{y} \left[
\frac{1}{\Delta{ _t}} \int \frac{d^d \vec{k}}{(2\pi)^{d}} \,
\frac{k^2 + m_{\text{eff}}^2}{k^2 + m_{\text{eff}}^2 + \frac{e^{-2t}}{\Delta{ _t}}} \,
e^{i \vec{k} \cdot (\vec{y} - {\vec{x}})}
\right] \varphi(\vec{y})
\\&= -\int d^d \vec{y} \; \mathcal{S}{ _t}(x,y) \varphi(\vec{y})
\end{align}
where we have defined:
\begin{equation}
\boxed{\mathcal{S}{ _t}(x, y) =  \int \frac{d^d \vec{k}}{(2\pi)^{d}} \,
\frac{1}{\Delta{ _t} + \frac{e^{-2t}}{k^2 + m^2_{\text{eff}}}} \,
e^{i \vec{k} \cdot (\vec{y} - \vec{x})}}.
\end{equation}
As expected, the score is linear and invariant under translation. It also has the right limit $t \rightarrow 0$, in that the inverse Fourier transform of the inverse of the Gaussian propagator is then recovered. 

Because the score is linear, we can compare this result with that from \cite[Eq.~(8)]{BiroliMezard2023}, which in the discrete case reads
\begin{equation}
W = \left[ (1 - e^{-2t}) \mathbb{I} + e^{-2t} C^0 \right]^{-1},
\end{equation}
where $W$ is the weight matrix implementing the linear score, $\mathbb{I}$ is the identity, and $C^0$ is the original covariance matrix of the data. Upon computing the equivalent of $WW^{-1} = \mathbb{I}$ for the continuous case, we indeed find
\begin{equation}
\int d^d \vec{y} \; \mathcal{S}_t(\vec{x}, \vec{y}) \left[ \Delta{ _t} \, \delta(\vec{y} - \vec{z}) + e^{-2t} \langle \varphi(\vec{y}) \varphi(\vec{z}) \rangle \right] 
\end{equation}
with the first term
\begin{equation}
\int d^d \vec{y} \; \mathcal{S}_t(\vec{x}, \vec{y}) \, \Delta{ _t} \, \delta(\vec{y} - \vec{z}) 
= \Delta{ _t} \, \mathcal{S}_t(\vec{x}, \vec{z})
\end{equation}
and the second term
\begin{align}
  e^{-2t} \int d^d \vec{y} \; \mathcal{S}_t(\vec{x}, \vec{y}) \langle \varphi(\vec{y}) \varphi(\vec{z}) \rangle
&= e^{-2t} \int d^d \vec{y} \int \frac{d^d \vec{k}}{(2\pi)^d} \; \frac{1}{\Delta{ _t} + \frac{e^{-2t}}{k^2 + m_{\text{eff}}^2}} \, e^{i \vec{k} \cdot (\vec{x} - \vec{y})}
\cdot \frac{1}{(2\pi)^d} \int d^d \vec{q} \; \frac{1}{q^2 + m_{\text{eff}}^2} \, e^{-i \vec{q} \cdot (\vec{y} + \vec{z})}\nonumber\\
&= e^{-2t} \, \int \frac{d^d\vec{k}}{(2\pi)^d} \;
\frac{1}{\Delta{ _t} + \frac{e^{-2t}}{k^2 + m_{\text{eff}}^2}} \,
e^{i \vec{k} \cdot (\vec{x} - \vec{z})} \,
\frac{1}{k^2 + m_{\text{eff}}^2}
\end{align}
where we have used the expression for the correlation function obtained in App.~\ref{app:lambda_eq_derivation} and the identity in Eq.~\eqref{eq:deltaidentity}.
Summing both terms gives
\begin{equation}
\int \frac{d^d\vec{k}}{(2\pi)^d} \; e^{i \vec{k} \cdot (\vec{x} - \vec{z})} \,
\frac{\Delta{ _t} (k^2 + m_{\text{eff}}^2) + e^{-2t}}{\Delta{ _t} (k^2 + m_{\text{eff}}^2) + e^{-2t}} = \int \frac{d^d\vec{k}}{(2\pi)^d} \; e^{i \vec{k} \cdot (\vec{x} - \vec{z})} = \delta(\vec{x} - \vec{z}),
\end{equation}
as it should.

\subsection{Saddle point computation}
\label{sec:check_exchange}

In the previous section, we have computed the score in the case of the Gaussian action, effectively first considering the \(n\to \infty\) limit and then the noising process by taking \(t>0\). We here show that the same result holds if the two operations are inverted, i.e. if we first take \(t>0\) and only then send \(n\to\infty\). In this case, the noised \(\psi\)-action, Eq.~\eqref{eq:action_intermediate}, reads
\begin{equation}\label{eq:noised_full_action}
\begin{gathered}
\mathfrak S_\psi[\psi;\varphi]
=
\frac{1}{2}
\sum_{a=1}^n
\frac{1}{(2\pi)^d}
\int d^d \vec{k}\,
\tilde{\psi}_a(\vec{k})
\left(
\vec{k}^{\,2}+m^2+\frac{e^{-2t}}{\Delta_t}
\right)
\tilde{\psi}_a(-\vec{k})
\\[0.4em]
-
\frac{e^{-t}}{\Delta_t}
\sum_{a=1}^n
\frac{1}{(2\pi)^d}
\int d^d \vec{k}\,
\tilde{\varphi}_a(\vec{k})
\tilde{\psi}_a(-\vec{k})
+
\frac{g}{4n}
\int d^d \vec{x}
\left(
\sum_{a=1}^n
\psi_a(\vec{x})^2
\right)^2 .
\end{gathered}
\end{equation}
At large \(n\), we introduce the parameter
\begin{equation}
\Lambda(\vec{x})
=
\frac{1}{n}
\sum_{a=1}^n
\psi_a(\vec{x})^2
\end{equation}
by inserting the functional identity
\begin{equation}\label{eq:delta_Lambda}
\begin{gathered}
1
=
\int \mathcal D\Lambda\,
\delta \left (
\Lambda(\vec{x})
-
\frac{1}{n}
\sum_{a=1}^n
\psi_a(\vec{x})^2
\right)
\\[0.4em]
=
\int \mathcal D\Lambda\,\mathcal D\hat{\Lambda}\,
\exp\left\{
\frac{i n}{2}
\int d^d \vec{x}\,
\hat{\Lambda}(\vec{x})
\left[
\Lambda(\vec{x})
-
\frac{1}{n}
\sum_{a=1}^n
\psi_a(\vec{x})^2
\right]
\right\},
\end{gathered}
\end{equation}
where \(\hat{\Lambda}(\vec{x})\) is the conjugate field to \(\Lambda(\vec{x})\). Then, the original quartic term in the action in Eq.~\eqref{eq:noised_full_action} is replaced by
\begin{equation}
\frac{g}{4n}
\int d^d \vec{x}
\left(
\sum_{a=1}^n
\psi_a(\vec{x})^2
\right)^2
\longrightarrow
\frac{ng}{4}
\int d^d \vec{x}\,
\Lambda(\vec{x})^2 ,
\end{equation}
while inserting Eq.~\eqref{eq:delta_Lambda} contributes to the action as
\begin{equation}
-\frac{i n}{2}
\int d^d \vec{x}\,
\hat{\Lambda}(\vec{x})\Lambda(\vec{x})
+
\frac{i}{2}
\sum_{a=1}^n
\int d^d \vec{x}\,
\hat{\Lambda}(\vec{x})\psi_a(\vec{x})^2 .
\end{equation}

By translational invariance, we impose
\begin{equation}
\Lambda(\vec{x})=\Lambda,
\qquad
\hat{\Lambda}(\vec{x})=\hat{\Lambda},
\end{equation}
(and noting that \(\int_V d^d \vec{x}=V\)), and the action then becomes
\begin{equation}
\begin{gathered}
\mathfrak S_\psi[\psi;\varphi,\Lambda,\hat{\Lambda}]
=
nV
\left(
\frac{g}{4}\Lambda^2
-
\frac{i}{2}\hat{\Lambda}\Lambda
\right)
+
\frac{1}{2}
\sum_{a=1}^n
\frac{1}{(2\pi)^d}
\int d^d \vec{k}\,
\tilde{\psi}_a(\vec{k})
\left(
\vec{k}^{\,2}+m^2+i\hat{\Lambda}+\frac{e^{-2t}}{\Delta_t}
\right)
\tilde{\psi}_a(-\vec{k})
\\[0.4em]
-
\frac{e^{-t}}{\Delta_t}
\sum_{a=1}^n
\frac{1}{(2\pi)^d}
\int d^d \vec{k}\,
\tilde{\varphi}_a(\vec{k})
\tilde{\psi}_a(-\vec{k}).
\end{gathered}
\end{equation}
We can now define
\begin{equation}
K_{\hat{\Lambda}}(\vec{k})
=
\vec{k}^{\,2}+m^2+i\hat{\Lambda}+\frac{e^{-2t}}{\Delta_t}
\end{equation}
and integrate over the \(\psi\) fields. Adding the additional $\varphi$-only quadratic term, $\frac{1}{2\Delta_t}
\sum_{a=1}^n
\frac{1}{(2\pi)^d}
\int d^d \vec{k}\,
\tilde{\varphi}_a(\vec{k})\tilde{\varphi}_a(-\vec{k})$, then yields the $\varphi$ action
\begin{equation}
\begin{gathered}
S[\varphi,\Lambda,\hat{\Lambda}]
=
nV
\left(
\frac{g}{4}\Lambda^2
-
\frac{i}{2}\hat{\Lambda}\Lambda
\right)
+
\frac{nV}{2}
\frac{1}{(2\pi)^d}
\int d^d \vec{k}\,
\log K_{\hat{\Lambda}}(\vec{k})
\\[0.4em]
+
\frac{1}{2}
\sum_{a=1}^n
\frac{1}{(2\pi)^d}
\int d^d \vec{k}\,
\tilde{\varphi}_a(\vec{k})
\left[
\frac{1}{\Delta_t}
-
\frac{e^{-2t}}{\Delta_t^2}
\frac{1}{K_{\hat{\Lambda}}(\vec{k})}
\right]
\tilde{\varphi}_a(-\vec{k}).
\end{gathered}
\end{equation}

We now introduce
\begin{equation}
I(\vec{k})
=
\frac{1}{nV}
\sum_{a=1}^n
\tilde{\varphi}_a(\vec{k})\tilde{\varphi}_a(-\vec{k})
=
\frac{1}{nV}
\sum_{a=1}^n
|\tilde{\varphi}_a(\vec{k})|^2,
\end{equation}
(where the factor $1/V$ is needed to remove the usual momentum-space volume factor from
 $(2\pi)^d\delta^{(d)}(\vec{0})=V$) by inserting
\begin{equation}
\begin{gathered}
1
=
\int \mathcal D I\,
\delta\left (
I(\vec{k})
-
\frac{1}{nV}
\sum_{a=1}^n
\tilde{\varphi}_a(\vec{k})\tilde{\varphi}_a(-\vec{k})
\right )
\\[0.4em]
=
\int \mathcal D I\,\mathcal D\hat I\,
\exp\left\{
i n V
\frac{1}{(2\pi)^d}
\int d^d \vec{k}\,
\hat I(\vec{k})
\left[
I(\vec{k})
-
\frac{1}{nV}
\sum_{a=1}^n
\tilde{\varphi}_a(\vec{k})\tilde{\varphi}_a(-\vec{k})
\right]
\right\}
\\[0.4em]
=
\int \mathcal D I\,\mathcal D\hat I\,
\exp\left\{
i n V
\frac{1}{(2\pi)^d}
\int d^d \vec{k}\,
\hat I(\vec{k})I(\vec{k})
-
i
\sum_{a=1}^n
\frac{1}{(2\pi)^d}
\int d^d \vec{k}\,
\hat I(\vec{k})
\tilde{\varphi}_a(\vec{k})\tilde{\varphi}_a(-\vec{k})
\right\}.
\end{gathered}
\end{equation}
After this insertion, the partition function becomes
\begin{equation}
\mathcal{Z}
=
\int
\mathcal D\varphi\ D I\,\mathcal D\hat I\,
\mathcal D\Lambda\,\mathcal D\hat{\Lambda}\,
\mathcal,
\exp\left\{
-
S[\varphi, I,\hat I,\Lambda,\hat{\Lambda}]
\right\},
\end{equation}
where
\begin{equation}
\begin{gathered}
S[\varphi, I,\hat I,\Lambda,\hat{\Lambda}]
=
nV
\left(
\frac{g}{4}\Lambda^2
-
\frac{i}{2}\hat{\Lambda}\Lambda
\right)
+
\frac{nV}{2}
\frac{1}{(2\pi)^d}
\int d^d \vec{k}\,
\log K_{\hat{\Lambda}}(\vec{k})
+
\frac{nV}{2}
\frac{1}{(2\pi)^d}
\int d^d \vec{k}\,
I(\vec{k})
\left[
\frac{1}{\Delta_t}
-
\frac{e^{-2t}}{\Delta_t^2}
\frac{1}{K_{\hat{\Lambda}}(\vec{k})}
\right]
\\[0.4em]
-
i n V
\frac{1}{(2\pi)^d}
\int d^d \vec{k}\,
\hat I(\vec{k})I(\vec{k})
+
i
\sum_{a=1}^n
\frac{1}{(2\pi)^d}
\int d^d \vec{k}\,
\hat I(\vec{k})
\tilde{\varphi}_a(\vec{k})\tilde{\varphi}_a(-\vec{k}) .
\end{gathered}
\end{equation}

The integral over the $\varphi$ fields then yields:
\begin{equation}
\begin{gathered}
\int \mathcal D\varphi\,
\exp\left[
-
i
\sum_{a=1}^n
\frac{1}{(2\pi)^d}
\int d^d \vec{k}\,
\hat I(\vec{k})
\tilde{\varphi}_a(\vec{k})\tilde{\varphi}_a(-\vec{k})
\right]
\propto
\exp\left[
-\frac{nV}{2}
\frac{1}{(2\pi)^d}
\int d^d \vec{k}\,
\log\left(i\hat I(\vec{k})\right)
\right].
\end{gathered}
\end{equation}
Therefore,
\begin{equation}
\hat{\mathcal{Z}}
=
\int
\mathcal D I\,\mathcal D\hat I\,
\mathcal D\Lambda\,\mathcal D\hat{\Lambda}\,
\exp\left[
-nV\,s[I,\hat I,\Lambda,\hat{\Lambda}]
\right],
\end{equation}
with
\begin{equation}
\begin{gathered}
s[I,\hat I,\Lambda,\hat{\Lambda}]
=
\frac{g}{4}\Lambda^2
-
\frac{i}{2}\hat{\Lambda}\Lambda
+
\frac{1}{2}
\frac{1}{(2\pi)^d}
\int d^d \vec{k}\,
\log K_{\hat{\Lambda}}(\vec{k})
+
\frac{1}{2}
\frac{1}{(2\pi)^d}
\int d^d \vec{k}\,
I(\vec{k})
\left[
\frac{1}{\Delta_t}
-
\frac{e^{-2t}}{\Delta_t^2}
\frac{1}{K_{\hat{\Lambda}}(\vec{k})}
\right]
\\[0.4em]
-
i
\frac{1}{(2\pi)^d}
\int d^d \vec{k}\,
\hat I(\vec{k})I(\vec{k})
+
\frac{1}{2}
\frac{1}{(2\pi)^d}
\int d^d \vec{k}\,
\log\left(i\hat I(\vec{k})\right).
\end{gathered}
\end{equation}
\end{widetext}
We now take the saddle point with respect to \(I(\vec{k})\), \(\hat I(\vec{k})\), \(\Lambda\), and \(\hat{\Lambda}\), denoting with $\star$ the parameter values at the saddle point. Variation with respect to \(\hat I(\vec{k})\) gives
\begin{equation}
\label{eq:I_hatI_relation}
I_\star(\vec{k})
=
\frac{1}{2i\hat I_\star(\vec{k})},
\end{equation}
and variation with respect to \(I(\vec{k})\) gives
\begin{equation}
\label{eq:I_hat_saddle}
2i\hat I_\star(\vec{k})
=
\frac{1}{\Delta_t}
-
\frac{e^{-2t}}{\Delta_t^2}
\frac{1}{K_{\hat{\Lambda}_\star}(\vec{k})}.
\end{equation}
Combining Eq.~\eqref{eq:I_hatI_relation} and Eq.~\eqref{eq:I_hat_saddle}, we find
\begin{align}\label{eq:I_star_Ct_hatLambda}
I_\star(\vec{k})
&=
\frac{1}{
\frac{1}{\Delta_t}
-
\frac{e^{-2t}}{\Delta_t^2}
\frac{1}{K_{\hat{\Lambda}_\star}(\vec{k})}
}\nonumber\\
&=
\Delta_t
+
\frac{e^{-2t}}{
\vec{k}^{\,2}+m^2+i\hat{\Lambda}_\star
},
\end{align}
which has the same form as Eq.~\eqref{eq:solution_score}. To determine \(\hat{\Lambda}_\star\), we solve the last two saddle point equations. Variation with respect to \(\Lambda\) gives
\begin{equation}\label{eq:lambda_lambdahat} 
i\hat{\Lambda}_\star=g\Lambda_\star, 
\end{equation}
while variation with respect to \(\hat{\Lambda}\) gives
\begin{align}
\label{eq:Lambda_hatLambda_saddle}
\Lambda_\star
=&
\frac{1}{(2\pi)^d}
\int d^d \vec{k}\,
\frac{1}{K_{\hat{\Lambda}_\star}(\vec{k})}
\\&+
\frac{1}{(2\pi)^d}
\int d^d \vec{k}\,
I_\star(\vec{k})
\frac{e^{-2t}}{\Delta_t^2}
\frac{1}{K_{\hat{\Lambda}_\star}(\vec{k})^2}.
\nonumber
\end{align}
At the saddle, using Eq.~\eqref{eq:I_star_Ct_hatLambda} and Eq.~\eqref{eq:lambda_lambdahat}, one finds
\begin{align}
\label{eq:gap_hatLambda}
\Lambda_\star
&=
\frac{1}{(2\pi)^d}
\int d^d \vec{k}\,
\frac{1}{\vec{k}^{\,2}+m^2+i\hat{\Lambda}_\star}\nonumber\\
&=
\frac{1}{(2\pi)^d}
\int d^d \vec{k}\,
\frac{1}{\vec{k}^{\,2}+m^2+g\Lambda_\star},
\end{align}
which is the same gap equation found when considering the target, \(t=0\), model; see Eq.~\eqref{eq:lambdadefinition}. We conclude that considering first the noised action and then the \(n \to \infty\) limit yields the same results as those presented in the main text.

\subsection{Stochastic differential equation (SDE) version}\label{app:SDE_treatment}

In the main text, the generation dynamics is considered under the deterministic ordinary differential equation, Eq.~\eqref{eq:solution_backward}. A similar computation is possible for its stochastic counterpart,
\begin{equation}\label{eq:SDE_backward_diffusion}
-\partial_t \varphi(\vec{x}, t) = \varphi(\vec{x}, t) + 2\mathcal{F}[\varphi(\vec{x}, t)] + \zeta(\vec{x}, t),
\end{equation}
where $\zeta(\vec{x}, t)$ is a Gaussian white noise term.
One can indeed write the Fokker--Plank equation associated to the backward diffusion (denoising) process in Fourier space as:
\begin{align}
\frac{\partial P_t(\ff(\vec{k}))}{\partial t}
=&
-\frac{\delta}{\delta \ff(\vec{k})}\Bigl[
P_t(\ff(\vec{k}))\,(2\tilde{\mathcal{S}}_t(\vec{k})-1)\,\ff(\vec{k})
\Bigr]\nonumber\\
&+\frac{1}{(2\pi)^d}\frac{\delta^2}{\delta \ff(\vec{k})^2}\Bigl[
P_t(\ff(\vec{k}))
\Bigr].
\end{align}
For $P(\ff)$, one can make a Gaussian ansatz, which holds if fields are Gaussianly distributed at the beginning of the denoising process,
\begin{equation}
P_t(\ff(\vec{k})) \propto
\frac{1}{\sqrt{\sigma_{\vec{k}}^2}}
\exp\Bigl[
-\frac{\ff(\vec{k})\ff(-\vec{k})}{2\sigma_{\vec{k}}^2}
\Bigr].
\end{equation}
The Fokker--Plank equation then reduces to an expression for the time evolution of the variance,
\begin{equation}\label{eq:var_sde}
    \frac{\partial \sigma^2_{\vec{k}}}{\partial t} = -2\left [\frac{1}{(2\pi)^d}-(2\tilde{\mathcal{S}}_t(\vec{k})-1)\sigma_{\vec{k}}^2 \right].
\end{equation}
One easily verifies that the variance found during the forward diffusion (noising) process,
\begin{equation}
   \sigma^2_\vk = \frac{1}{(2\pi)^d}\frac{\Delta_t(k^2+m_\mathrm{eff}^2)+ e^{-2t}}{(k^2+m_\mathrm{eff}^2)},
\end{equation}
is indeed the solution to Eq.~\eqref{eq:var_sde} with initial condition $\sigma^2_{\vec{k}}(t=0) = 1/[(2 \pi)^d(k^2+m_\mathrm{eff}^2)]$.

\section{Dependence of the error on discretization}\label{app:discretizations}
In this appendix, we consider the impact of numerical discretization on the diffusion process by controlling the resulting errors.

\subsection{Finite training dataset}

To investigate the dependence of the training on the number of data samples $M$, we consider a discretization of space, such that a configuration of the system is given in terms of $N$ variables. Using the linearity of the score, we get that the best approximation $S_t$ of the exact score kernel $\mathcal{S}_t$ is obtained by estimating the empirical correlation matrix $\hat{C}_0$ and then using 
\cite[Eq.~(8)]{BiroliMezard2023};
\begin{equation}\label{eq:exact_linear_score}
S_t = \left[ (1 - e^{-2t}) \mathbb{I} + e^{-2t} \hat{C}_0 \right]^{-1},
\end{equation}
thus effectively considering the training time $\bar{t} \to \infty$. 

The problem then becomes one of estimating a $N \times N$ matrix starting from $M$ data points. First, we note that in the regime $N > M$, the correlation matrix is non-invertible, so Eq.~\eqref{eq:exact_linear_score} relies on the diagonal term $(1-e^{-2t})\mathbb{I}$ in order not to cause problems. {A similar behavior has also been observed in Ref.~\cite{maillard2026memorisationconvergencegeneralisationgenerative}.} This leads to huge errors in the estimation of $\mathcal{S}_t$ when $t \simeq 0$. In the $M>N$ regime, the error on the single element of the matrix is $\mathcal{O} \left ({\frac{1}{\sqrt{M}}} \right )$. The total error then depends on $N$ and on the type of error measure chosen. For instance, the $L_{1,1}$ norm, which corresponds to the sum of the absolute errors on the single entries
\begin{equation}
\label{eq:EL11}
    E_{1,1} =||S_t-\mathcal{S}_t||_{L_{1,1}} = \sum_{x,y}|S_t(x,y)-\mathcal{S}_t(x,y)|,
\end{equation}
yields a behavior for the form
\begin{equation}
    E_{1,1} \sim \frac{N^2}{\sqrt{M}},
\end{equation}
while the Frobenius norm
\begin{equation}
\label{eq:EF}
    E_F =||S_t-\mathcal{S}_t||_{F} = \sqrt{\sum_{x,y}|S_t(x,y)-\mathcal{S}_t(x,y)|^2}
\end{equation}
yields
\begin{equation}
    E_{F} \sim \frac{N}{\sqrt{M}}.
\end{equation}
These behaviors are depicted in Fig.~\ref{fig:data_comparison}.

Real-space translational invariance makes the diffusion kernel $\tilde{\mathcal{S}}$ diagonal in momentum space. One can then simply train a diagonal kernel $\tilde{S}_t$, thus reducing the number of training parameters from $N^2$ to $\mathcal{O}(N)$. (The precise number varies slightly if one considers the $k\leftrightarrow -k$ symmetry.) The errors are then of the form:
\begin{equation}
    E_{1,1} \sim \frac{N}{\sqrt{M}}
\end{equation}
and
\begin{equation}
    E_{F} \sim \sqrt\frac{N}{M}.
\end{equation}
Moreover, the problems arising from a non-invertible correlation matrix in the $M < N$ regime then disappear. Examples are shown in Fig.~\ref{fig:data_comparison_momentum}. The above reasoning also applies when one tries to learn a translational invariant score kernel, for which, therefore, the relative error at fixed $M$ does not depend on the size of the system.

\begin{figure*}[t]
    \centering
    \includegraphics[width=0.85\linewidth]{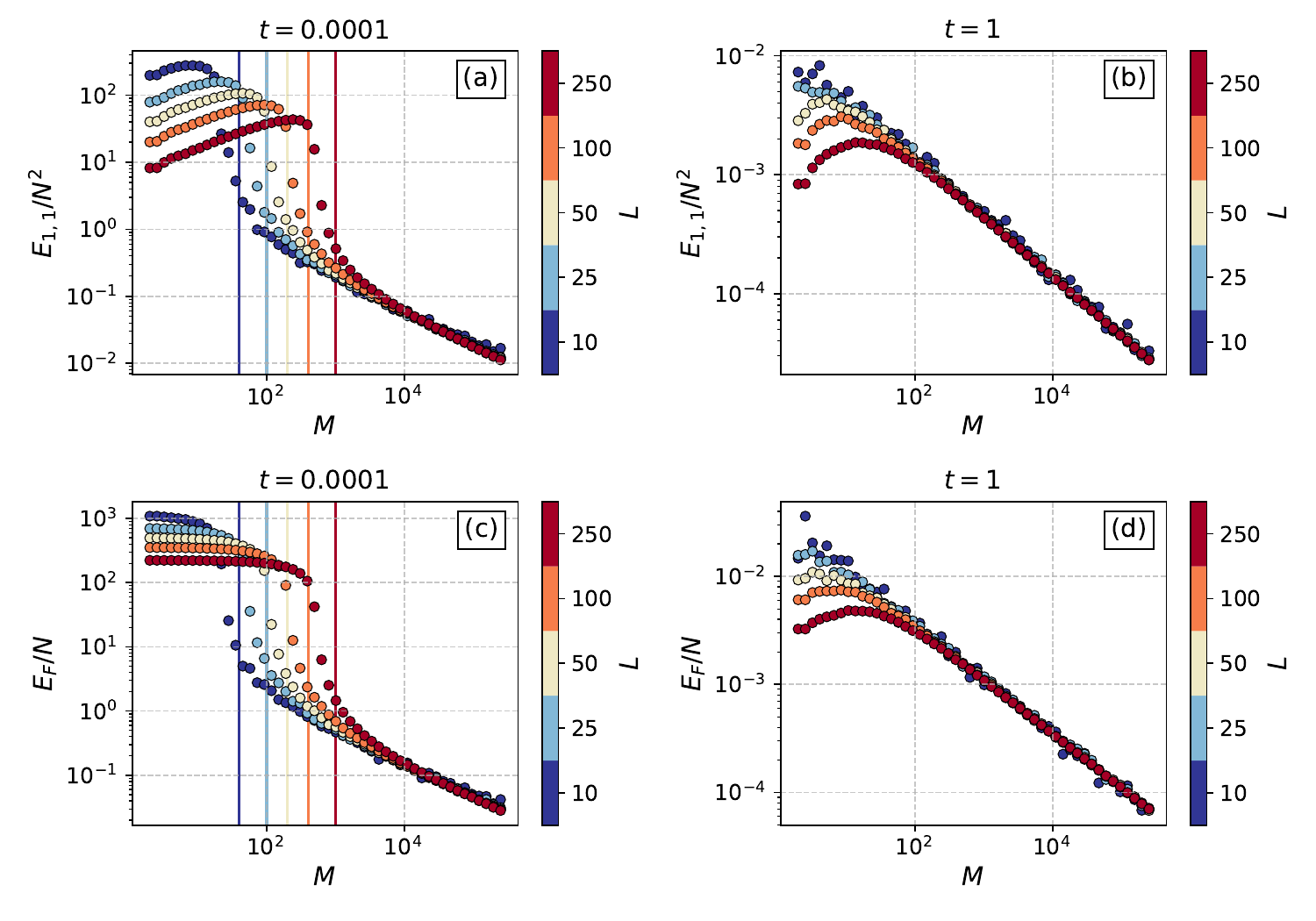}
    \caption{Error of the trained real-space score as a function of the training dataset size $M$ (points) for different forward diffusion times $t$: $L_{1,1}$ norm, Eq.~\eqref{eq:EL11}, for \textbf{(a)} $t=10^{-4}$ and \textbf{(b)} $t=1$; Frobenius norm, Eq.~\eqref{eq:EF} for \textbf{(c)} $t=10^{-4}$ and \textbf{(d)} $t=1$.  The inferred $S_t$ is obtained by estimating the empirical correlation matrix and then using Eq.~\eqref{eq:exact_linear_score} (effectively considering the training time $\bar{t}\to\infty$). Full vertical lines denote $M=N$, where $N=2L/a$ is the number of points that discretize the space. \textbf{Setting:} $d=1$, $a=0.5$, and $m_\mathrm{eff}=1$.}
    \label{fig:data_comparison}
\end{figure*}

\begin{figure*}[t]
    \centering
    \includegraphics[width=0.85\linewidth]{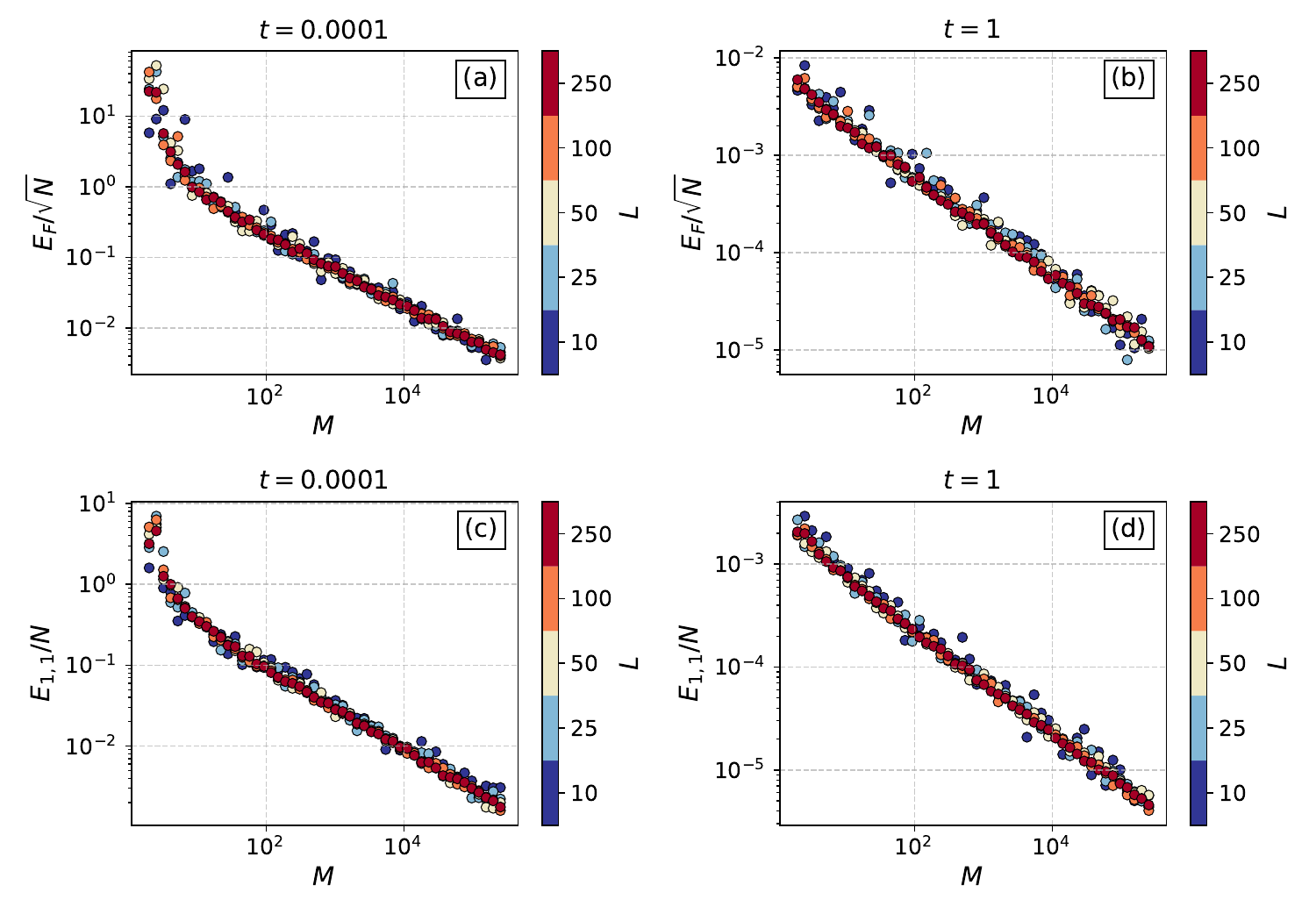}
    \caption{Error of the trained momentum-space score as a function of the training dataset size $M$ (points) for different forward diffusion times $t$: $L_{1,1}$ norm, Eq.~\eqref{eq:EL11}, for \textbf{(a)} $t=10^{-4}$ and \textbf{(b)} $t=1$; Frobenius norm, Eq.~\eqref{eq:EF} for \textbf{(c)} $t=10^{-4}$ and \textbf{(d)} $t=1$.  The inferred $S_t$ is obtained by estimating the empirical correlation matrix and then using Eq.~\eqref{eq:exact_linear_score} (effectively considering the training time $\bar{t}\to\infty$). Full vertical lines denote $M=N$, where $N=2L/a$ is the number of points that discretize the space. \textbf{Setting:} $d=1$, $m_\mathrm{eff}=1$, and $a=0.5$.}
    \label{fig:data_comparison_momentum}
\end{figure*}

\subsection{Backward diffusion (denoising) process}

\begin{figure*}[t!]
    \centering
    \includegraphics[width=0.75\linewidth]{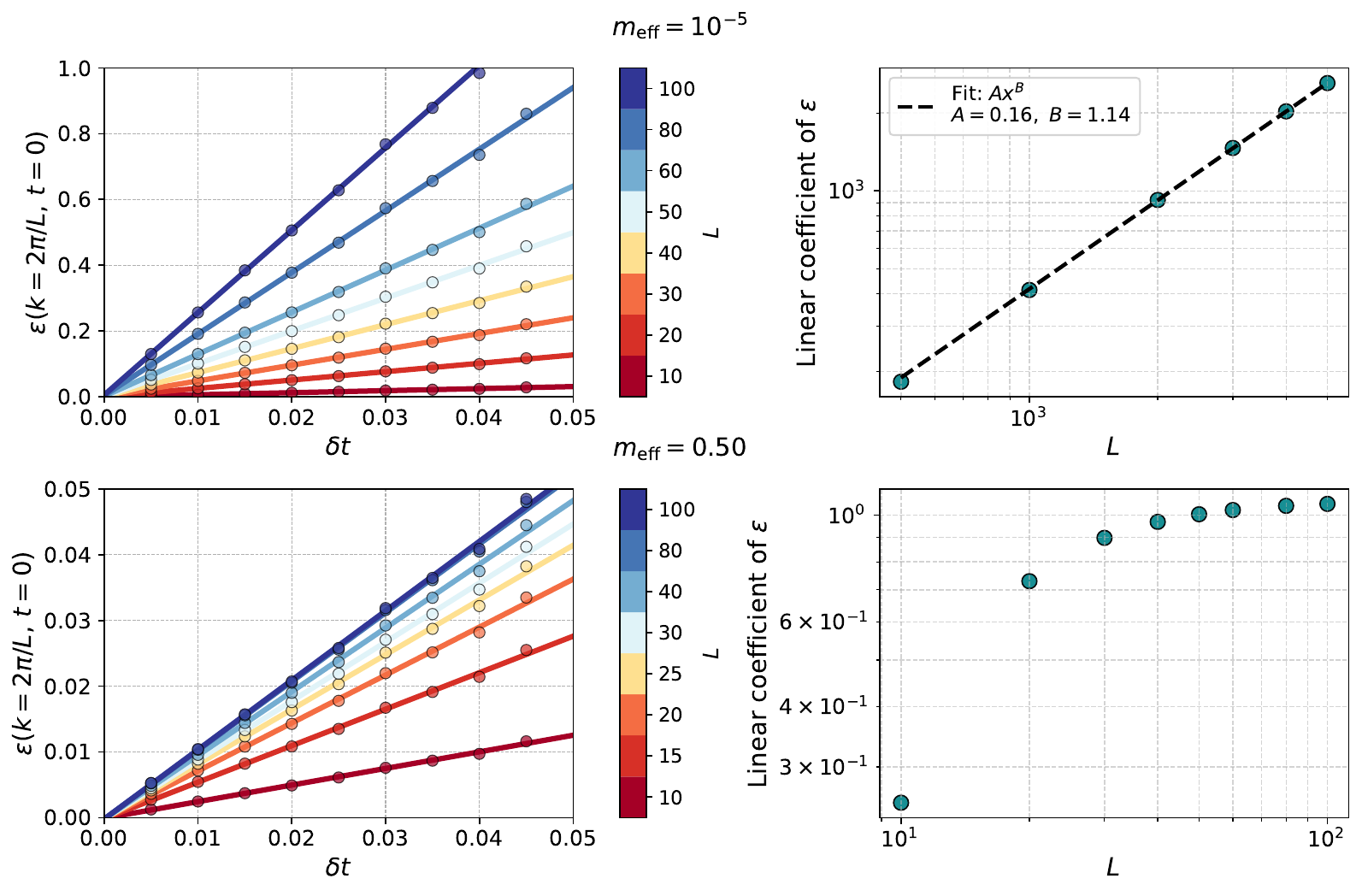}
    \caption{Discretization error for the standard deviation estimate of the $k=2\pi/L$ mode of a $d=1$ system at $t=0$, $\varepsilon(t)=\sigma_\mathrm{dis}(\pi/L,0)-\sigma_\mathrm{ex}(\pi/L,0)$. \textbf{Left column:} error behavior for different linear sizes $L$. \textbf{Right column:} linear coefficients of the fits from the left panel (for a larger set of $L$) as a function of $L$. \textbf{Top row}: $m_\mathrm{eff}=0.10^{-5}$. \textbf{Bottom row}: $m_\mathrm{eff}=0.5$. In the top right plot a power-law fit highlights that the error scales as $L$, while in the bottom right plot the error saturates given the relatively large value of the mass. \textbf{Setting:} $a=1$ and $t_\mathrm{max}\simeq 10$ (adjusted to be a nearby multiple of $\delta t$). }
    \label{fig:disc_error_LandM}
\end{figure*}

We now consider the error made by approximating the continuous ODE for the backward diffusion (denoising) process in Eq.~\eqref{eq:diffusion_ode_exact} by a discrete dynamical process. Consider a simple Euler integration scheme of step $\delta t$ and the perfect score given by Eq.~\eqref{eq:sigma}. The total error is then $\mathcal{O}(\delta t)$. A higher precision is possible by considering the ODE in momentum space discretized as
\begin{equation}
    \frac{\tilde{\varphi}(\vec{k}, t-\delta t)-\tilde{\varphi}(\vec{k}, t)}{\delta t} = \tilde{\varphi}(\vec{k}, t) + (2 \pi)^d\tilde{\varphi}(\vec{k}, t)\tilde{\mathcal{S}}_t(\vec{k}).
\end{equation}
This equation can be rewritten as a dynamical system
\begin{equation}\label{eq:discretized_ODE_mometumspace}
    \tilde{\varphi}(\vec{k}, t-\delta t) = \tilde{\varphi}(\vec{k}, t) \left[ 1+ \delta t \left(1-(2 \pi)^d\tilde{\mathcal{S}}_t(\vec{k})\right) \right].
\end{equation}
Because the fields are originally Gaussian and all operations are linear, fields remain Gaussian at all times; the backward diffusion (denoising) process only changes their variance. Because of the discretization, however, this variance differs from the exact one obtained by solving explicitly the continuous ODE as done in Sec.~\ref{sec:generation_exact_score}. 
We can write down an explicit recursive equation for the error in the standard deviation (either of the real or the imaginary part) $\varepsilon(t) = \sigma_\mathrm{dis}(\vec{k}, t)-\sigma_\mathrm{ex}(\vec{k}, t)$, where $\sigma_\mathrm{dis}(\vec{k}, t)$ and $\sigma_\mathrm{ex}(\vec{k}, t)$ are the standard deviations of the fields in momentum space obtained via the discretized and the exact backward diffusion (denoising) process, respectively. We find
\begin{widetext}
\begin{equation}\label{eq:ds_error_discretization}
\varepsilon(\vec{k}, t - \delta t) = 
- \sigma_{\text{ex}}(\vec{k}, t - \delta t) +
\sigma_{\text{ex}}(\vec{k}, t)
\left[ 
1 + \delta t \left( 1 - (2 \pi)^d \tilde{\mathcal{S}}_t(\vec{k}) 
\right)
\right]
+
\varepsilon(\vec{k},t)
\left[ 
1 + \delta t \left( 1 - (2 \pi)^d \tilde{\mathcal{S}}_t(\vec{k})
\right)
\right].
\end{equation}
which can then be solved numerically to obtain the error as a function of diffusion time $t$. 
\end{widetext}
In Fig.~\ref{fig:disc_error_LandM} the final generation error $\varepsilon(k = 2\pi/L, t = 0)$ (for a $d=1$ system) is shown as a function of the discretization step $\delta t$ for different linear system size $L$ and effective mass $m_\mathrm{eff}$. As expected, the final error is $\mathcal{O}(\delta t)$. At fixed $\delta t$, when the minimum momentum is controlled by the system size, i.e., $|\vec{k}| \sim 1/L$, the error grows---for large enough $L$ and sufficiently small masses---linearly in $L$ (top row). For $m_\mathrm{eff} \gg 1/L$, the error becomes independent of the system size (bottom row). 

This error can be made still smaller by using higher-order integration schemes. For instance, the  4th-order Runge-Kutta algorithm used in Fig.~\ref{fig:complete} requires a discretization step that scales as $L^{1/4}$, while only requiring 4 times more calls to the score. Higher-order schemes can further reduce the error \cite{hairer2015runge}. Therefore, as long as an integration scheme is available, the scaling with $L$ can be made arbitrary small while the number of calls to the score is multiplied by a factor that does not scale with $L$. Adaptive schemes could also be considered to further decrease the integration error.

\section{Non-local score assessment}\label{app:local_score}

As discussed at the end of Sec.~\ref{sec:twolayers} and the beginning of Sec.~\ref{sec:fourier}, standard convolutional layers whose kernel spans the whole system have a forward pass complexity of $N^2$. It might therefore be tempting to reduce their complexity by introducing a \textit{local} score, whose kernel does not scale as the system size. Indeed, the role of locality in diffusion models has recently begun to be studied \cite{lukoianov2026locality}, particularly in connection with generalization properties \cite{kamb2025an} and even more recently in statistical-physics settings, in relation with the  $k$-SAT and $k$-XORSAT problems~\cite{bhatt2026generating}. This approach indeed works for finite values of the effective mass $m_\mathrm{eff}$, for which the correlation length stays finite, but fails when $m_\mathrm{eff} \to 0$. Interestingly one can still have a finite error in the total score while experiencing this failure, because the effect only impacts the score small momenta. In this appendix, we first show that a (moderately growing) local score is sufficient to control the point-wise total error in the estimated score, and then show that such a local score nevertheless fails to capture the critical behavior of the physical system. We conclude that looking solely at the standard error on the score is a poor diagnostic for the critical behavior of the generated configurations. 

\subsection{Single site total error}

\subsubsection{Error analysis}
\begin{figure*}
    \centering
    \includegraphics[width=0.85\linewidth]{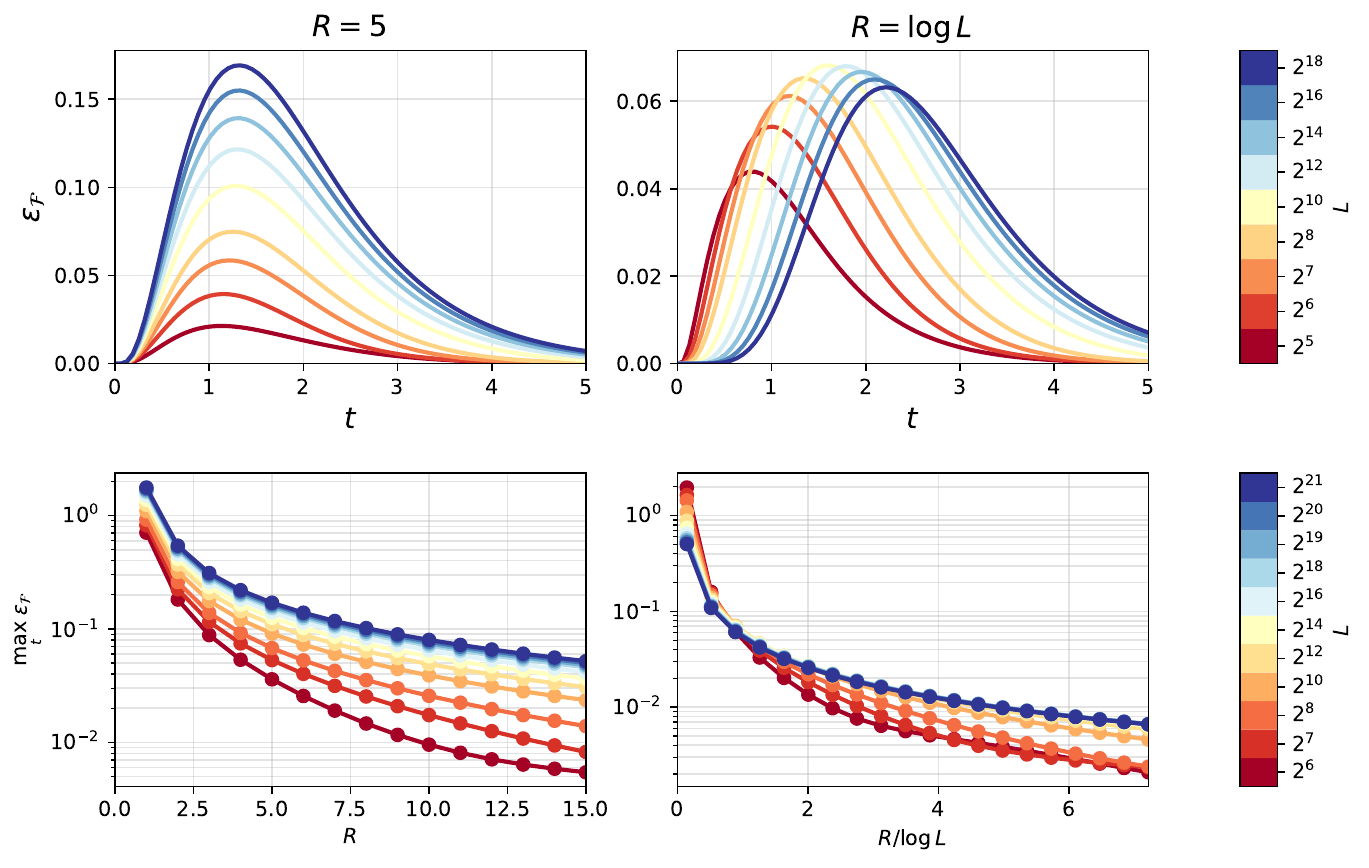}
    \caption{\textbf{Top row:} backward diffusion (denoising) time evolution of the relative error $\varepsilon_\mathcal{F}$, Eq.~\eqref{eq:epsilonF}, for different system sizes $L$ with (\textbf{left}) fixed kernel size cutoff radius $R = 5$ and (\textbf{right}) $R \sim \log L$. For $R = 5$, the relative error increases with $L$; for $R = \log L$ it saturates at sufficiently large $L$. \textbf{Bottom row:} peak error of $\varepsilon_\mathcal{F}(t)$ (for $t \in [0.02, 5]$) for different $L$, as a function of (\textbf{left}) the cutoff radius $R$ and (\textbf{right}) the rescaled cutoff radius $R/\log L$. \textbf{Setting:} $d=2$, $m_\mathrm{eff} = 0$, and $a =1$.}
    \label{fig:analytical_relative_error_local}
\end{figure*}
Note that, from the definition of the action in  Eq.~\ref{eq:lambda_action}, the score kernel in real space is local at $t = 0$,  $\mathcal{S}_0(\vec{x},\vec{y})=-\nabla^2 + m_\mathrm{eff}^2$, because the Laplacian is a local operator; it is also local at $t = \infty$, because it is then proportional to the identity matrix. The truncation difficulties therefore take places at intermediate times. We here specifically consider approximating the score with a (local) kernel that vanishes for {$|\vec{x}-\vec{y}|>\mathcal{K}/2$ (where we define as $\mathcal{K}$ the total size of the kernel window)}.\footnote{For a discretized version of the system and with its $k=0$ component removed, the score kernel acquires (for $m_\mathrm{eff}\neq 0$) a non-local component. The total contribution is nevertheless independent of $L$. Removing the zero mode introduces an error $\mathcal{O}(1/L^d)$ on each of the score kernel weights, and a local score kernel would neglect $\mathcal{O}(L^d)$ of these elements. Therefore, the total contribution of the neglected elements remains finite as $L$ grows.}

We first approximate the score  with the same form as Eq.~\eqref{eq:exact_score}, but with a score kernel that is Eq.~\eqref{eq:sigma} for $|\vec{x}-\vec{y}|<R = \mathcal{K}/2$  and 0 otherwise. The error in the score for a given field configuration $\varphi_t$ and a given site (without loss of generality, chosen at the origin, $\vec{x} = 0$) is then
\begin{equation}\label{eq:error_deltaF}
    \delta \mathcal{F}_t^{(R)}
    =
    \mathcal{F}_t-\mathcal{F}_{t}^{(R)}
    =
    -\int_{|\vec y|>R} d^2\vec y \, \mathcal{S}_t(0,\vec y)\,\varphi_t(\vec y).
\end{equation}
For $\varphi_t \sim P_t(\varphi)$, the error is then a Gaussian random variable of zero mean and variance:
\begin{equation}\label{eq:integral_variance}
\begin{gathered}
    \mathrm{Var}\!\left[\delta \mathcal{F}_t^{(R)}\right]
    =
    \iint_{|\vec y|>R,\,|\vec y'|>R}
    \!\!\!\!\!\!\!\!\!\!\!\!\!\!\!\!\!\!\!\!\!\!\!\!\!\!\!d^2\vec y\, d^2\vec y'\,
    \mathcal{S}_t(0,\vec y)\,
    C_t(|\vec y-\vec y'|)\,
    \mathcal{S}_t(0,\vec y').
\end{gathered}
\end{equation}
where $C_t$ is the covariance of $P_t$.

A crude scaling argument starting from Eq.~\ref{eq:integral_variance} offers a solution for general $d$. Introducing the adimensional quantities $\vec{\xi} = \vec{y}/R$ and $\vec{\xi}' = \vec{y}'/R$ gives
\begin{equation}\label{eq:real_space_variance_analysis}
\begin{gathered}
\mathrm{Var}\!\left[\delta \mathcal{F}_t^{(R)}\right]
\simeq
\iint_{|\vec \xi|>1,\,|\vec \xi'|>1}\!\!\!\!\!\!\!\!\!\!\!\!\!\!\!\!\!\!\!\!\!\!\!\!
d^d\vec \xi\, d^d\vec \xi'\,
g_d(|\vec \xi|)\,
C_t\!\bigl(R|\vec \xi-\vec \xi'|\bigr)\,
g_d(|\vec \xi'|),
\end{gathered}
\end{equation}
where $\mathcal{S}_t\sim R^{-d}g_d(|\vec y|/R)$ with
\begin{equation}
g_d(\rho)
=
\frac{1}{(2\pi)^{d/2}}
\left(\frac{M_tR}{\rho}\right)^{\frac d2-1}
K_{\frac d2-1}(M_tR \rho).
\end{equation}
For $d \geq 3$, $C_t(R|\vec{\xi}-\vec{\xi'}|) \sim R^{-d}|\vec{\xi}-\vec{\xi'}|^{-(d-2)}$ and Eq.~\eqref{eq:real_space_variance_analysis} gives:
\begin{equation}
\mathrm{Var}\!\left[\delta \mathcal{F}_t^{(R)}\right]
\sim
R^{-d},
\end{equation}
with no leading $L$ dependence. For $d = 2$, $C_t(R|\vec{\xi}-\vec{\xi'}|) \sim R^{-2}\log\!\left(\frac{L}{R|\vec{\xi}-\vec{\xi'}|}\right)$, and therefore the error scales as 
\begin{equation}
\mathrm{Var}\!\left[\delta \mathcal{F}_t^{(R)}\right]
\sim
\frac{\log(L)}{R^2}.
\end{equation}
Finally, for $d = 1$, $C_t(R|\vec{\xi}-\vec{\xi'}|) \sim \frac{L}{R^2}$, so that
\begin{equation}
\mathrm{Var}\!\left[\delta \mathcal{F}_t^{(R)}\right]
\sim
\frac{L}{R^2}.
\end{equation}
In summary, a local score kernel that grows at most weakly with system size suffices to control the error \eqref{eq:error_deltaF}.

The above argument can be refined to facilitate its numerical validation. For instance, in $d=2$ Eq.~\ref{eq:error_deltaF} becomes
\begin{equation}
    \delta \mathcal{F}_t^{(R)}
    =
    \frac{e^{-2t}}{2\pi \Delta_t^2}
    \int_{|\vec y|>R} d^2\vec y \,
    K_0\!\left(M_t |\vec y|\right)\,\varphi_t(\vec y),
\end{equation}
where we have used the explicit form of the kernel,
\begin{equation}
    \mathcal{S}_t(0,\vec y)
    =
    \frac{1}{\Delta_t}\,\delta^{(2)}(\vec y)
    -
    \frac{e^{-2t}}{2\pi \Delta_t^2}
    K_0\!\left(M_t |\vec y|\right).
\end{equation}
This expression can be simplified, yielding---after introducing infrared $k_{\min}= 2\pi/L$ and UV $k_{\max}= \pi/a$ cutoffs
\begin{equation}\label{eq:variance}
\begin{gathered}
    \mathrm{Var}\!\left[\delta \mathcal{F}_t^{(R)}\right]
    =\\
    \frac{e^{-4t}}{\Delta_t^4}
    \int_0^\infty \frac{k\,dk}{2\pi}\,
    \widetilde C_t(k)\,
    \left[
    \int_R^\infty dr\, r\, J_0(kr)\,K_0(M_t r)
    \right]^2,\\
    =    \frac{e^{-4t}}{\Delta_t^4}
    \int_{k_{\min}}^{k_{\max}} \frac{k\,dk}{2\pi}\,
    \tilde{C}_t(k)
    I^2(R,t).
\end{gathered}
\end{equation}
where $J_\nu$ is the ordinary Bessel function of the first kind of order $\nu$ and its integration gives \begin{equation}\label{eq:integral_for_variance}
    I(R,t) = \frac{R}{k^2+M_t^2}
    M_t K_1(M_tR)J_0(kR)-kK_0(M_tR)J_1(kR).
\end{equation}
The relative error at diffusion time $t$ is then
\begin{equation}
    \varepsilon_\mathcal{F} = \left ( \frac{\mathrm{Var}\!\left[\delta \mathcal{F}_t^{(R)}\right]}{\mathrm{Var}\!\left[\mathcal{F}_t\right]} \right )^\frac{1}{2},
\end{equation}
with
\begin{equation}
\mathrm{Var}\!\left[\mathcal{F}_t\right]
=
\int_{k_{\min}}^{k_{\max}} \frac{k\,dk}{2\pi}\,
\frac{k^2+m^2}{\Delta_t\!\left(k^2+m^2\right)+e^{-2t}}.
\end{equation}

As expected from the general scaling argument, taking $R = O(1)$ gives a maximum error that increases with $L$ (see top left panel Fig.~\ref{fig:analytical_relative_error_local}), while taking $R \sim \log L$  controls the error (see top right panel of Fig.~\ref{fig:analytical_relative_error_local}). In particular, the maximum of the error in diffusion time $t$ of the different curves (bottom left panel of Fig.~\ref{fig:analytical_relative_error_local}) collapses when plotted as a function of $R/\log L$. A logarithmic scaling of the kernel size therefore controls the error. Repeating this analysis gives that $R$ should grow as $\sqrt{L}$ in $d = 1$ and stay finite in $d \geq 3$ to control the error (see Fig.~\ref{fig:d1_analytical}). Additionally, the above argument can be reproduced as a momentum space analysis. Indeed, in $d$ dimensions, Eqs.~\eqref{eq:variance} and \eqref{eq:integral_for_variance} become
\begin{align}\label{eq:general_d_variance}
\mathrm{Var}\!\left[\delta \mathcal{F}_t^{(R)}\right]
&=\frac{e^{-4t}}{\Delta_t^4}\,
\frac{M_t^{\,d-2}}{2^{\,d-1}\pi^{d/2}\Gamma(d/2)}\times\\
&
\int_{k_{\min}}^{k_{\max}} dk\, k\,\widetilde C_t(k)\,
\left[
I_{\frac d2-1}(R,t;k)
\right]^2\nonumber
\end{align}
and
\begin{align}
&I_{\frac d2-1}(R,t;k)
=\frac{R}{k^2+M_t^2}\times\\
&
\left[
M_t K_{\frac d2}(M_tR)\,J_{\frac d2-1}(kR)
-
k K_{\frac d2-1}(M_tR)\,J_{\frac d2}(kR)
\right],\nonumber
\end{align}
respectively. The $L$-dependence in the integral in Eq.~\eqref{eq:general_d_variance} then comes from the low-$k$ behavior for $R \ll L$, such that $kR \ll 1$ and $RM_{t^*} = \mathcal{O}(1)$. For the time $t^*$ at which the error is maximal, one finds that the relative error scales as
\begin{equation}
    \mathrm{Var}\!\left[\delta \mathcal{F}^{(R)}_t(0)\right] \simeq R^{-2}\int_{2\pi/L}^{c/R} dk\, k^{d-3} + \tilde{c}(R), 
\end{equation}
where $c$ is a suitably chosen constant and $\tilde{c}(R)$ is an $L$-independent function that decreases as $R$ increases.
Moreover, $\mathrm{Var}\!\left[\mathcal{F}_t\right]$ does not depend on $L$ nor $R$ at leading order. From this result, one gets that to bound the error:
\begin{itemize}
    \item for $d = 1$, one needs $R \sim \sqrt{L}$;
    \item for $d = 2$, one needs a power of $\log L$---the precise power cannot be obtained from this analysis;
    \item for $d \geq 3$, $R$ need not grow with $L$.
\end{itemize}

\begin{figure*}
    \centering
    \includegraphics[width=1.0\linewidth]{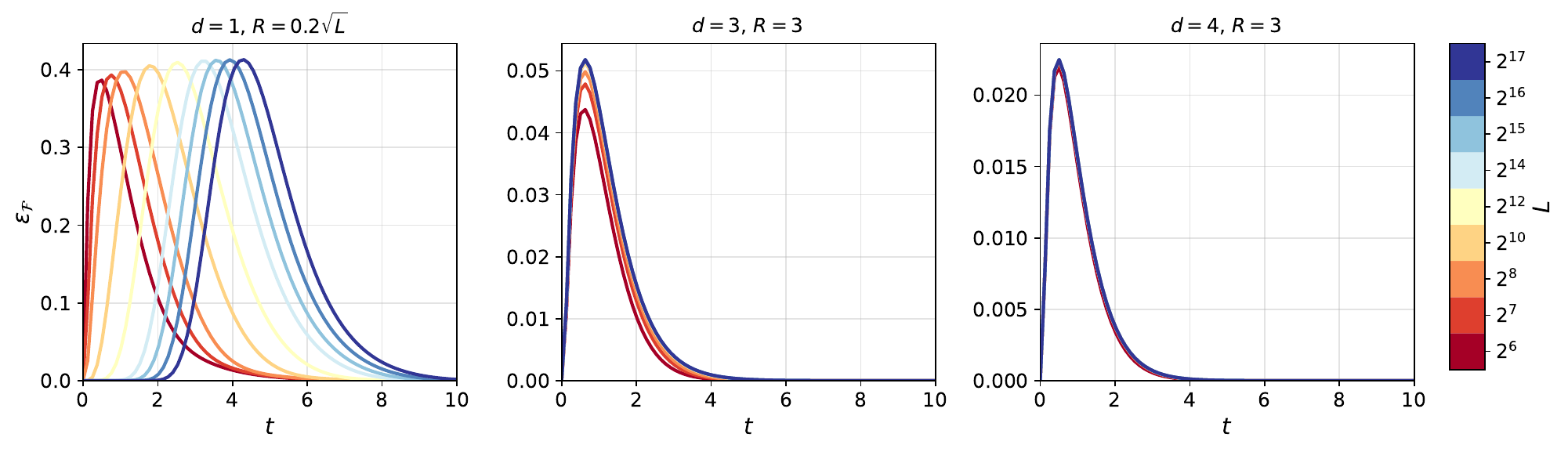}
    \caption{Relative error $\varepsilon_\mathcal{F}$ as a function of time for different system sizes $L$ and varying cutoff radius $R = 0.2\sqrt{L}$, $d = 1$ for (\textbf{left}), fixed cutoff radius $R = 3$ for (\textbf{center}) $d = 3$ and (\textbf{right}) $d=4$. For a one-dimensional system, the kernel size should grow as $\sqrt{L} $ in order for the error to remain bounded for increasing system sizes. For $d \geq 3$, even if the kernel size remains constant the relative error does not increase with system size. \textbf{Setting:} $m_\mathrm{eff} = 0$ and $a =1$.}
    \label{fig:d1_analytical}
\end{figure*}

\begin{figure*}
    \centering
    \includegraphics[width=0.8\linewidth]{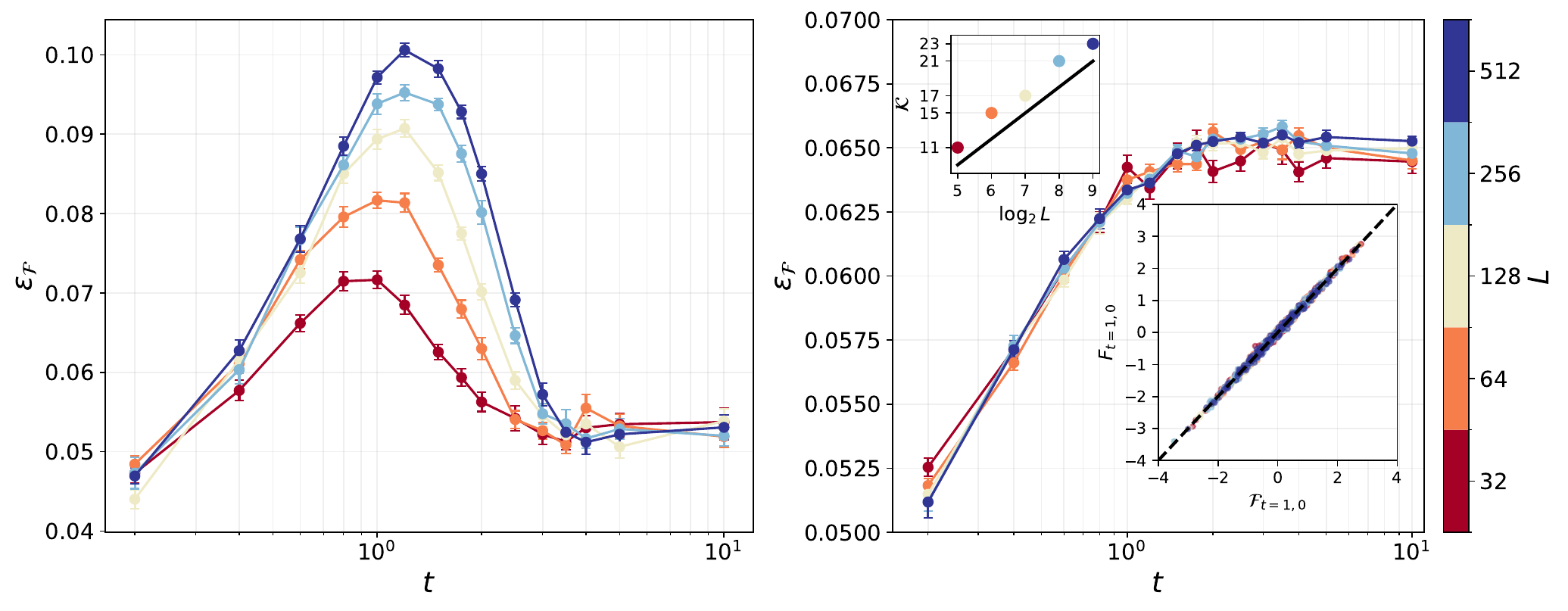}
    \caption{Backward diffusion (denoising) time evolution of the relative error $\varepsilon_\mathcal{F}$, Eq.~\eqref{eq:epsilonF}, for a neural network (see text for architectural details) trained on systems of different size $L$ with \textbf{(left)} a fixed kernel size $\mathcal{K}= 3$ and \textbf{(right)} a logarithmically growing kernel, $\mathcal{K}= 2\lfloor 3/2\log_2 L - 2 \rfloor+1$, i.e., an odd integer.       For the former, the peak error around $t \simeq 1$ grows with $L$; for the latter, $\varepsilon_\mathcal{F}$ is essentially independent of $L$. The error is controlled and relatively small, even though the training is only performed with a small (10) number of epochs. Error bars denote $\pm 1$ standard deviation. \textbf{Top inset}: Scaling of the kernel size with increasing system size $L$ (scatter points) together with the logarithmic trend line $\sim \log L$. \textbf{Bottom inset}: First component of the trained score $F_{t=1; 1}$ at time $t=1$ plotted against the first component of the exact score $\mathcal{F}_{t=1;1}$ for different system sizes. The correlation is nearly perfect. \textbf{Setting:} $d=2$, $m_\mathrm{eff} = 0$, and $a =1$. {Training was performed for 10 epochs on 1024 configurations with batch size 256 and learning rate $\eta = 0.13$.}}   
    \label{fig:error_score_multinetwork}
\end{figure*}

\subsubsection{Numerical analysis} 
We test the above analysis in $d=2$ using a network with three linear convolutional layers and four channels, all with circular padding. As in the main text, we use a different model (e.g. a different set of parameters) for each diffusion time $t$.
Figure~\ref{fig:error_score_multinetwork} shows the relative error of the real-space score kernel
\begin{equation}
\label{eq:epsilonF}
\varepsilon_{\mathcal F}
=
\left(
\frac{\overline{\|F-\mathcal{F}\|_2^2}}
{\overline{\|\mathcal{F}\|_2^2}}
\right)^{1/2},
\end{equation}
where $\|\cdot\|_2$ denotes the Euclidean norm over all lattice sites, with slight notational abuse. As expected, fixing the kernel size for all $L$, results in a growing peak height---and hence relative error---with $L$, but taking $\mathcal{K} \sim \log L$  fully controls the relative error. 

This analysis shows that one can control the error on the score, as characterized by the variance associated to the corresponding random variable, by a moderately small kernel. However, this does not imply that all the modes are learned correctly, as shown next.

\subsection{Effects of the score truncation}

We show here what is the effect of truncating the score. We consider the exact score \eqref{eq:exact_score} in $d = 2$, but set it to zero when $|\vec{x}-\vec{y}| > \mathcal{K}/2$. We use this cropped score to produce configurations by numerically integrating the backward ODE, as done in the main text (Sec.~\ref{sec:fourier}). We then plot the corresponding power spectrum $P(k)$ \eqref{eq:powerspectrum} and compare it with the correct one. As shown in Fig.~\ref{fig:cropped}, the effect of introducing a cutoff is similar to using a finite effective mass, and therefore a finite correlation length. In particular, as long as $m_\mathrm{eff}$ is large enough, and correspondingly the correlation length is small enough, the cropped score suffices to produce configurations whose $P(k)$ matches the exact one. However, as the effective mass is decreased, the discrepancy between the real and the generated curves increases more. A local kernel, therefore, although sufficient to capture the physics for finite effective masses, is incapable of correctly reproducing the critical behavior of the model.

\begin{figure*}
    \centering
    \includegraphics[width=1\linewidth]{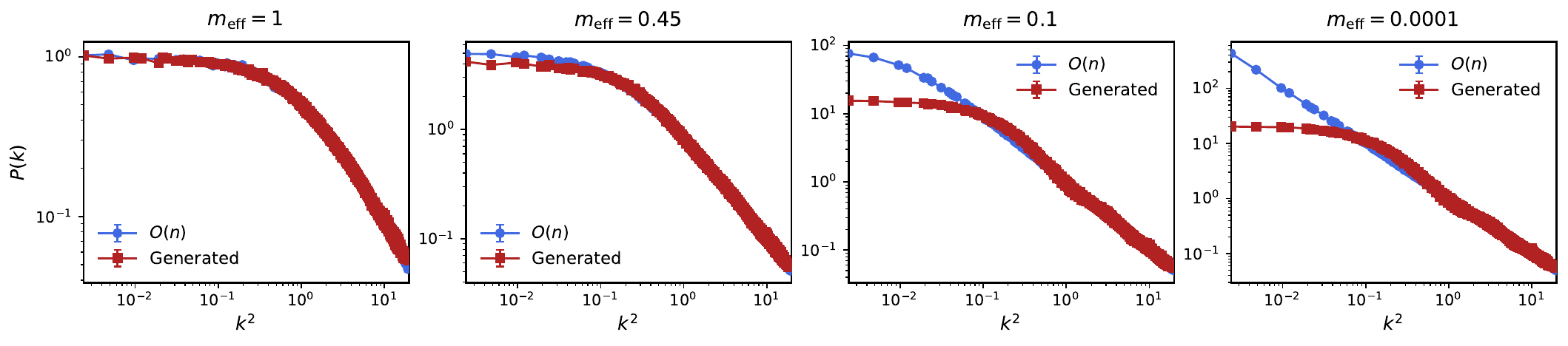}
    \caption{Comparison between the power spectra $P(k)$ computed on exact (blue) and generated (red) configurations, for $L = 128$ and different values of $m_\mathrm{eff}$. The generated configurations are produced using a kernel $\mathcal{K} = 9$. As long as $m_\mathrm{eff}$ is large enough, the generated configurations have the same power spectrum as the target, but the two start to differ at low $k$ when the effective mass is decreased, thus highlighting the non-locality of the score kernel near criticality. All the spectra have been estimated using 1000 configurations. Lines are just a guide for the eye.}
    \label{fig:cropped}
\end{figure*}

\clearpage

\bibliography{refs}

\clearpage
\onecolumngrid
\FloatBarrier

\renewcommand{\thesection}{S\arabic{section}} \renewcommand{\appendixname}{}
\renewcommand{\thefigure}{S\arabic{figure}}
\setcounter{figure}{0}
\setcounter{section}{0}

\pagenumbering{gobble}
\newpage

\centerline{\large\bf Supplementary Material for ``The critical slowing down in diffusion models''}

\section{Additional figures}\label{app:addit_figures}

\subsection{{Non-convergence of Adam and Adagrad in the one layer linear setting}}

{In Figure \ref{fig:nonconvergence}, we plot the error in the score kernel ${\varepsilon_{\mathcal{S}}}$ as a function of the number of training epochs for GD, Adam and Adagrad. As can be observed from the figure, both Adam and Adagrad plateau to a non-zero error ${\varepsilon_{\mathcal{S}}}$ in the score kernel. On the other hand, gradient descent converges exponentially fast, as long as the learning rate is small enough to guarantee convergence.}

\begin{figure}[ht]
    \centering
    \includegraphics[width=1\linewidth]{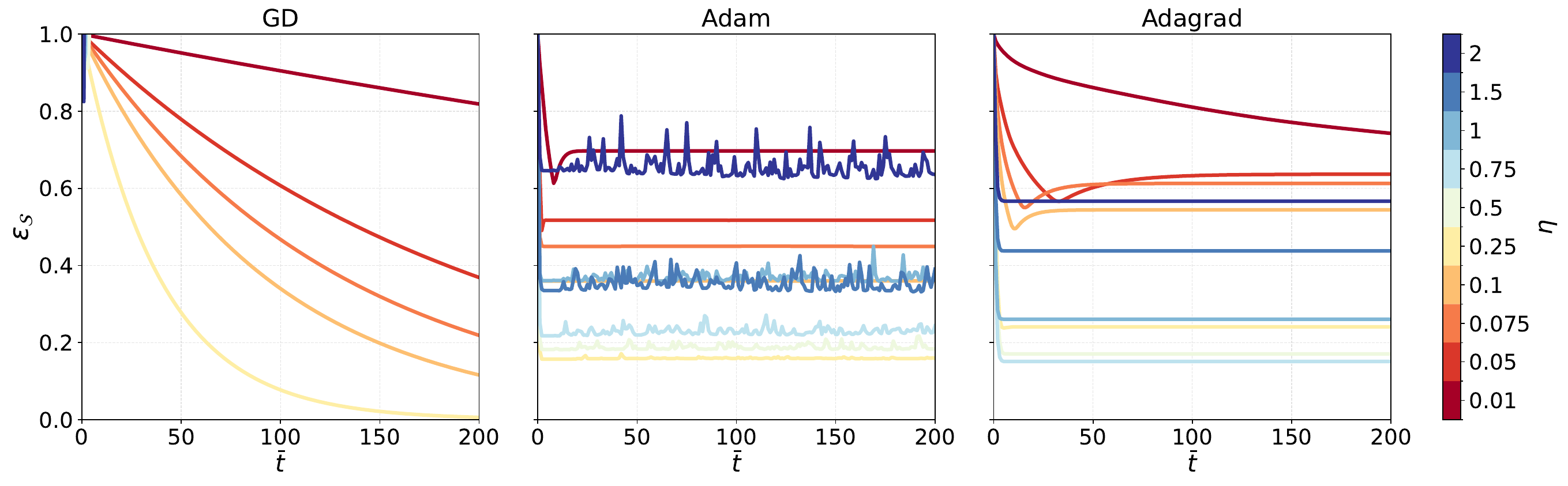}
    \caption{{Relative error in the score kernel ${\varepsilon_{\mathcal{S}}}$ as a function of the training epoch $\bar{t}$ for three optimization algorithms: GD (\textbf{left}), Adam (\textbf{center}), and Adagrad (\textbf{right}). The differnt curves correspon to different learning rates $\eta$. GD converges to the optimum, corresponding to zero error, exponentially fast, as long as the learning rate is small enough to guarantee convergence. On the other hand, both Adam and Adagrad plateau to non-zero values. \textbf{Setting}: $d = 1$, $a=1$, $t = 10^{-4}$, $m_{\mathrm{eff}}=0.05$. $M=5000$ training configurations were used, together with a batch size equal to 5000 for GD and to 100 for Adam and Adagrad. Both Adam and Adagrad use the default \texttt{PyTorch} hyperparameters.}}
    \label{fig:nonconvergence}
\end{figure}

\newpage

\subsection{Critical slowing down in configuration generation: Wasserstein error}

In this section, we consider the Wasserstein error $W_1$ (Fig.~\ref{fig:curves_wasserstein_analytical}) for the same systems as in Fig.~\ref{fig:curves_frob_analytical}. As expected, the two quantities behave qualitatively similarly.

\begin{figure}[ht]
    \centering
    \includegraphics[width=\columnwidth]{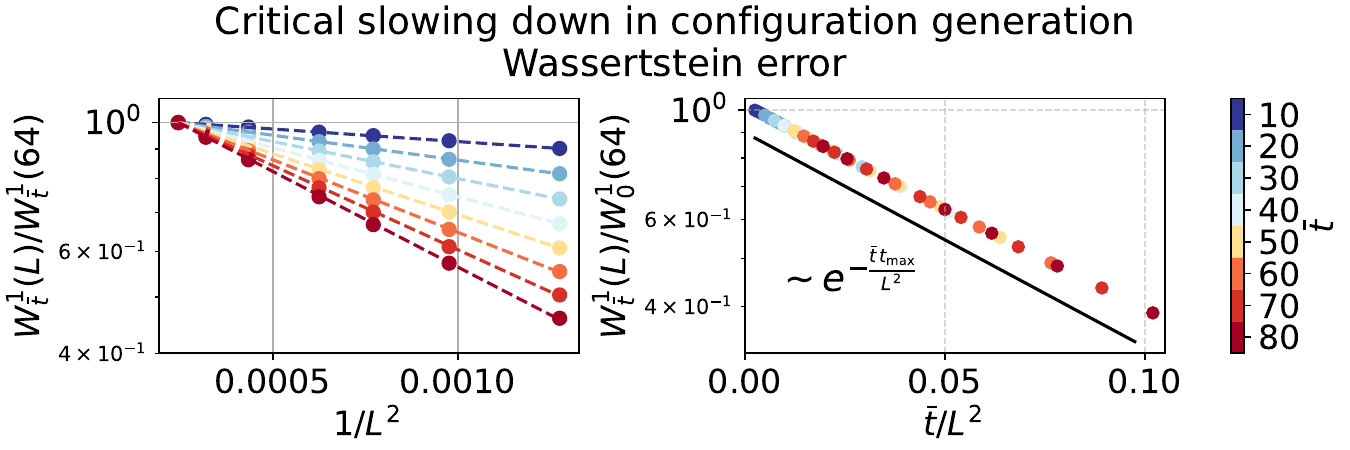}
    \caption{Same as Fig.~\ref{fig:curves_frob_analytical}, but using the Wasserstein error $W^1$ instead of the Frobenius error in the covariance matrix.  \textbf{Left:} System size evolution of the $W^1$ error, $W_1$, of the matrix for different training times rescaled by the result for the point ($\bar{t} = 10$, $L = 64$). Results are fitted to the form $Ae^{-C/L^2}$ (dotted line). \textbf{Right}: Same  results rescaled as $\bar{t}/L^2$. The ensuing collapse follows the predicted scaling $\sim e^{-\frac{\bar{t}\, t_{\mathrm{max}}}{L^2}}$ (black curve). Small discrepancies are due to the various approximations. \textbf{Setting:} $d=2$, $m_\mathrm{eff} = 0.001$, and $a = 1$. }
    \label{fig:curves_wasserstein_analytical}
\end{figure}

\newpage
\subsection{Training results for the same layer repeated twice}

In this section, we consider the case in which in the neural network architecture consists of a single layer repeated twice (Fig.~\ref{fig:1vs2layers_sim_app}), instead of two distinct layers (Fig.~\ref{fig:1vs2layers_sim}). As expected, the two models behave qualitatively similarly.

\begin{figure}[ht]
    \centering
    \includegraphics[width=\columnwidth]{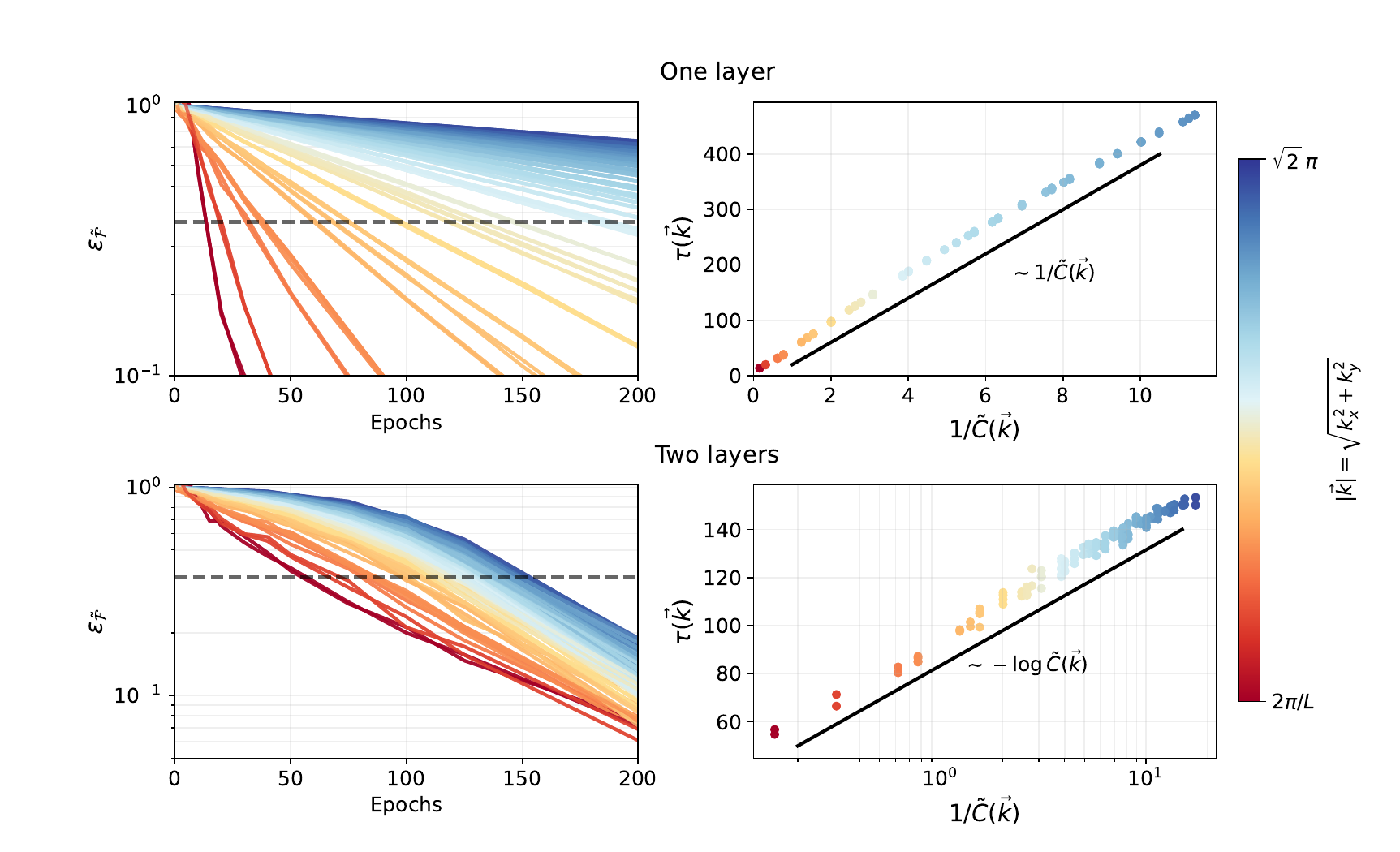}
    \caption{{Error analysis for the one-layer network architecture (\textbf{top row}) and the architecture with a single layer repeated twice (\textbf{bottom row})}. \textbf{Left column:} relative error in the real part of the score, $\varepsilon_{\widetilde{\mathcal F}}(\vec{k})$ (Eq. \eqref{eq:fourier_score_error}), for different Fourier modes as a function of training time. {The dashed black lines corresponds to an error $e^{-1} \simeq 0.37$.}  \textbf{Right column:} Characteristic decay time $\tau(\vec{k})$ of the covariance eigenvalue for different modes, $\widetilde C(\vec{k}) = 1/k^{2}$. In the one-layer case, the decay time scales linearly with the covariance eigenvalue; in the two-layer case, that scaling is logarithmic, {(as evidenced by the reference full black lines)}. \textbf{Setting:} $d=2$, $t = 0.01$, $L = 16$, and (top) $\eta = 0.01$ and (bottom) $\eta = 0.005$. Optimization is performed using gradient descent over 1024 training configurations, and the average is taken over 500 simulations.}
    \label{fig:1vs2layers_sim_app}
\end{figure}

\end{document}